\documentclass[aps,prl,10pt,twocolumn,nofootinbib,superscriptaddress]{revtex4-1}

\usepackage{amsmath}
\usepackage{amssymb}
\usepackage{mathrsfs}
\usepackage{graphicx}
\usepackage{xcolor}
\usepackage{hyperref}
\usepackage{orcidlink}
\usepackage{comment}
\usepackage{placeins}

\newcommand{\ud}{\mathrm{d}}
\newcommand{\RR}{\mathbb{R}}

\newcommand{\scL}{\mathscr{L}}

\newcommand{\mcM}{\mathcal{M}}
\newcommand{\mcP}{\mathcal{P}}
\newcommand{\mcC}{\mathcal{C}}
\newcommand{\mcD}{\mathcal{D}}
\newcommand{\mcZ}{\mathcal{Z}}
\newcommand{\mcE}{\mathcal{E}}
\newcommand{\mcB}{\mathcal{B}}
\newcommand{\ui}{\mathrm{i}}
\newcommand{\Lie}{\mathcal{L}}
\newcommand{\blue}[1]{#1}

\providecommand{\blue}[1]{\textcolor{blue}{#1}}
\providecommand{\Dl}[1]{\frac{\mathrm{D}#1}{\ud s_h}}
\providecommand{\rD}{\mathrm{D}}

\begin{document}

\title{On the existence of constants of motion for multipolar test-body dynamics}

\author{Paul Ramond\,\orcidlink{0000-0001-7123-0039}}
\email{ramond@lpccaen.in2p3.fr}
\affiliation{Universit\'{e} de Caen Normandie, ENSICAEN, CNRS/IN2P3, LPC Caen UMR6534, F-14000 Caen, France}

\begin{abstract}
We show that the conservative equations of motion of an extended test body, with force and torque sourced by spin-induced and adiabatic tidal multipoles of arbitrary order, are the flow of the Hamiltonian $H=-\tfrac12\mu^2+\mu U$, where $\mu$ is the body's kinematic mass and all multipoles enter through a covariant scalar potential $U$. For such bodies, a certain notion of mass and a spin norm are shown to be exactly conserved, at all multipolar orders and in any background. We then apply our framework to the Kerr spacetime, and construct extensions of the Carter and R\"udiger constants through hexadecapolar order (quartic in spin). They are found to exist if, and only if, the body's multipoles are those of a Kerr black hole. Together with the energy and angular momentum, they make the dynamics of a spinning test black hole in Kerr perturbatively integrable through that order. Lastly, at the next, quintic-in-spin order, an analytic criterion implemented numerically reveals that neither constant admits a continuation, for any conservative multipoles. In the conservative, test-body limit, black-hole binaries thus carry a `dynamical' hidden symmetry shared by no other compact objects, which persists to high multipolar order but ultimately breaks.
\end{abstract}

\maketitle

\emph{Introduction.---} Consider the free-falling motion of a small test body around a black hole. To a very good approximation, its trajectory is a geodesic of the Kerr spacetime \cite{Ei.15,Ke.63}. In 1968, B. Carter showed that this motion is remarkably regular, in fact \emph{integrable} in the sense of Hamiltonian mechanics, by discovering the eponymous constant of motion \cite{Carter.68}. This regularity was later traced back to a Killing-Yano (KY) tensor \cite{WalkPen.70,Floyd.73} that generates the ``hidden symmetry'' of the Kerr geometry \cite{HughSomm.73,Frolov.17}. However, a geodesic is only the leading-order approximation of the body's trajectory. Beyond it, and neglecting back-reaction effects, finite-size features like the body's spin, quadrupole and higher-order multipole moments add degrees of freedom and couple to spacetime curvature. They make the trajectory deviate from a geodesic, and can potentially lead to a loss of Carter's constant of motion and integrability. Such a loss would matter beyond the formal question of integrability: in the modeling of extreme-mass-ratio inspirals, the constants of motion of the test-body problem provide the fundamental frequencies of the multiscale expansion, the flux-balance laws that drive the inspiral, and the trajectories along which the fluxes are computed \cite{Mi.03,HinFle.08,IsoAl.19,PoWa.21}. 

Fortunately, integrability is known to survive the leading finite-size effects. At dipolar order, Rüdiger found a new constant and a generalization of Carter's \cite{Rudiger.I.81,Rudiger.II.83}, rendering the dynamics integrable at that (linear-in-spin) order \cite{KuCa.12,WitzHJ.19,SkouWitz.25,Ra.Iso.IntegO1.26}. At quadrupolar order, Compère, Druart and Vines could extend the Rüdiger and Carter constants, but only when the body's spin-induced quadrupole moment is exactly that of a Kerr black hole \cite{ComDru.22,ComDruVin.23}. Integrability of both systems was shown recently (\cite{WitzHJ.19} and \cite{Ra.CQG.24}, respectively) by formulating the dynamics in a Hamiltonian framework. In that context, it appears that Kerr's hidden symmetry still provides the constants of motion at quadrupolar order, but only when the orbiting object is also a black-hole itself. Whether this property persists beyond quadratic order in spin has since been raised repeatedly \cite{ComDruVin.23,Ra.CQG.24,Ra.Iso.Dru.IntegO2.26,BeKoLuRoTe.23,AkFeKrRuZe.25,AkFeKrSmZe.25,AkGo.25,deFiVi.26,KiLe.26,Scheopner:2026sok}. In this work, we determine to which multipolar order this property propagates, and whether black holes are indeed singled out among conservative test bodies.

Answering this question required four ingredients that are interesting and useful on their own: (i) a Hamiltonian formulation of the dynamics valid at all multipolar orders for conservative moments; (ii) a systematic way to find new constants of motion in Kerr; (iii) a way to handle the (currently) incomplete knowledge of the nonlinear multipole moments of a black hole secondary; and (iv) a method to establish that given constants of motion admit no higher-order continuation. Accordingly, our results are as follows. (i) For any conservative multipoles, in any spacetime, the laws of motion of a test body are the flow of a Hamiltonian $H=-\frac12\mu^2+\mu U$ (kinetic plus potential) on a ten-dimensional phase space with explicit Poisson brackets. It corresponds to an exactly conserved mass $m^2=\mu^2-2\mu U$, with $\mu$ the kinematic mass and $U$ the potential generating the conservative multipole moments. (ii) In Kerr, the Poisson brackets close on a finite algebra of covariant scalars built solely from the KY tensor, making the Hamiltonian calculations very easy to perform. (iii) Carter's and Rüdiger's constants extend through quartic order in spin, making the Hamiltonian integrable through that order. (iv) At the next multipolar (quintic-in-spin) order, we find that neither Carter's nor Rüdiger's constant extends, for any conservative, adiabatic, spinning test body. Throughout, we use geometric units $G=c=1$, Wald's conventions \cite{Wald} for geometry and indices, and defer technical details to the End Matter and to the Supplemental Material (Secs.~\ref{sec:conv}--\ref{sec:num}).

\emph{Relativistic dynamics.---} In a background spacetime with metric $g_{ab}$, let $\scL$ be the representative worldline of a test body with linear momentum $p_a$ and spin tensor $S^{ab}$ satisfying the condition $C^a:=S^{ab}p_b=0$ to fix $\scL$ uniquely \cite{CoNa.15,Scheopner:2026sok}. We define $\mu^2:=-p_ap^a$, $t^a:=p^a/\mu$, the spin norm $s^2:=\tfrac12S_{ab}S^{ab}$ and the spin vector $s_a:=\frac{1}{2}\varepsilon_{bcda}S^{bc}t^d$. The momenta solve the Mathisson-Papapetrou-Dixon (MPD) equations \cite{Ma.37,Pa.51,Tu.59,Di.74,Scheopner:2026sok}
\begin{equation}\label{MPD}
\nabla_v p_a = R_{abcd}S^{bc}v^d+F_a\,,\quad \nabla_vS^{ab}=2p^{[a}v^{b]}+N^{ab}\,,
\end{equation}
with $\nabla_v:=v^a\nabla_a$, $v^a$ tangent to $\scL$ associated with some (not necessarily affine) parameter $\tilde{s}$, and the force $F_a$ and torque $N^{ab}$ sourced by finite-size effects. Both are encoded into Harte's generalized force \cite{Ha.12,Ha.15}
\begin{equation}\label{genF}
\!F_a k^a + \tfrac{1}{2} N^{ab} \nabla_a k_b = \sum_{n\ge2} a_n\! \, J^{c_1\cdots c_{n-1}ac_nb}\!\Lie_k g_{ab,c_1\cdots c_n},
\end{equation}
which defines $(F_a,N^{ab})$ uniquely (\ref{sec:dyn}). In \eqref{genF}, $a_n:=2(n-1)/(n+1)!$, $k^a$ is any generalized Killing field, $g_{(n)}:=g_{ab,c_1\cdots c_n}$ are the (metric tensor) extensions \cite{Ha.12}, and $J^{(n)}:=J^{c_1\ldots c_{n-1} a c_n b}$ is the Dixon-Harte multipole moment of order $n$ (the $2^n$-pole) \cite{Di.15,Ha.15}. We consider a \emph{test} body, i.e., neglect back-reaction effects. 

\begin{figure}[t]
\centering
\includegraphics[width=\columnwidth]{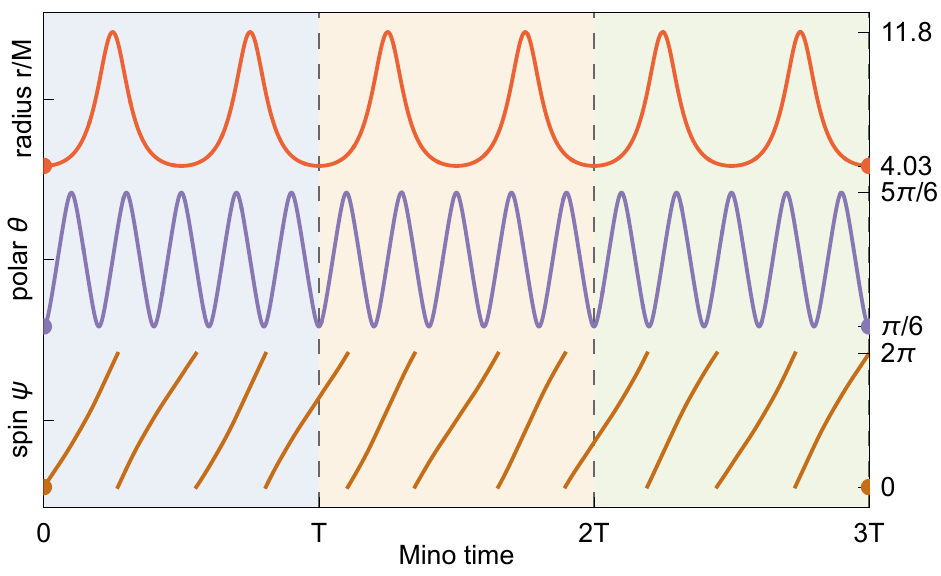}
\caption{A resonant spin-orbit solution of the test-gyroscope limit of \eqref{MPD} in Kerr with $a=0.6M$, $x=1/2$, $p\simeq6.0M$, $e\simeq0.49$, (the torus $T_2$ of Table~\ref{tab:tori}). The radius $r$, polar angle $\theta$ and spin phase $\psi$ are plotted against Mino time $\lambda$ over $3T$, one color per period $T$ (blue, orange, green). The radial and polar frequencies are in ratio $2:5$, so $r$ and $\theta$ return to their initial values after each $T$ (two radial and five polar oscillations). The spin phase advances by $11/3$ turns per $T$: after $T$ and $2T$ the orbit has closed but the spin has not (it is rotated by $4\pi/3$ and $2\pi/3$, respectively), whereas after $3T$ (11 turns) all three curves are back at their starting values (dots). Only then is the whole state, orbit and spin, closed.}
\label{fig:torus}
\end{figure}

\emph{Test-gyroscope limit.---} Setting the right-hand sides of \eqref{MPD} to zero corresponds to the test-gyroscope limit of the MPD equations, with $\scL$ a geodesic and $S^{ab}$ parallel-transported along it. In Kerr this is exactly solvable thanks to the parallel-transported Marck tetrad \cite{Marck.83,VdM.20}: the geodesic is labeled by the energy $E_0$, the axial angular momentum $L_0$ and the Carter constant $Q_0$, and, in the Marck tetrad, the spin has a constant component along one leg and rotates about it by the Marck angle $\psi$ (\ref{sec:geo}). The motion is integrable and, when quasi-periodic, admits four frequencies: radial, polar, azimuthal and spin precession (Fig.~\ref{fig:torus}). Spin and multipolar forces and torques in \eqref{MPD} perturb this order-zero case, and the question is whether constants of motion persist with them.

\emph{Conservative moments.---} We assume that the system is \emph{conservative} in the sense that the body's moments $J^{(n)}$ are encoded in a scalar potential $U$, built solely from $(p_a,S^{ab},g_{ab},\varepsilon_{abcd},g_{(n)})$, such that $J^{(n)}=-(\partial U/\partial g_{(n)})/a_n$. We grade each moment by its degree $j$ in the extensions $g_{(m)}$: moments typically used in the literature are the quadrupole and octupole ($n=2,3$), either spin-induced ($j=0$) or linear tidal ($j=1$) \cite{Hen.al.20,Ma.15,RahShaPouMat.26}, and our framework covers any $n$ and any $j$. Conversely, Euler's theorem rebuilds $U$ from the moments,
\begin{equation} \label{potential}
U=-\sum_{n\ge 2} a_n\, g_{(n)}\!\cdot \mathcal J^{(n)}\,,\quad \mathcal J^{(n)}:=\sum_j J^{(n)}_{[j]}/(j+1)\,,
\end{equation}
where $J^{(n)}_{[j]}$ is the $2^n$-pole of degree $j$ in the extensions (\ref{sec:dyn}). In practice, one either inserts a body's moments into \eqref{potential}, or, as done below, writes $U$ directly as a combination of multipole$\times$curvature couplings at a given order, the moments then following from their definition. 

\emph{Conserved mass and spin.---} In terms of $U$ in \eqref{potential}, the force and the torque of \eqref{MPD} are found to be (\ref{sec:mass}) 
\begin{equation}\label{FN}
F_a=-\nabla_aU\,,\quad N^{ab}=2\mu^{-1}(G^{[a}p^{b]}+G^{[a}{}_cS^{b]c})\,,
\end{equation}
with $G^a:=\partial(\mu U)/\partial p_a$, $G_{ab}:=2\,\partial(\mu U)/\partial S^{ab}$, and $\nabla_a U$ the variation of $U$ through the extensions only. Two mass parameters $m$ and $\nu$ emerge naturally as
\begin{equation}\label{mass}
m^2:=\mu^2-2\mu U\,,\quad \nu:=\mu-t_a G^a\,,
\end{equation}
and we fix the tangent vector $v^a$ in \eqref{MPD} by setting $v^ap_a=-\nu$. Our first result is that, for any conservative moments $J^{(n)}$, $m$ and $s$ are \emph{exactly conserved} along \eqref{MPD} (see the proof in Sec.~\ref{sec:mass}), just like the Killing charges $p_a k^a+\frac{1}{2} S^{ab} \nabla_a k_b$ for any Killing vector $k^a$ \cite{Ha.12,Di.15}. Previous studies, including our own, found only approximate conservation laws for $(m,s)$ because of a different choice of tangent vector $v^a$ in \eqref{MPD}.

\emph{Hamiltonian formulation.---} We now formulate the MPD system \eqref{MPD}-\eqref{FN} with conservative moments \eqref{potential} as a Hamiltonian system. This requires a phase space $\mcP$, a Poisson bracket $\{,\}^{\rm D}$ and a Hamiltonian $H$ defined as follows and adapted from our previous works \cite{Ra.CQG.24,Ra.NoCarter.26,Ra.Iso.Dru.IntegO2.26}. First, one introduces a 14-dimensional phase space $\mcM \ni(x^\alpha,p_\alpha,S^{\alpha\beta})$ with degenerate SO(1,3) Poisson brackets
$\{x^\alpha,p_\beta\}=\delta^\alpha_\beta$, $\{p_\alpha,p_\beta\}=R_{\alpha \gamma \delta \beta }S^{\gamma\delta}$, $\{p_\alpha,S^{\beta\gamma}\} = 2\Gamma^{[\gamma}_{\alpha\delta} S^{\beta]\delta}$, $\{S^{\alpha\beta},S^{\gamma\delta}\}=4g^{[\alpha[\delta}S^{\gamma]\beta]}$.
Second, the structure is restricted to the 10-dimensional symplectic submanifold $\mcP$ where $p_\alpha S^{\alpha\beta}=0$ holds identically and $s^2$ is fixed (as a Casimir of $\{,\}$), via the Dirac procedure \cite{Bro.22,Ra.Iso.IntegO1.26}. The resulting brackets on $\mcP$ are
\begin{equation}\label{Dirac}
\{F,G\}^{\rm D}:=\{F,G\}- (\mcD s^2)^{-1} \{F,C^\alpha\}S_{\alpha\beta}\{C^\beta,G\},
\end{equation}
with $\mcD := \mu^2 + \frac14 R_{abcd} S^{ab} S^{cd}$. Then, the Hamiltonian
\begin{equation}\label{H}
H:=-\tfrac12\mu^2+\mu U,
\end{equation}
along with Hamilton's equations $\ud F/\ud s_h=\{F,H\}^{\rm D}$ describe the MPD system \eqref{MPD}-\eqref{FN} exactly, with the unique evolution parameter $s_h$ such that $\ud\tilde s=\mu\,\ud s_h$ (\ref{sec:ham}). The Hamiltonian \eqref{H} has a familiar ``kinetic + potential'' form: $-\mu^2/2$ generates the universal, pole-dipole dynamics (spin-curvature term and hidden momentum) \cite{CoNa.15,WiStLu.19,Ra.Iso.IntegO1.26}, while $\mu U$ encodes all body-specific multipole$\times$curvature couplings and generates all forces and torques of \eqref{FN}. This framework therefore describes any conservative multipolar model of the class considered here, on any fixed background (no back-reaction), and makes the standard tools of Hamiltonian mechanics available, such as the Arnold-Liouville theorem \cite{Arn} and resonant averaging \cite{Poin}, both used below.

\emph{Order-counting of multipoles.---} A body is specified by its conservative moments $J^{(n)}_{[j]}$ via \eqref{potential}, but the extensions $g_{(n)}$ in \eqref{FN} mix derivatives and products of the curvature (e.g., $g_{(4)}\sim\nabla\nabla R+R^2$). A systematic, consistent multipole-counting scheme is thus required. We label each multipole$\times$curvature coupling by its order $N$, defined as the number of derivatives of the metric it contains, and write it with the power of $s^2$ that makes it of degree $N$ in the spin, so that $N$ is also the order $O(S^N)$ in the spin used throughout (Sec.~\ref{sec:couplings}). For instance, $\mcE_{ss}:=\mcE_{ab}s^as^b$, with $\mcE_{ab}:=R_{acbd}t^ct^d$ the electric part of the curvature (and $\mcB_{ab}:=R^\star_{acbd}t^ct^d$ its magnetic part), contains two derivatives of the metric, so $N=2$, and it is already quadratic in the spin; $\mathrm{tr}\,\mcE^2:=\mcE_{ab}\mcE^{ab}$ contains four, so $N=4$, but it is of degree $0$ in the spin, so it is written $s^4\,\mathrm{tr}\,\mcE^2$. Similarly, the usual tidal quadrupole and octupole \cite{Hen.al2.20} are $N=4$ and $N=6$ multipoles, respectively. Grading by $N$ makes the list of multipole$\times$curvature couplings at each multipolar order finite, whereas grading by, say, the spin power alone never closes, as any tidal multipole could be added to it (Sec.~\ref{sec:couplings}).

\emph{A basis for scalars.---}
We now apply our Hamiltonian formalism to a Kerr background with parameters $(M,a)$ and with Boyer-Lindquist coordinates $x^\alpha=(t,r,\theta,\phi)$. The Hamiltonian itself and the Killing charges of the stationary Killing vector $\xi^a$ and of the axial Killing vector of Kerr provide three constants of motion $(H,E,L)$, exactly conserved at any multipolar order \cite{Di.64,Di.74,Ra.Iso.IntegO1.26}. Since $\mcP$ is 10-dimensional, two extra Poisson-commuting constants of motion (denoted $Q$ and $K$ below) must be found to ensure integrability. We look for them order by order in the multipole expansion index $N$ defined above. Finding constants of motion in Hamiltonian mechanics is a difficult problem with no generic strategy, but Kerr's hidden symmetry not only yields unexpected constants of motion; it also provides a systematic way to search for them at high multipolar order. We build on the pioneering work \cite{ComDruVin.23} to exploit the fact that any geometric tensor in Kerr can be expressed solely in terms of the KY tensor $f_{ab}$ \cite{Ra.Iso.Dru.IntegO2.26}, or, more practically, three intermediary (complex) tensors $\mcC,\xi^a$ and $Z^{ab}$.\footnote{$\mcC$ is a power of the Weyl scalar and $\mcC Z_{ab}$ the complex conformal KY tensor.} All three descend from $f_{ab}$ alone \cite{Ra.Iso.Dru.IntegO2.26}, and the metric, Levi-Civita tensor and Riemann tensor are all algebraic in them (e.g., $g_{ab}=Z_{ac}Z^{c}_{\phantom{c}b}$), while their derivatives close on themselves (e.g., $\nabla_a\mcC = -\ui Z_{ab}\xi^b$), see details in  Sec.~\ref{sec:kerr}. 
Contracting $(\xi_a,Z_{ab})$ with the body's momenta $(t^a,s^a)$ in all possible ways gives nineteen scalars. Any \emph{scalar} function on $\mcP$ (masses, Hamiltonians, potentials, constants of motion) is a \emph{rational combination} of them (Sec.~\ref{sec:kerr}), and the Poisson bracket \eqref{Dirac} of any two of them closes on them. For example,
\begin{equation}\label{CCbar}
    \{\mathcal{C} , \bar{\mathcal{C}}\}^{\rm D} = \ui\,\mathcal{D}^{-1}\left(E_s\,\mcZ_{t\xi}-E_0\,\mcZ_{s\xi}\right),
\end{equation}
with $E_0=-\xi^a t_a$, $E_s=-\xi^a s_a$, $\mcZ_{ab}:=Z_{ac}\bar Z^c_{\phantom{c}b}$, and subscripts denoting contractions, e.g., $\mcZ_{s\xi}:=\mcZ_{ab}s^a\xi^b$. 
A constant of motion at each multipolar order can thus be searched for as an unknown combination of monomials of this algebra of scalars, and the equations to solve, namely $\{K,H\}^{\rm D}=\{Q,H\}^{\rm D}=\{K,Q\}^{\rm D}=0$, once expanded by the Leibniz rule and brackets like \eqref{CCbar}, become a \emph{linear system} for their unknown Ansatz coefficients. The Ansatz space is made tractable thanks to a fine analysis of the powers allowed by mass and spin gradings. We performed all this symbolically and then solved the linear systems exactly by modular arithmetic (mainly to avoid a blow-up of coefficients) at random rational phase-space points with rational reconstruction, as detailed in Sec.~\ref{sec:KQ}.

\emph{Solution at $O(S^3)$.---} At orders $N=2$ and $3$, only one multipole coupling is possible: the well-known spin-induced quadrupole and octupole, respectively \cite{Ma.15}. In \eqref{H}, they are generated by the following potentials 
\begin{equation}\label{Ukerr}
U_2=-\frac{\kappa_2}{2\mu}\,\mcE_{ab}s^as^b,\quad U_3=\frac{\kappa_3}{6\mu^2}\,\nabla_a \mcB_{bc}s^a s^b s^c\,,
\end{equation}
where the coupling constants $\kappa_2,\kappa_3$ depend on the body's internal properties (equation of state, microphysics, etc.). For a Kerr black hole, $\kappa_2=\kappa_3=1$ \cite{ScheopnerVines.24}. 

At $N=2$, the system is known to be integrable if $\kappa_2=1$ \cite{Ra.CQG.24}, thanks to two constants of motion: Rüdiger's constant $K_{\leq2}$ and a (generalized) Carter constant $Q_{\leq2}$, both discovered in \cite{ComDruVin.23}.
At $N=3$ no such result was known. As explained above, in our framework, looking for cubic-in-spin corrections $K_3$ and $Q_3$ to the quadrupole constants boils down to solving a linear system. We found a solution if and only if $\kappa_3=1$. The Rüdiger constant at this order reads $K_{\leq3}:=K_{\leq2}+K_3$ with $K_{\leq2}=\mathrm{Re}(2\bar B\bar\mcC)$ and a cubic correction 
\begin{equation}\label{Kstructure}
K_3:=\frac{2M}{\mu^2}\,\mathrm{Im}\!\left[\frac{B(B^2-s^2)}{\mcC^2}+\frac{B\bar\mcC(B\bar B-\mcZ_{ss})}{2\mcC^3}\right], 
\end{equation}
where $B:=Z_{ab}s^at^b$ and $\mcZ_{ss}:=\mcZ_{ab}s^{a}s^{b}$. The Carter constant $Q_{\leq 3}:=Q_{\leq 2}+Q_3$ admits a similar form (see End Matter). We verified explicitly that $\{K_{\leq 3},Q_{\leq 3}\}^{\rm D}=0$ at this order, and more generally that $(H,E,L,Q_{\leq 3},K_{\leq 3})$ are functionally independent and in involution. This guarantees integrability at octupolar order, for a test black hole in Kerr.

\emph{Couplings at $O(S^4)$.---}
At $N=4$ the situation changes. 
The extension $g_{(4)}$ in \eqref{genF} contains both $\nabla\nabla R$ and $R\,R$. As a result, two different moments yield  multipole$\times$curvature couplings quadratic in the curvature: the spin-induced hexadecapole, through the $R\,R$ part of $g_{(4)}$ (a form of spin-dependent tidal response), and the tidal quadrupole, linear in $R$ and contracted with $g_{(2)}\sim R$. The split between spin-induced and tidal is mere bookkeeping, because the laws of motion only see the  multipole$\times$curvature couplings in \eqref{genF}, not the moments alone. We resolve this ambiguity by listing and including \emph{all} possible conservative multipole$\times$curvature couplings of order $N=4$.

These couplings are the covariant scalars built from $R_{abcd}$, $\nabla_a$, $t^a$, $s^a$ and $\varepsilon_{abcd}$ that are non-dissipative (even under time reversal). They reduce to two linear-in-curvature invariants, $\ell_1=\nabla_a\!\nabla_b\mcE_{cd}s^as^bs^cs^d$ and $\ell_2=s^2\nabla_a\!\nabla_b\mcE_{cd}s^cs^d\,t^at^b$, and six quadratic ones, $I_{1,2,3}=\mcE_{ss}^2,\ s^2(\mcE^2)_{ss},\ s^4\mathrm{tr}\mcE^2$ and $I_{4,5,6}$, the same with $\mcB$ in place of $\mcE$. Among them, $I_3$ and $I_6$ are the well-known static tidal quadrupole, $\lambda_2^{\mcE}\,\mathrm{tr}\mcE^2$ and $\lambda_2^{\mcB}\,\mathrm{tr}\mcB^2$ (with $s^2$ conserved), and the other $I_i$ are spin-dependent tidal responses (Sec.~\ref{sec:couplings}). The most general conservative Hamiltonian at this order is thus $H=-\tfrac12\mu^2+\mu\,(U_2+U_3+U_4)$ with $(U_2,U_3)$ in \eqref{Ukerr} and hexadecapolar potential
\begin{equation}\label{H4}
U_4=\frac{\kappa_4}{24\mu^3} \nabla_{a}\!\nabla_b \mcE_{cd}s^as^bs^cs^d +\frac{1}{\mu^3}\Big(\lambda_4\ell_2+\sum_{i=1}^{6}b_iI_i\Big),
\end{equation}
with eight coupling parameters $(\kappa_4,\lambda_4,b_1,\ldots,b_6)$. For a Kerr secondary, the linear-in-curvature couplings $\kappa_2=\kappa_3=\kappa_4=1$ and $\lambda_4=-\tfrac1{40}$ are fixed by Dixon moments \cite{ScheopnerVines.24}, and the $b_i$ by other means, as discussed below. 

\emph{Solution at $O(S^4)$.---} As before, we looked for quartic corrections $(K_4,Q_4)$ to the cubic Rüdiger and Carter constants, now leaving the eight couplings $(\kappa_4,\lambda_4,b_1,\ldots,b_6)$ free. The conditions $\{K_{\leq4},H\}^{\rm D}=\{Q_{\leq4},H\}^{\rm D}=0$ form a linear system in the Ansatz coefficients and the couplings, and within our Ansatz, these two brackets and $\{K_{\leq4},Q_{\leq4}\}^{\rm D}$ all vanish if and only if
\begin{equation}\label{O4result}
\kappa_4=1\,,\quad b_1=\cdots=b_6=0\,,
\end{equation}
and arbitrary $\lambda_4$. The coupling $\ell_2$ is a total derivative along the unperturbed flow (Sec.~\ref{sec:couplings}), so it does not affect integrability at this order. See End Matter for the expressions of ($H$, $K$, $Q$).

The conditions \eqref{O4result} single out a unique integrable test body at this order with the spin-induced hexadecapole $\ell_1$ (the first term in \eqref{H4} with $\kappa_4=1$) and no tidal response (vanishing $b_1,\ldots,b_6$). By matching with the literature,\footnote{The couplings $b_1,\ldots,b_6$ are Wilson coefficients of the dynamical mass function \cite{ScheopnerVines.24}. Spin exponentiation of the Compton amplitude \cite{GuOcVi.19} sets them all to zero, whatever the value of the coefficient $E_{4b}$, which the amplitude does not determine; shift symmetry, Teukolsky matching \cite{BGKV.23} and effective field theory \cite{BeSh.23} agree (Sec.~\ref{sec:couplings}).} we found that the couplings \eqref{O4result} are exactly those expected for a test Kerr black hole at quartic order in spin. Kerr is thus, once again, ``selected'' by integrability, even at this order involving non-linearities in curvature.

\emph{Failure at $O(S^5)$.---} At $N=5$, the  multipole$\times$curvature couplings are linear in $\nabla^3R$ (spin-induced $2^5$-pole) or bilinear in $(R,\nabla R)$ (tidal responses mixing spin quadrupole and octupole). Again, we consider all possible cases as for $N=4$ and reduce them, leaving a complete 24-dimensional space of coefficients $c_i$ of multipole$\times$curvature couplings $X_i$ (see End Matter). The Kerr black hole lies at a currently unsettled point of this space: its Dixon moments fix the linear couplings (among which $\kappa_5=1$) \cite{ScheopnerVines.24}, but the principles that fix the quadratic ones at $N=4$ are no longer mutually consistent at $N=5$ \cite{BGKV.23,ScheopnerVines.24,Scheopner:2026sok}. Regardless, our method found no combination of parameters such that $K_5$ exists. But this search relies on an Ansatz. We therefore devised a test that requires none and applies to both $K$ and $Q$: it is inspired by Poincar\'e's resonant obstruction \cite{Poin,Kozlov.83} and described below.

\emph{Resonant averages.---} Writing $Q_{\leq 5}=Q_{\le4}+Q_5$ and using integrability at lower orders, the $O(S^5)$ part of $\{Q_{\leq 5},H_{\leq 5}\}^{\rm D}=0$ reads
\begin{equation}\label{O5cond}
\{Q_5,\!H_0\}^{\rm D}_5+ \underbrace{\{Q_{\le4},\!H_{\le4}\}^{\rm D}_5+\sum_{i=1}^{24}c_i\{Q_{\le4},\!X_i\}^{\rm D}_5 }_{:=\,\sigma_5} =0,
\end{equation}
where $\{\cdot,\cdot\}^{\rm D}_5$ denotes the $O(S^5)$ part of the Poisson bracket \eqref{Dirac}, and $H_{\le 4}$ is the integrable Hamiltonian with couplings \eqref{O4result} and $H_5:=\sum_{i=1}^{24}c_iX_i$. Since $H_0$ does not depend on the spin, the first term in \eqref{O5cond} is $\ud Q_5/\ud s_h$ along the $O(S^0)$-\emph{unperturbed motion} $\scL_0$, a Kerr geodesic with parallel-transported spin, as depicted in Fig.~\ref{fig:torus}. If $Q_5$ exists, Eq.~\eqref{O5cond} holds at every point of $\mcP$; but $Q_5$ is unknown, so it cannot be checked directly. We remove it by averaging: along a \emph{closed} orbit of $\scL_0$ (i.e., one whose full position-momentum-spin state returns to its initial value), $\{Q_5,H_0\}^{\rm D}_5=\ud Q_5/\ud s_h$ averages to zero whatever $Q_5$, and the remaining condition, $\langle\sigma_5\rangle=0$, is necessary and no longer involves $Q_5$.

\emph{Closed orbits.---} Building closed orbits for the average takes two steps. First, we take a resonant geodesic, such that $n_r\Lambda_r=n_\theta\Lambda_\theta=:T$ in Mino time $\lambda$, with integers $(n_r,n_\theta)$ and radial and polar periods $(\Lambda_r,\Lambda_\theta)$: position and momentum then return to their initial values after $T$.
Second, as the spin rotates about one leg of Marck's tetrad with angle $\psi$, after $T$ it is rotated by some $2\pi w$, with $w$ the fractional part of $T/\Lambda_\psi$ and $\Lambda_\psi$ the mean precession period. We then move along the $n_r:n_\theta$ resonance (tuning eccentricity $e$ and semi-latus rectum $p$) until $w$ is rational with denominator $q$. The full orbit (including spin) is then closed after $qT$ (in Fig.~\ref{fig:torus}, $(w,q)=(2/3,3)$). 

\emph{Non-existence criterion.---} Each closed orbit lies on a \emph{resonant torus} of phase space. Starting at $t=\phi=0$ at periapsis, it is labeled by its initial polar and spin phases $(\varphi_0,\psi_0)$ (Sec.~\ref{sec:geo}). Rather than integrating over the full period $qT$, separately for each $\psi_0$, we integrate over a single period, at spin order $n$,
\begin{equation}\label{harm}
\langle \sigma_n\rangle(\varphi_0,\psi_0):=\frac1T\int_0^T\!\sigma_n\,\Sigma\,\ud\lambda,
\end{equation}
with $\sigma_n$ the analogue of $\sigma_5$ and $\Sigma:=r^2+a^2\cos^2\theta$ so that $\Sigma\,\ud\lambda=\mu\,\ud s_h$, and treat $\psi_0$ exactly. Since $\sigma_n$ has degree $n$ in the spin, which rotates by $\psi$ in Marck's tetrad, one has $\langle\sigma_n\rangle=\sum_{|m|\le n}\langle\sigma_n\rangle_m(\varphi_0)\,e^{\ui m\psi_0}$ (Sec.~\ref{sec:num}). After a time $jT$, the state is that of the orbit started at $\psi_0+2\pi jw$, so the average over the closed orbit, $\frac1q\sum_{j=0}^{q-1}\langle\sigma_n\rangle(\varphi_0,\psi_0+2\pi jw)$, keeps only the harmonics with $mw\in\mathbb Z$. Requiring it to vanish for every $\psi_0$ gives our final necessary condition for the existence of a constant of motion:
\begin{equation}\label{test}
\langle\sigma_n\rangle_m(\varphi_0)=0\quad\text{for all $(m,\varphi_0)$ with } mw\in\mathbb Z\,,
\end{equation}
i.e.\ $m$ a multiple of $q$ ($m=0,\pm3$ in Fig.~\ref{fig:torus}).

\emph{Numerical implementation.---} In practice, we use the BHP Toolkit \cite{BHPToolkit} for geodesics and the analytic ODE for the Marck angle \cite{VdM.20} to tune the closed orbits and evaluate the necessary integrals over a single period $[0,T]$ like \eqref{harm}, and then look at the spin harmonics of the result. We evaluate \eqref{harm} in high-precision arithmetic on twelve resonant tori (Table~\ref{tab:tori}), with varying Kerr parameter $a$, a variety of orbits (prograde, retrograde and inclined), five orbital resonances and six values of $w$, sampling $\varphi_0$ on each. For each torus, order $n$ and constant of motion, we condense the conditions \eqref{test} into a single number $\mathcal R_n$, their root-mean-square relative to the mean of $|\sigma_n\Sigma|$ (see End Matter for definitions and explanations). If a constant of motion exists, $\mathcal R_n$ vanishes up to numerical accuracy, the latter measured on each torus with a combination of the sources whose average should vanish identically. If not, $\mathcal R_n$ is nonzero, though generally $\ll1$ since the oscillating integrand still largely cancels: the obstruction of a constant of motion thus shows as a value of $\mathcal{R}_n$ many decades above the numerical zero. Lastly, since $\sigma_n$ is affine in the couplings, so are its amplitudes: on enough closed orbits, \eqref{test} is an overdetermined linear system for the $N=5$ couplings $c_1,\ldots,c_{24}$, and if it has no solution, no coupling combination yields a continuation of $Q$ or $K$.

\begin{figure*}[ht!]
\centering
\includegraphics[width=\textwidth]{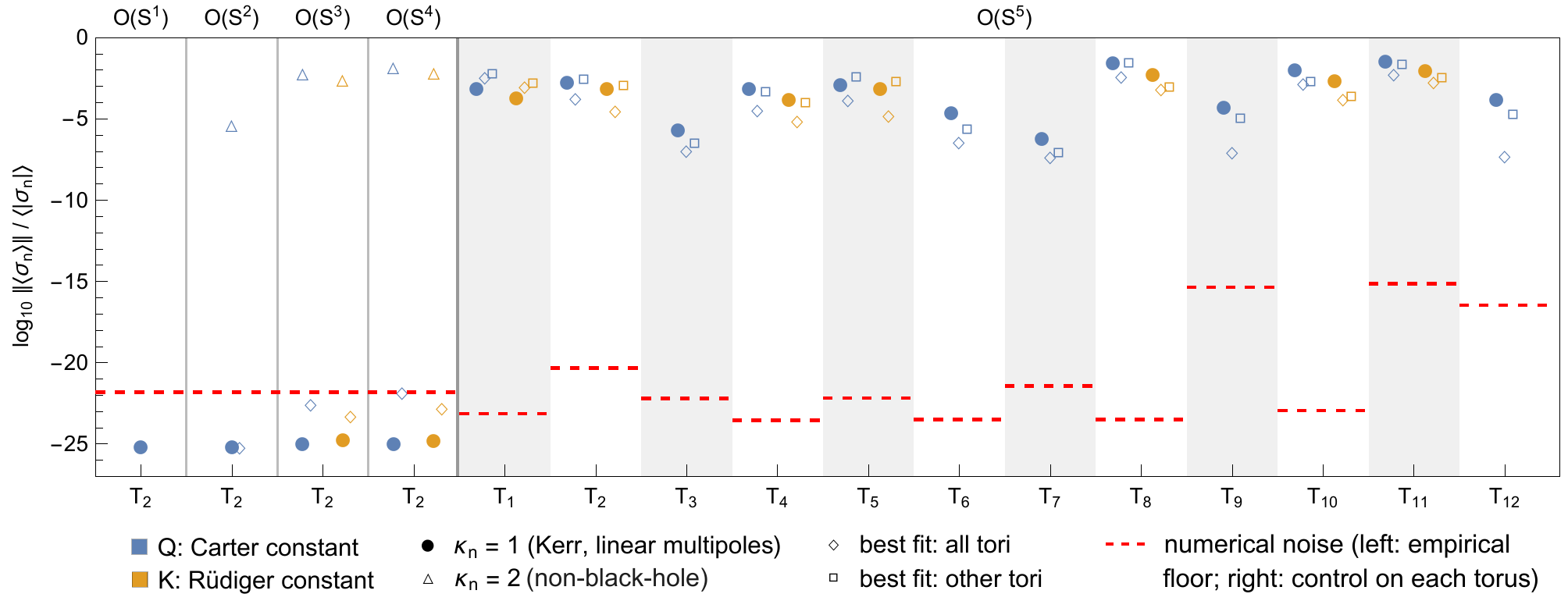}
\caption{Relative size $\mathcal R_n$ of the resonant averages \eqref{test} for $Q$ (blue) and $K$ (orange), plotted as $\log_{10}\mathcal R_n$: the root-mean-square of the conditions over the sampled orbits, relative to the typical size of the integrand (see details in End Matter).
\emph{Left:} $O(S^1)$--$O(S^4)$ on the torus $T_2$ of Fig.~\ref{fig:torus}. Circles: black-hole couplings, $\kappa_n=1$. Triangles: one non-black-hole coupling, $\kappa_n=2$. Diamonds: least-squares refit of the couplings to these non-black-hole values.
\emph{Right:} $O(S^5)$ on the twelve tori $T_1,\dots,T_{12}$ of Table~\ref{tab:tori}. Circles: Kerr's linear couplings alone ($\kappa_5=1$, all other couplings zero). Diamonds: best fit of the 24 couplings to all tori. Squares: the same fit with the plotted torus left out. $K$ is omitted where it is insensitive to the resonant harmonics. Dashed red lines: numerical zero, i.e., the smallest $\mathcal R_n$ the computation resolves. Were $Q$ or $K$ to extend for some couplings, every single diamond and square on the right would fall on or below the red lines, as they do on the left.}
\label{fig:data}
\end{figure*}

\emph{Benchmark through $O(S^{4})$.---} We first benchmarked the method at the lower orders $n\le4$, known to be integrable for the black-hole couplings, on the torus of Fig.~\ref{fig:torus}. On the left of Fig.~\ref{fig:data}, one can see that the test is \emph{specific} (no false alarms): for $\kappa_n=1$, $\mathcal R_n$ sits at the numerical zero, $10^{-25}$ (circles); and \emph{sensitive} (it detects an obstruction): setting $\kappa_n=2$ raises $\mathcal R_n$ by about twenty decades (triangles). Moreover, treating $\kappa_2,\kappa_3,\kappa_4$ as unknowns and starting from the (non-integrable) values $\kappa_n=2$, a least-squares fit of the resonant averages finds the values that make them vanish: it recovers $\kappa_n=1$ to within $10^{-20}$, and returns $\mathcal R_n$ to the numerical zero (diamonds). In other words, the test confidently reports when a constant does not exist, and when it does, a fit can find it.

\emph{Results at $O(S^5)$.---} Applied at $O(S^5)$, the natural choice of Kerr's linear couplings alone ($\kappa_5=1$ and all others zero) gives $\mathcal R_5$ between $10^{-6.2}$ and $10^{-1.4}$ depending on the tori and the constant tested (Fig.~\ref{fig:data}, right). Fitting all 24 couplings for $Q$ and $K$ simultaneously on all tori gains little, with $\mathcal R_5$ between $10^{-7.3}$ and $10^{-2.2}$ (diamonds), i.e., at the level of the non-black-hole benchmark. Fitting the couplings to $Q$ or $K$ separately, or to any combination thereof, gives the same conclusion (Sec.~\ref{sec:num}). Couplings fitted on eleven tori fail on the twelfth too (squares). Every tested case lies at least fourteen decades above the integrable benchmark and at least eight above the numerical zero of its own torus.\footnote{And at least fourteen decades above it on all but three tori $(T_9,T_{11},T_{12})$, where it is higher (Sec.~\ref{sec:num}).} Hence, within our numerical precision, no Hamiltonian of the family makes all resonant averages vanish: neither the Carter nor the Rüdiger constant, nor any function of them, extends to $O(S^5)$, for any spinning, conservative, adiabatic body. 

\emph{Schwarzschild limit.---} In Schwarzschild ($a=0$), we found that $Q_{\le4}=-L^2$, with $L^2$ the squared total angular momentum, conserved by spherical symmetry\footnote{Since $L^2$ is quadratic in the spin, $Q_{\le2}=-L^2$ and $Q_3=Q_4=0$ identically (Sec.~\ref{sec:schw}).}, so that $Q$ extends at every order. As an additional consistency check, the source $\sigma_5$ of $Q$ computed by our pipeline vanishes at every point of the Schwarzschild orbits, for any couplings. For $K$, we repeated the full resonant analysis on three Schwarzschild tori (Table~\ref{tab:schw}), with the same 24 couplings and fits as for Kerr: $\mathcal R_5$ lies between $10^{-6.8}$ and $10^{-3.2}$, at least seventeen decades above the numerical zero of each torus. Hence $K$ does not extend to $O(S^5)$ in Schwarzschild either, whatever the couplings (Sec.~\ref{sec:schw}).

\emph{Interpretation.---} For generic non-black-hole bodies, the Carter and Rüdiger constants fail. This is somewhat expected, since the multipoles of such bodies are not universal and depend on the equation of state. That they survive for a black hole through $O(S^4)$, at a unique point of the coupling space, calls instead for a finer understanding of the symmetries at play, in line with the spin-shift symmetry of Kerr amplitudes \cite{BeKoLuRoTe.23,BGKV.23}, and with asymptotic integrability \cite{AkGo.25}, both of which hold through quartic order and break beyond. At $O(S^5)$, their absence even for a black hole points either to the genuine breaking of a symmetry, or to a model of the secondary that is not realistic enough. The latter is supported by the Kerr Compton amplitude, which at that order involves absorptive, horizon-dependent terms \cite{BGKV.23,ScheopnerVines.24}, beyond the reach of any conservative, adiabatic model like ours. This possibility could be tested within the Hamiltonian formulations of non-conservative dynamics \cite{Galley.13,Aykroyd:2025yzd}, which accommodate such effects.

\emph{Applications.---} The Hamiltonian \eqref{H} and the exact conservation of $m$ and $s$ apply to any study of the MPD equations with conservative finite-size effects, on any background: they replace the approximate conservation laws used so far by exact ones, and make symplectic integrators directly applicable \cite{Sey.Lu.12,Wu.al.21}. In Kerr, through hexadecapolar order (quartic-in-spin), the explicit constants $(H,E,L,Q,K)$ given here make the full machinery of integrable Hamiltonian systems available for a test, spinning, extended black hole orbiting the arbitrarily strong field of a Kerr background: action-angle variables and fundamental frequencies in closed form \cite{Arn,Ra.Iso.IntegO1.26,WitzHJ.19,SkouWitz.25}, and analytic solutions of the equations of motion, all including quartic-in-spin effects in full general relativity. In turn, these will provide valuable strong-field data for exact and hybrid waveform models of compact binaries \cite{HinFle.08,PoWa.21,LISAWHITE.23,Sk.26}, new high-order post-Newtonian coefficients and matching information for the calibration of phenomenological models, and a first law of binary mechanics at this order \cite{Le.al.12,Bl.al.13,ALT.14,Le.15,RaLe.22,GoLePo.25}. At the dawn of the next generation of gravitational wave observatories \cite{LISAWHITE.23, Mag.al.20} that will focus on finite-size effects and Kerr uniqueness tests, the aforementioned results are timely.

\bibliography{ListeRef.bib}

\begin{thebibliography}{76}%
\makeatletter
\providecommand \@ifxundefined [1]{%
 \@ifx{#1\undefined}
}%
\providecommand \@ifnum [1]{%
 \ifnum #1\expandafter \@firstoftwo
 \else \expandafter \@secondoftwo
 \fi
}%
\providecommand \@ifx [1]{%
 \ifx #1\expandafter \@firstoftwo
 \else \expandafter \@secondoftwo
 \fi
}%
\providecommand \natexlab [1]{#1}%
\providecommand \enquote  [1]{``#1''}%
\providecommand \bibnamefont  [1]{#1}%
\providecommand \bibfnamefont [1]{#1}%
\providecommand \citenamefont [1]{#1}%
\providecommand \href@noop [0]{\@secondoftwo}%
\providecommand \href [0]{\begingroup \@sanitize@url \@href}%
\providecommand \@href[1]{\@@startlink{#1}\@@href}%
\providecommand \@@href[1]{\endgroup#1\@@endlink}%
\providecommand \@sanitize@url [0]{\catcode `\\12\catcode `\$12\catcode `\&12\catcode `\#12\catcode `\^12\catcode `\_12\catcode `\%12\relax}%
\providecommand \@@startlink[1]{}%
\providecommand \@@endlink[0]{}%
\providecommand \url  [0]{\begingroup\@sanitize@url \@url }%
\providecommand \@url [1]{\endgroup\@href {#1}{\urlprefix }}%
\providecommand \urlprefix  [0]{URL }%
\providecommand \Eprint [0]{\href }%
\providecommand \doibase [0]{http://dx.doi.org/}%
\providecommand \selectlanguage [0]{\@gobble}%
\providecommand \bibinfo  [0]{\@secondoftwo}%
\providecommand \bibfield  [0]{\@secondoftwo}%
\providecommand \translation [1]{[#1]}%
\providecommand \BibitemOpen [0]{}%
\providecommand \bibitemStop [0]{}%
\providecommand \bibitemNoStop [0]{.\EOS\space}%
\providecommand \EOS [0]{\spacefactor3000\relax}%
\providecommand \BibitemShut  [1]{\csname bibitem#1\endcsname}%
\let\auto@bib@innerbib\@empty
\bibitem [{\citenamefont {Einstein}(1915)}]{Ei.15}%
  \BibitemOpen
  \bibfield  {author} {\bibinfo {author} {\bibfnamefont {A.}~\bibnamefont {Einstein}},\ }\href@noop {} {\bibfield  {journal} {\bibinfo  {journal} {Sitzber. Preuss. Akad. Wiss.}\ ,\ \bibinfo {pages} {831}} (\bibinfo {year} {1915})}\BibitemShut {NoStop}%
\bibitem [{\citenamefont {Kerr}(1963)}]{Ke.63}%
  \BibitemOpen
  \bibfield  {author} {\bibinfo {author} {\bibfnamefont {R.~P.}\ \bibnamefont {Kerr}},\ }\href@noop {} {\bibfield  {journal} {\bibinfo  {journal} {Phys. Rev. Lett.}\ }\textbf {\bibinfo {volume} {11}},\ \bibinfo {pages} {237} (\bibinfo {year} {1963})}\BibitemShut {NoStop}%
\bibitem [{\citenamefont {Carter}(1968)}]{Carter.68}%
  \BibitemOpen
  \bibfield  {author} {\bibinfo {author} {\bibfnamefont {B.}~\bibnamefont {Carter}},\ }\href@noop {} {\bibfield  {journal} {\bibinfo  {journal} {Physical Review}\ }\textbf {\bibinfo {volume} {174}},\ \bibinfo {pages} {1559} (\bibinfo {year} {1968})}\BibitemShut {NoStop}%
\bibitem [{\citenamefont {Walker}\ and\ \citenamefont {Penrose}(1970)}]{WalkPen.70}%
  \BibitemOpen
  \bibfield  {author} {\bibinfo {author} {\bibfnamefont {M.}~\bibnamefont {Walker}}\ and\ \bibinfo {author} {\bibfnamefont {R.}~\bibnamefont {Penrose}},\ }\href@noop {} {\bibfield  {journal} {\bibinfo  {journal} {Communications in Mathematical Physics}\ }\textbf {\bibinfo {volume} {18}},\ \bibinfo {pages} {265} (\bibinfo {year} {1970})}\BibitemShut {NoStop}%
\bibitem [{\citenamefont {Floyd}(1973)}]{Floyd.73}%
  \BibitemOpen
  \bibfield  {author} {\bibinfo {author} {\bibfnamefont {R.}~\bibnamefont {Floyd}},\ }\href@noop {} {\bibfield  {journal} {\bibinfo  {journal} {PhD Thesis, University of London}\ } (\bibinfo {year} {1973})}\BibitemShut {NoStop}%
\bibitem [{\citenamefont {Hughston}\ and\ \citenamefont {Sommers}(1973)}]{HughSomm.73}%
  \BibitemOpen
  \bibfield  {author} {\bibinfo {author} {\bibfnamefont {L.~P.}\ \bibnamefont {Hughston}}\ and\ \bibinfo {author} {\bibfnamefont {P.}~\bibnamefont {Sommers}},\ }\href@noop {} {\bibfield  {journal} {\bibinfo  {journal} {Communications in Mathematical Physics}\ }\textbf {\bibinfo {volume} {33}},\ \bibinfo {pages} {129} (\bibinfo {year} {1973})}\BibitemShut {NoStop}%
\bibitem [{\citenamefont {Frolov}\ \emph {et~al.}(2017)\citenamefont {Frolov}, \citenamefont {Krtou{\v{s}}},\ and\ \citenamefont {Kubiz{\v{n}}{\'a}k}}]{Frolov.17}%
  \BibitemOpen
  \bibfield  {author} {\bibinfo {author} {\bibfnamefont {V.~P.}\ \bibnamefont {Frolov}}, \bibinfo {author} {\bibfnamefont {P.}~\bibnamefont {Krtou{\v{s}}}}, \ and\ \bibinfo {author} {\bibfnamefont {D.}~\bibnamefont {Kubiz{\v{n}}{\'a}k}},\ }\href@noop {} {\bibfield  {journal} {\bibinfo  {journal} {Living reviews in relativity}\ }\textbf {\bibinfo {volume} {20}},\ \bibinfo {pages} {1} (\bibinfo {year} {2017})}\BibitemShut {NoStop}%
\bibitem [{\citenamefont {Mino}(2003)}]{Mi.03}%
  \BibitemOpen
  \bibfield  {author} {\bibinfo {author} {\bibfnamefont {Y.}~\bibnamefont {Mino}},\ }\href@noop {} {\bibfield  {journal} {\bibinfo  {journal} {Phys. Rev. D}\ }\textbf {\bibinfo {volume} {67}},\ \bibinfo {pages} {084027} (\bibinfo {year} {2003})},\ \Eprint {http://arxiv.org/abs/arXiv:gr-qc/0302075} {arXiv:gr-qc/0302075} \BibitemShut {NoStop}%
\bibitem [{\citenamefont {Hinderer}\ and\ \citenamefont {Flanagan}(2008)}]{HinFle.08}%
  \BibitemOpen
  \bibfield  {author} {\bibinfo {author} {\bibfnamefont {T.}~\bibnamefont {Hinderer}}\ and\ \bibinfo {author} {\bibfnamefont {E.~E.}\ \bibnamefont {Flanagan}},\ }\href@noop {} {\bibfield  {journal} {\bibinfo  {journal} {Physical Review D}\ }\textbf {\bibinfo {volume} {78}},\ \bibinfo {pages} {064028} (\bibinfo {year} {2008})},\ \Eprint {http://arxiv.org/abs/0805.3337} {arXiv:0805.3337 [gr-qc]} \BibitemShut {NoStop}%
\bibitem [{\citenamefont {Isoyama}\ \emph {et~al.}(2019)\citenamefont {Isoyama}, \citenamefont {Fujita}, \citenamefont {Nakano}, \citenamefont {Sago},\ and\ \citenamefont {Tanaka}}]{IsoAl.19}%
  \BibitemOpen
  \bibfield  {author} {\bibinfo {author} {\bibfnamefont {S.}~\bibnamefont {Isoyama}}, \bibinfo {author} {\bibfnamefont {R.}~\bibnamefont {Fujita}}, \bibinfo {author} {\bibfnamefont {H.}~\bibnamefont {Nakano}}, \bibinfo {author} {\bibfnamefont {N.}~\bibnamefont {Sago}}, \ and\ \bibinfo {author} {\bibfnamefont {T.}~\bibnamefont {Tanaka}},\ }\href {\doibase 10.1093/ptep/pty136} {\bibfield  {journal} {\bibinfo  {journal} {PTEP}\ }\textbf {\bibinfo {volume} {2019}},\ \bibinfo {pages} {013E01} (\bibinfo {year} {2019})},\ \Eprint {http://arxiv.org/abs/1809.11118} {arXiv:1809.11118 [gr-qc]} \BibitemShut {NoStop}%
\bibitem [{\citenamefont {Pound}\ and\ \citenamefont {Wardell}(2021)}]{PoWa.21}%
  \BibitemOpen
  \bibfield  {author} {\bibinfo {author} {\bibfnamefont {A.}~\bibnamefont {Pound}}\ and\ \bibinfo {author} {\bibfnamefont {B.}~\bibnamefont {Wardell}},\ }\href@noop {} {\enquote {\bibinfo {title} {Black hole perturbation theory and gravitational self-force},}\ } (\bibinfo {year} {2021}),\ \bibinfo {note} {invited chapter for "Handbook of Gravitational Wave Astronomy" (Eds. C. Bambi, S. Katsanevas, and K. Kokkotas; Springer, Singapore, 2021)},\ \Eprint {http://arxiv.org/abs/2101.04592} {arXiv:2101.04592 [gr-qc]} \BibitemShut {NoStop}%
\bibitem [{\citenamefont {R{\"u}diger}(1981)}]{Rudiger.I.81}%
  \BibitemOpen
  \bibfield  {author} {\bibinfo {author} {\bibfnamefont {R.}~\bibnamefont {R{\"u}diger}},\ }\href@noop {} {\bibfield  {journal} {\bibinfo  {journal} {Proceedings of the Royal Society of London. A. Mathematical and Physical Sciences}\ }\textbf {\bibinfo {volume} {375}},\ \bibinfo {pages} {185} (\bibinfo {year} {1981})}\BibitemShut {NoStop}%
\bibitem [{\citenamefont {R{\"u}diger}(1983)}]{Rudiger.II.83}%
  \BibitemOpen
  \bibfield  {author} {\bibinfo {author} {\bibfnamefont {R.}~\bibnamefont {R{\"u}diger}},\ }\href@noop {} {\bibfield  {journal} {\bibinfo  {journal} {Proceedings of the Royal Society of London. A. Mathematical and Physical Sciences}\ }\textbf {\bibinfo {volume} {385}},\ \bibinfo {pages} {229} (\bibinfo {year} {1983})}\BibitemShut {NoStop}%
\bibitem [{\citenamefont {Kubiz{\v{n}}{\'a}k}\ and\ \citenamefont {Cariglia}(2012)}]{KuCa.12}%
  \BibitemOpen
  \bibfield  {author} {\bibinfo {author} {\bibfnamefont {D.}~\bibnamefont {Kubiz{\v{n}}{\'a}k}}\ and\ \bibinfo {author} {\bibfnamefont {M.}~\bibnamefont {Cariglia}},\ }\href@noop {} {\bibfield  {journal} {\bibinfo  {journal} {Physical Review Letters}\ }\textbf {\bibinfo {volume} {108}},\ \bibinfo {pages} {051104} (\bibinfo {year} {2012})}\BibitemShut {NoStop}%
\bibitem [{\citenamefont {Witzany}(2019)}]{WitzHJ.19}%
  \BibitemOpen
  \bibfield  {author} {\bibinfo {author} {\bibfnamefont {V.}~\bibnamefont {Witzany}},\ }\href@noop {} {\bibfield  {journal} {\bibinfo  {journal} {Physical Review D}\ }\textbf {\bibinfo {volume} {100}},\ \bibinfo {pages} {104030} (\bibinfo {year} {2019})},\ \Eprint {http://arxiv.org/abs/1903.03651} {arXiv:1903.03651 [gr-qc]} \BibitemShut {NoStop}%
\bibitem [{\citenamefont {Skoup{\'y}}\ and\ \citenamefont {Witzany}(2025)}]{SkouWitz.25}%
  \BibitemOpen
  \bibfield  {author} {\bibinfo {author} {\bibfnamefont {V.}~\bibnamefont {Skoup{\'y}}}\ and\ \bibinfo {author} {\bibfnamefont {V.}~\bibnamefont {Witzany}},\ }\href {\doibase 10.1103/PhysRevLett.134.171401} {\bibfield  {journal} {\bibinfo  {journal} {Phys. Rev. Lett.}\ }\textbf {\bibinfo {volume} {134}},\ \bibinfo {pages} {171401} (\bibinfo {year} {2025})},\ \Eprint {http://arxiv.org/abs/2411.16855} {arXiv:2411.16855 [gr-qc]} \BibitemShut {NoStop}%
\bibitem [{\citenamefont {Ramond}\ and\ \citenamefont {Isoyama}(2026)}]{Ra.Iso.IntegO1.26}%
  \BibitemOpen
  \bibfield  {author} {\bibinfo {author} {\bibfnamefont {P.}~\bibnamefont {Ramond}}\ and\ \bibinfo {author} {\bibfnamefont {S.}~\bibnamefont {Isoyama}},\ }\href@noop {} {\  (\bibinfo {year} {2026})},\ \Eprint {http://arxiv.org/abs/2210.03866} {arXiv:2210.03866} \BibitemShut {NoStop}%
\bibitem [{\citenamefont {Comp{\`e}re}\ and\ \citenamefont {Druart}(2022)}]{ComDru.22}%
  \BibitemOpen
  \bibfield  {author} {\bibinfo {author} {\bibfnamefont {G.}~\bibnamefont {Comp{\`e}re}}\ and\ \bibinfo {author} {\bibfnamefont {A.}~\bibnamefont {Druart}},\ }\href@noop {} {\bibfield  {journal} {\bibinfo  {journal} {SciPost Physics}\ }\textbf {\bibinfo {volume} {12}},\ \bibinfo {pages} {012} (\bibinfo {year} {2022})},\ \Eprint {http://arxiv.org/abs/2105.12454} {arXiv:2105.12454 [gr-qc]} \BibitemShut {NoStop}%
\bibitem [{\citenamefont {Compère}\ \emph {et~al.}(2023)\citenamefont {Compère}, \citenamefont {Druart},\ and\ \citenamefont {Vines}}]{ComDruVin.23}%
  \BibitemOpen
  \bibfield  {author} {\bibinfo {author} {\bibfnamefont {G.}~\bibnamefont {Compère}}, \bibinfo {author} {\bibfnamefont {A.}~\bibnamefont {Druart}}, \ and\ \bibinfo {author} {\bibfnamefont {J.}~\bibnamefont {Vines}},\ }\href {\doibase 10.21468/SciPostPhys.15.6.226} {\bibfield  {journal} {\bibinfo  {journal} {SciPost Phys.}\ }\textbf {\bibinfo {volume} {15}},\ \bibinfo {pages} {226} (\bibinfo {year} {2023})},\ \Eprint {http://arxiv.org/abs/2302.14549} {arXiv:2302.14549 [gr-qc]} \BibitemShut {NoStop}%
\bibitem [{\citenamefont {Ramond}(2025)}]{Ra.CQG.24}%
  \BibitemOpen
  \bibfield  {author} {\bibinfo {author} {\bibfnamefont {P.}~\bibnamefont {Ramond}},\ }\href {\doibase 10.1088/1361-6382/adb197} {\bibfield  {journal} {\bibinfo  {journal} {Class. Quant. Grav.}\ }\textbf {\bibinfo {volume} {42}},\ \bibinfo {pages} {065019} (\bibinfo {year} {2025})},\ \Eprint {http://arxiv.org/abs/2402.02670} {arXiv:2402.02670 [gr-qc]} \BibitemShut {NoStop}%
\bibitem [{\citenamefont {Ramond}\ \emph {et~al.}(2026)\citenamefont {Ramond}, \citenamefont {Isoyama},\ and\ \citenamefont {Druart}}]{Ra.Iso.Dru.IntegO2.26}%
  \BibitemOpen
  \bibfield  {author} {\bibinfo {author} {\bibfnamefont {P.}~\bibnamefont {Ramond}}, \bibinfo {author} {\bibfnamefont {S.}~\bibnamefont {Isoyama}}, \ and\ \bibinfo {author} {\bibfnamefont {A.}~\bibnamefont {Druart}},\ }\href@noop {} {\  (\bibinfo {year} {2026})},\ \Eprint {http://arxiv.org/abs/2601.06416} {arXiv:2601.06416} \BibitemShut {NoStop}%
\bibitem [{\citenamefont {Bern}\ \emph {et~al.}(2023)\citenamefont {Bern}, \citenamefont {Kosmopoulos}, \citenamefont {Luna}, \citenamefont {Roiban},\ and\ \citenamefont {Teng}}]{BeKoLuRoTe.23}%
  \BibitemOpen
  \bibfield  {author} {\bibinfo {author} {\bibfnamefont {Z.}~\bibnamefont {Bern}}, \bibinfo {author} {\bibfnamefont {D.}~\bibnamefont {Kosmopoulos}}, \bibinfo {author} {\bibfnamefont {A.}~\bibnamefont {Luna}}, \bibinfo {author} {\bibfnamefont {R.}~\bibnamefont {Roiban}}, \ and\ \bibinfo {author} {\bibfnamefont {F.}~\bibnamefont {Teng}},\ }\href {\doibase 10.1103/PhysRevLett.130.201402} {\bibfield  {journal} {\bibinfo  {journal} {Phys. Rev. Lett.}\ }\textbf {\bibinfo {volume} {130}},\ \bibinfo {pages} {201402} (\bibinfo {year} {2023})},\ \Eprint {http://arxiv.org/abs/2203.06202} {arXiv:2203.06202 [hep-th]} \BibitemShut {NoStop}%
\bibitem [{\citenamefont {Akpinar}\ \emph {et~al.}(2025{\natexlab{a}})\citenamefont {Akpinar}, \citenamefont {Febres~Cordero}, \citenamefont {Kraus}, \citenamefont {Ruf},\ and\ \citenamefont {Zeng}}]{AkFeKrRuZe.25}%
  \BibitemOpen
  \bibfield  {author} {\bibinfo {author} {\bibfnamefont {D.}~\bibnamefont {Akpinar}}, \bibinfo {author} {\bibfnamefont {F.}~\bibnamefont {Febres~Cordero}}, \bibinfo {author} {\bibfnamefont {M.}~\bibnamefont {Kraus}}, \bibinfo {author} {\bibfnamefont {M.~S.}\ \bibnamefont {Ruf}}, \ and\ \bibinfo {author} {\bibfnamefont {M.}~\bibnamefont {Zeng}},\ }\href {\doibase 10.1007/JHEP03(2025)126} {\bibfield  {journal} {\bibinfo  {journal} {JHEP}\ }\textbf {\bibinfo {volume} {03}},\ \bibinfo {pages} {126} (\bibinfo {year} {2025}{\natexlab{a}})},\ \Eprint {http://arxiv.org/abs/2407.19005} {arXiv:2407.19005} \BibitemShut {NoStop}%
\bibitem [{\citenamefont {Akpinar}\ \emph {et~al.}(2025{\natexlab{b}})\citenamefont {Akpinar}, \citenamefont {Febres~Cordero}, \citenamefont {Kraus}, \citenamefont {Smirnov},\ and\ \citenamefont {Zeng}}]{AkFeKrSmZe.25}%
  \BibitemOpen
  \bibfield  {author} {\bibinfo {author} {\bibfnamefont {D.}~\bibnamefont {Akpinar}}, \bibinfo {author} {\bibfnamefont {F.}~\bibnamefont {Febres~Cordero}}, \bibinfo {author} {\bibfnamefont {M.}~\bibnamefont {Kraus}}, \bibinfo {author} {\bibfnamefont {A.}~\bibnamefont {Smirnov}}, \ and\ \bibinfo {author} {\bibfnamefont {M.}~\bibnamefont {Zeng}},\ }\href {\doibase 10.1103/c2dh-tj4v} {\bibfield  {journal} {\bibinfo  {journal} {Phys. Rev. Lett.}\ }\textbf {\bibinfo {volume} {135}},\ \bibinfo {pages} {041602} (\bibinfo {year} {2025}{\natexlab{b}})},\ \Eprint {http://arxiv.org/abs/2502.08961} {arXiv:2502.08961 [hep-th]} \BibitemShut {NoStop}%
\bibitem [{\citenamefont {Akpinar}\ \emph {et~al.}(2026)\citenamefont {Akpinar}, \citenamefont {Brown}, \citenamefont {Gonzo},\ and\ \citenamefont {Zeng}}]{AkGo.25}%
  \BibitemOpen
  \bibfield  {author} {\bibinfo {author} {\bibfnamefont {D.}~\bibnamefont {Akpinar}}, \bibinfo {author} {\bibfnamefont {G.~R.}\ \bibnamefont {Brown}}, \bibinfo {author} {\bibfnamefont {R.}~\bibnamefont {Gonzo}}, \ and\ \bibinfo {author} {\bibfnamefont {M.}~\bibnamefont {Zeng}},\ }\href {\doibase 10.1103/dd5t-8dn5} {\bibfield  {journal} {\bibinfo  {journal} {Phys. Rev. D}\ }\textbf {\bibinfo {volume} {113}},\ \bibinfo {pages} {L101501} (\bibinfo {year} {2026})},\ \Eprint {http://arxiv.org/abs/2508.10761} {arXiv:2508.10761 [hep-th]} \BibitemShut {NoStop}%
\bibitem [{\citenamefont {de~Firmian}\ and\ \citenamefont {Vines}(2026)}]{deFiVi.26}%
  \BibitemOpen
  \bibfield  {author} {\bibinfo {author} {\bibfnamefont {C.}~\bibnamefont {de~Firmian}}\ and\ \bibinfo {author} {\bibfnamefont {J.}~\bibnamefont {Vines}},\ }\href {\doibase 10.1007/JHEP05(2026)055} {\bibfield  {journal} {\bibinfo  {journal} {JHEP}\ }\textbf {\bibinfo {volume} {05}},\ \bibinfo {pages} {055} (\bibinfo {year} {2026})},\ \Eprint {http://arxiv.org/abs/2602.18790} {arXiv:2602.18790 [gr-qc]} \BibitemShut {NoStop}%
\bibitem [{\citenamefont {Kim}\ and\ \citenamefont {Lee}(2026)}]{KiLe.26}%
  \BibitemOpen
  \bibfield  {author} {\bibinfo {author} {\bibfnamefont {S.}~\bibnamefont {Kim}}\ and\ \bibinfo {author} {\bibfnamefont {S.}~\bibnamefont {Lee}},\ }\href {\doibase 10.1007/JHEP10(2026)065} {\bibfield  {journal} {\bibinfo  {journal} {JHEP}\ }\textbf {\bibinfo {volume} {10}},\ \bibinfo {pages} {065} (\bibinfo {year} {2026})},\ \Eprint {http://arxiv.org/abs/2604.26544} {arXiv:2604.26544 [hep-th]} \BibitemShut {NoStop}%
\bibitem [{\citenamefont {Scheopner}\ and\ \citenamefont {Lichtner}(2026)}]{Scheopner:2026sok}%
  \BibitemOpen
  \bibfield  {author} {\bibinfo {author} {\bibfnamefont {T.}~\bibnamefont {Scheopner}}\ and\ \bibinfo {author} {\bibfnamefont {F.}~\bibnamefont {Lichtner}},\ }\href@noop {} {\  (\bibinfo {year} {2026})},\ \Eprint {http://arxiv.org/abs/2611.10794} {arXiv:2611.10794} \BibitemShut {NoStop}%
\bibitem [{\citenamefont {Wald}(1984)}]{Wald}%
  \BibitemOpen
  \bibfield  {author} {\bibinfo {author} {\bibfnamefont {R.~M.}\ \bibnamefont {Wald}},\ }\href@noop {} {\emph {\bibinfo {title} {General relativity}}}\ (\bibinfo  {publisher} {University of Chicago Press},\ \bibinfo {address} {Chicago},\ \bibinfo {year} {1984})\BibitemShut {NoStop}%
\bibitem [{\citenamefont {Costa}\ and\ \citenamefont {Nat{\'a}rio}(2015)}]{CoNa.15}%
  \BibitemOpen
  \bibfield  {author} {\bibinfo {author} {\bibfnamefont {L.~F.~O.}\ \bibnamefont {Costa}}\ and\ \bibinfo {author} {\bibfnamefont {J.}~\bibnamefont {Nat{\'a}rio}},\ }\href@noop {} {\bibfield  {journal} {\bibinfo  {journal} {Fund. Theor. Phys.}\ }\textbf {\bibinfo {volume} {179}},\ \bibinfo {pages} {215} (\bibinfo {year} {2015})},\ \Eprint {http://arxiv.org/abs/1410.6443} {arXiv:1410.6443 [gr-qc]} \BibitemShut {NoStop}%
\bibitem [{\citenamefont {Mathisson}(1937)}]{Ma.37}%
  \BibitemOpen
  \bibfield  {author} {\bibinfo {author} {\bibfnamefont {M.}~\bibnamefont {Mathisson}},\ }\href@noop {} {\bibfield  {journal} {\bibinfo  {journal} {Acta Phys. Polon.}\ }\textbf {\bibinfo {volume} {6}},\ \bibinfo {pages} {163} (\bibinfo {year} {1937})},\ \bibinfo {note} {translated in English by A.~Ehlers and reprinted in Gen. Rel. Grav. \textbf{42}, 1011 (2010)}\BibitemShut {NoStop}%
\bibitem [{\citenamefont {Papapetrou}(1951)}]{Pa.51}%
  \BibitemOpen
  \bibfield  {author} {\bibinfo {author} {\bibfnamefont {A.}~\bibnamefont {Papapetrou}},\ }\href@noop {} {\bibfield  {journal} {\bibinfo  {journal} {Proc. R. Soc. Lond. A}\ }\textbf {\bibinfo {volume} {209}},\ \bibinfo {pages} {248} (\bibinfo {year} {1951})}\BibitemShut {NoStop}%
\bibitem [{\citenamefont {Tulczyjew}(1959)}]{Tu.59}%
  \BibitemOpen
  \bibfield  {author} {\bibinfo {author} {\bibfnamefont {W.}~\bibnamefont {Tulczyjew}},\ }\href@noop {} {\bibfield  {journal} {\bibinfo  {journal} {Acta Phys. Polon.}\ }\textbf {\bibinfo {volume} {18}},\ \bibinfo {pages} {393} (\bibinfo {year} {1959})}\BibitemShut {NoStop}%
\bibitem [{\citenamefont {Dixon}(1974)}]{Di.74}%
  \BibitemOpen
  \bibfield  {author} {\bibinfo {author} {\bibfnamefont {W.~G.}\ \bibnamefont {Dixon}},\ }\href@noop {} {\bibfield  {journal} {\bibinfo  {journal} {Phil. Trans. R. Soc. Lond. A}\ }\textbf {\bibinfo {volume} {277}},\ \bibinfo {pages} {59} (\bibinfo {year} {1974})}\BibitemShut {NoStop}%
\bibitem [{\citenamefont {Harte}(2012)}]{Ha.12}%
  \BibitemOpen
  \bibfield  {author} {\bibinfo {author} {\bibfnamefont {A.~I.}\ \bibnamefont {Harte}},\ }\href@noop {} {\bibfield  {journal} {\bibinfo  {journal} {Class. Quant. Grav.}\ }\textbf {\bibinfo {volume} {29}},\ \bibinfo {pages} {055012} (\bibinfo {year} {2012})},\ \Eprint {http://arxiv.org/abs/1103.0543} {arXiv:1103.0543 [gr-qc]} \BibitemShut {NoStop}%
\bibitem [{\citenamefont {Harte}(2015)}]{Ha.15}%
  \BibitemOpen
  \bibfield  {author} {\bibinfo {author} {\bibfnamefont {A.~I.}\ \bibnamefont {Harte}},\ }\href@noop {} {\bibfield  {journal} {\bibinfo  {journal} {Fund. Theor. Phys.}\ }\textbf {\bibinfo {volume} {179}},\ \bibinfo {pages} {327} (\bibinfo {year} {2015})},\ \Eprint {http://arxiv.org/abs/1405.5077} {arXiv:1405.5077 [gr-qc]} \BibitemShut {NoStop}%
\bibitem [{\citenamefont {Dixon}(2015)}]{Di.15}%
  \BibitemOpen
  \bibfield  {author} {\bibinfo {author} {\bibfnamefont {W.~G.}\ \bibnamefont {Dixon}},\ }in\ \href@noop {} {\emph {\bibinfo {booktitle} {Equations of Motion in Relativistic Gravity}}}\ (\bibinfo  {publisher} {Springer},\ \bibinfo {year} {2015})\ pp.\ \bibinfo {pages} {1--66}\BibitemShut {NoStop}%
\bibitem [{\citenamefont {Marck}(1983)}]{Marck.83}%
  \BibitemOpen
  \bibfield  {author} {\bibinfo {author} {\bibfnamefont {J.-A.}\ \bibnamefont {Marck}},\ }\href@noop {} {\bibfield  {journal} {\bibinfo  {journal} {Proceedings of the Royal Society of London. A. Mathematical and Physical Sciences}\ }\textbf {\bibinfo {volume} {385}},\ \bibinfo {pages} {431} (\bibinfo {year} {1983})}\BibitemShut {NoStop}%
\bibitem [{\citenamefont {Van~de Meent}(2020)}]{VdM.20}%
  \BibitemOpen
  \bibfield  {author} {\bibinfo {author} {\bibfnamefont {M.}~\bibnamefont {Van~de Meent}},\ }\href@noop {} {\bibfield  {journal} {\bibinfo  {journal} {Classical and Quantum Gravity}\ }\textbf {\bibinfo {volume} {37}},\ \bibinfo {pages} {145007} (\bibinfo {year} {2020})},\ \Eprint {http://arxiv.org/abs/1906.05090} {arXiv:1906.05090 [gr-qc]} \BibitemShut {NoStop}%
\bibitem [{\citenamefont {{Henry}}\ \emph {et~al.}(2020)\citenamefont {{Henry}}, \citenamefont {{Faye}},\ and\ \citenamefont {{Blanchet}}}]{Hen.al.20}%
  \BibitemOpen
  \bibfield  {author} {\bibinfo {author} {\bibfnamefont {Q.}~\bibnamefont {{Henry}}}, \bibinfo {author} {\bibfnamefont {G.}~\bibnamefont {{Faye}}}, \ and\ \bibinfo {author} {\bibfnamefont {L.}~\bibnamefont {{Blanchet}}},\ }\href {\doibase 10.1103/PhysRevD.102.044033} {\bibfield  {journal} {\bibinfo  {journal} {Phys. Rev. D}\ }\textbf {\bibinfo {volume} {102}},\ \bibinfo {eid} {044033} (\bibinfo {year} {2020})},\ \Eprint {http://arxiv.org/abs/2005.13367} {arXiv:2005.13367 [gr-qc]} \BibitemShut {NoStop}%
\bibitem [{\citenamefont {Marsat}(2015)}]{Ma.15}%
  \BibitemOpen
  \bibfield  {author} {\bibinfo {author} {\bibfnamefont {S.}~\bibnamefont {Marsat}},\ }\href@noop {} {\bibfield  {journal} {\bibinfo  {journal} {Class. Quant. Grav.}\ }\textbf {\bibinfo {volume} {32}},\ \bibinfo {pages} {085008} (\bibinfo {year} {2015})},\ \Eprint {http://arxiv.org/abs/1411.4118} {arXiv:1411.4118 [gr-qc]} \BibitemShut {NoStop}%
\bibitem [{\citenamefont {Rahman}\ \emph {et~al.}(2026)\citenamefont {Rahman}, \citenamefont {Shahzadi}, \citenamefont {Pound},\ and\ \citenamefont {Mathews}}]{RahShaPouMat.26}%
  \BibitemOpen
  \bibfield  {author} {\bibinfo {author} {\bibfnamefont {M.}~\bibnamefont {Rahman}}, \bibinfo {author} {\bibfnamefont {M.}~\bibnamefont {Shahzadi}}, \bibinfo {author} {\bibfnamefont {A.}~\bibnamefont {Pound}}, \ and\ \bibinfo {author} {\bibfnamefont {J.}~\bibnamefont {Mathews}},\ }\href {\doibase 10.1103/wmsn-pl9n} {\bibfield  {journal} {\bibinfo  {journal} {Phys. Rev. D}\ }\textbf {\bibinfo {volume} {114}},\ \bibinfo {pages} {064034} (\bibinfo {year} {2026})},\ \Eprint {http://arxiv.org/abs/2606.28937} {arXiv:2606.28937} \BibitemShut {NoStop}%
\bibitem [{\citenamefont {Ramond}(2026)}]{Ra.NoCarter.26}%
  \BibitemOpen
  \bibfield  {author} {\bibinfo {author} {\bibfnamefont {P.}~\bibnamefont {Ramond}},\ }\href@noop {} {\  (\bibinfo {year} {2026})},\ \Eprint {http://arxiv.org/abs/2607.27129} {arXiv:2607.27129} \BibitemShut {NoStop}%
\bibitem [{\citenamefont {Brown}(2022)}]{Bro.22}%
  \BibitemOpen
  \bibfield  {author} {\bibinfo {author} {\bibfnamefont {J.~D.}\ \bibnamefont {Brown}},\ }\href@noop {} {\bibfield  {journal} {\bibinfo  {journal} {Universe}\ }\textbf {\bibinfo {volume} {8}},\ \bibinfo {pages} {171} (\bibinfo {year} {2022})},\ \Eprint {http://arxiv.org/abs/2201.06558} {arXiv:2201.06558} \BibitemShut {NoStop}%
\bibitem [{\citenamefont {Witzany}\ \emph {et~al.}(2019)\citenamefont {Witzany}, \citenamefont {Steinhoff},\ and\ \citenamefont {Lukes-Gerakopoulos}}]{WiStLu.19}%
  \BibitemOpen
  \bibfield  {author} {\bibinfo {author} {\bibfnamefont {V.}~\bibnamefont {Witzany}}, \bibinfo {author} {\bibfnamefont {J.}~\bibnamefont {Steinhoff}}, \ and\ \bibinfo {author} {\bibfnamefont {G.}~\bibnamefont {Lukes-Gerakopoulos}},\ }\href@noop {} {\bibfield  {journal} {\bibinfo  {journal} {Classical and Quantum Gravity}\ }\textbf {\bibinfo {volume} {36}},\ \bibinfo {pages} {075003} (\bibinfo {year} {2019})},\ \Eprint {http://arxiv.org/abs/1808.06582} {arXiv:1808.06582 [gr-qc]} \BibitemShut {NoStop}%
\bibitem [{\citenamefont {Arnold}\ \emph {et~al.}(2006)\citenamefont {Arnold}, \citenamefont {Kozlov},\ and\ \citenamefont {Neishtadt}}]{Arn}%
  \BibitemOpen
  \bibfield  {author} {\bibinfo {author} {\bibfnamefont {V.}~\bibnamefont {Arnold}}, \bibinfo {author} {\bibfnamefont {V.}~\bibnamefont {Kozlov}}, \ and\ \bibinfo {author} {\bibfnamefont {A.}~\bibnamefont {Neishtadt}},\ }\href@noop {} {\emph {\bibinfo {title} {Mathematical aspects of classical and celestial mechanics}}},\ Vol.~\bibinfo {volume} {3}\ (\bibinfo  {publisher} {Springer},\ \bibinfo {year} {2006})\BibitemShut {NoStop}%
\bibitem [{\citenamefont {Poincar\'{e}}(2 99)}]{Poin}%
  \BibitemOpen
  \bibfield  {author} {\bibinfo {author} {\bibfnamefont {H.}~\bibnamefont {Poincar\'{e}}},\ }\href@noop {} {\emph {\bibinfo {title} {Les m\'{e}thodes nouvelles de la m\'{e}canique c\'{e}leste, Tome 1-3}}}\ (\bibinfo  {publisher} {Gauthiers-Vilard},\ \bibinfo {address} {Paris},\ \bibinfo {year} {1892-99})\BibitemShut {NoStop}%
\bibitem [{\citenamefont {Henry}\ \emph {et~al.}(2020)\citenamefont {Henry}, \citenamefont {Faye},\ and\ \citenamefont {Blanchet}}]{Hen.al2.20}%
  \BibitemOpen
  \bibfield  {author} {\bibinfo {author} {\bibfnamefont {Q.}~\bibnamefont {Henry}}, \bibinfo {author} {\bibfnamefont {G.}~\bibnamefont {Faye}}, \ and\ \bibinfo {author} {\bibfnamefont {L.}~\bibnamefont {Blanchet}},\ }\href {\doibase 10.1103/PhysRevD.101.064047} {\bibfield  {journal} {\bibinfo  {journal} {Physical Review D}\ }\textbf {\bibinfo {volume} {101}},\ \bibinfo {pages} {064047} (\bibinfo {year} {2020})},\ \Eprint {http://arxiv.org/abs/1912.01920} {arXiv:1912.01920 [gr-qc]} \BibitemShut {NoStop}%
\bibitem [{\citenamefont {Dixon}(1964)}]{Di.64}%
  \BibitemOpen
  \bibfield  {author} {\bibinfo {author} {\bibfnamefont {W.~G.}\ \bibnamefont {Dixon}},\ }\href@noop {} {\bibfield  {journal} {\bibinfo  {journal} {Il Nuovo Cimento}\ }\textbf {\bibinfo {volume} {34}},\ \bibinfo {pages} {317} (\bibinfo {year} {1964})}\BibitemShut {NoStop}%
\bibitem [{\citenamefont {Scheopner}\ and\ \citenamefont {Vines}(2024)}]{ScheopnerVines.24}%
  \BibitemOpen
  \bibfield  {author} {\bibinfo {author} {\bibfnamefont {T.}~\bibnamefont {Scheopner}}\ and\ \bibinfo {author} {\bibfnamefont {J.}~\bibnamefont {Vines}},\ }\href {\doibase 10.1007/JHEP12(2024)060} {\bibfield  {journal} {\bibinfo  {journal} {JHEP}\ }\textbf {\bibinfo {volume} {12}},\ \bibinfo {pages} {060} (\bibinfo {year} {2024})},\ \Eprint {http://arxiv.org/abs/2311.18421} {arXiv:2311.18421 [gr-qc]} \BibitemShut {NoStop}%
\bibitem [{\citenamefont {Guevara}\ \emph {et~al.}(2019)\citenamefont {Guevara}, \citenamefont {Ochirov},\ and\ \citenamefont {Vines}}]{GuOcVi.19}%
  \BibitemOpen
  \bibfield  {author} {\bibinfo {author} {\bibfnamefont {A.}~\bibnamefont {Guevara}}, \bibinfo {author} {\bibfnamefont {A.}~\bibnamefont {Ochirov}}, \ and\ \bibinfo {author} {\bibfnamefont {J.}~\bibnamefont {Vines}},\ }\href@noop {} {\bibfield  {journal} {\bibinfo  {journal} {JHEP}\ }\textbf {\bibinfo {volume} {09}},\ \bibinfo {pages} {056} (\bibinfo {year} {2019})},\ \Eprint {http://arxiv.org/abs/1812.06895} {arXiv:1812.06895 [hep-th]} \BibitemShut {NoStop}%
\bibitem [{\citenamefont {Bautista}\ \emph {et~al.}(2023)\citenamefont {Bautista}, \citenamefont {Guevara}, \citenamefont {Kavanagh},\ and\ \citenamefont {Vines}}]{BGKV.23}%
  \BibitemOpen
  \bibfield  {author} {\bibinfo {author} {\bibfnamefont {Y.~F.}\ \bibnamefont {Bautista}}, \bibinfo {author} {\bibfnamefont {A.}~\bibnamefont {Guevara}}, \bibinfo {author} {\bibfnamefont {C.}~\bibnamefont {Kavanagh}}, \ and\ \bibinfo {author} {\bibfnamefont {J.}~\bibnamefont {Vines}},\ }\href@noop {} {\bibfield  {journal} {\bibinfo  {journal} {JHEP}\ }\textbf {\bibinfo {volume} {05}},\ \bibinfo {pages} {211} (\bibinfo {year} {2023})},\ \Eprint {http://arxiv.org/abs/2212.07965} {arXiv:2212.07965 [hep-th]} \BibitemShut {NoStop}%
\bibitem [{\citenamefont {Ben-Shahar}(2024)}]{BeSh.23}%
  \BibitemOpen
  \bibfield  {author} {\bibinfo {author} {\bibfnamefont {M.}~\bibnamefont {Ben-Shahar}},\ }\href@noop {} {\bibfield  {journal} {\bibinfo  {journal} {JHEP}\ }\textbf {\bibinfo {volume} {03}},\ \bibinfo {pages} {108} (\bibinfo {year} {2024})},\ \Eprint {http://arxiv.org/abs/2311.01430} {arXiv:2311.01430 [hep-th]} \BibitemShut {NoStop}%
\bibitem [{\citenamefont {Kozlov}(1983)}]{Kozlov.83}%
  \BibitemOpen
  \bibfield  {author} {\bibinfo {author} {\bibfnamefont {V.~V.}\ \bibnamefont {Kozlov}},\ }\href@noop {} {\bibfield  {journal} {\bibinfo  {journal} {Russian Mathematical Surveys}\ }\textbf {\bibinfo {volume} {38}},\ \bibinfo {pages} {1} (\bibinfo {year} {1983})}\BibitemShut {NoStop}%
\bibitem [{BHP()}]{BHPToolkit}%
  \BibitemOpen
  \href@noop {} {\enquote {\bibinfo {title} {Black hole perturbation toolkit},}\ }\bibinfo {howpublished} {\href{http://bhptoolkit.org}{http://bhptoolkit.org}}\BibitemShut {NoStop}%
\bibitem [{\citenamefont {Galley}(2013)}]{Galley.13}%
  \BibitemOpen
  \bibfield  {author} {\bibinfo {author} {\bibfnamefont {C.~R.}\ \bibnamefont {Galley}},\ }\href@noop {} {\bibfield  {journal} {\bibinfo  {journal} {Phys. Rev. Lett.}\ }\textbf {\bibinfo {volume} {110}},\ \bibinfo {pages} {174301} (\bibinfo {year} {2013})},\ \Eprint {http://arxiv.org/abs/1210.2745} {arXiv:1210.2745} \BibitemShut {NoStop}%
\bibitem [{\citenamefont {Aykroyd}\ \emph {et~al.}(2025)\citenamefont {Aykroyd}, \citenamefont {Bourgoin},\ and\ \citenamefont {Le~Poncin-Lafitte}}]{Aykroyd:2025yzd}%
  \BibitemOpen
  \bibfield  {author} {\bibinfo {author} {\bibfnamefont {C.}~\bibnamefont {Aykroyd}}, \bibinfo {author} {\bibfnamefont {A.}~\bibnamefont {Bourgoin}}, \ and\ \bibinfo {author} {\bibfnamefont {C.}~\bibnamefont {Le~Poncin-Lafitte}},\ }\href@noop {} {\  (\bibinfo {year} {2025})},\ \Eprint {http://arxiv.org/abs/2512.11947} {arXiv:2512.11947 [gr-qc]} \BibitemShut {NoStop}%
\bibitem [{\citenamefont {Seyrich}\ and\ \citenamefont {Lukes-Gerakopoulos}(2012)}]{Sey.Lu.12}%
  \BibitemOpen
  \bibfield  {author} {\bibinfo {author} {\bibfnamefont {J.}~\bibnamefont {Seyrich}}\ and\ \bibinfo {author} {\bibfnamefont {G.}~\bibnamefont {Lukes-Gerakopoulos}},\ }\href {\doibase 10.1103/PhysRevD.86.124013} {\bibfield  {journal} {\bibinfo  {journal} {Phys. Rev. D}\ }\textbf {\bibinfo {volume} {86}},\ \bibinfo {pages} {124013} (\bibinfo {year} {2012})}\BibitemShut {NoStop}%
\bibitem [{\citenamefont {Wu}\ \emph {et~al.}(2021)\citenamefont {Wu}, \citenamefont {Wang}, \citenamefont {Sun},\ and\ \citenamefont {Liu}}]{Wu.al.21}%
  \BibitemOpen
  \bibfield  {author} {\bibinfo {author} {\bibfnamefont {X.}~\bibnamefont {Wu}}, \bibinfo {author} {\bibfnamefont {Y.}~\bibnamefont {Wang}}, \bibinfo {author} {\bibfnamefont {W.}~\bibnamefont {Sun}}, \ and\ \bibinfo {author} {\bibfnamefont {F.}~\bibnamefont {Liu}},\ }\href {\doibase 10.3847/1538-4357/abfc45} {\bibfield  {journal} {\bibinfo  {journal} {Astrophys. J.}\ }\textbf {\bibinfo {volume} {914}},\ \bibinfo {pages} {63} (\bibinfo {year} {2021})},\ \Eprint {http://arxiv.org/abs/2106.12356} {arXiv:2106.12356 [gr-qc]} \BibitemShut {NoStop}%
\bibitem [{\citenamefont {Afshordi}\ \emph {et~al.}(2025)\citenamefont {Afshordi} \emph {et~al.}}]{LISAWHITE.23}%
  \BibitemOpen
  \bibfield  {author} {\bibinfo {author} {\bibfnamefont {N.}~\bibnamefont {Afshordi}} \emph {et~al.} (\bibinfo {collaboration} {LISA Consortium Waveform Working Group}),\ }\href {\doibase 10.1007/s41114-025-00056-1} {\bibfield  {journal} {\bibinfo  {journal} {Living Rev. Rel.}\ }\textbf {\bibinfo {volume} {28}},\ \bibinfo {pages} {9} (\bibinfo {year} {2025})},\ \Eprint {http://arxiv.org/abs/2311.01300} {arXiv:2311.01300 [gr-qc]} \BibitemShut {NoStop}%
\bibitem [{\citenamefont {Skoup{\'y}}(2026)}]{Sk.26}%
  \BibitemOpen
  \bibfield  {author} {\bibinfo {author} {\bibfnamefont {V.}~\bibnamefont {Skoup{\'y}}},\ }\href {\doibase 10.1103/kpj2-bftm} {\bibfield  {journal} {\bibinfo  {journal} {Phys. Rev. D}\ }\textbf {\bibinfo {volume} {114}},\ \bibinfo {pages} {084030} (\bibinfo {year} {2026})},\ \Eprint {http://arxiv.org/abs/2603.13482} {arXiv:2603.13482} \BibitemShut {NoStop}%
\bibitem [{\citenamefont {{Le Tiec}}\ \emph {et~al.}(2012)\citenamefont {{Le Tiec}}, \citenamefont {Blanchet},\ and\ \citenamefont {Whiting}}]{Le.al.12}%
  \BibitemOpen
  \bibfield  {author} {\bibinfo {author} {\bibfnamefont {A.}~\bibnamefont {{Le Tiec}}}, \bibinfo {author} {\bibfnamefont {L.}~\bibnamefont {Blanchet}}, \ and\ \bibinfo {author} {\bibfnamefont {B.~F.}\ \bibnamefont {Whiting}},\ }\href@noop {} {\bibfield  {journal} {\bibinfo  {journal} {Phys. Rev. D}\ }\textbf {\bibinfo {volume} {85}},\ \bibinfo {pages} {064039} (\bibinfo {year} {2012})},\ \Eprint {http://arxiv.org/abs/1111.5378} {arXiv:1111.5378 [gr-qc]} \BibitemShut {NoStop}%
\bibitem [{\citenamefont {Blanchet}\ \emph {et~al.}(2013)\citenamefont {Blanchet}, \citenamefont {Buonanno},\ and\ \citenamefont {{Le Tiec}}}]{Bl.al.13}%
  \BibitemOpen
  \bibfield  {author} {\bibinfo {author} {\bibfnamefont {L.}~\bibnamefont {Blanchet}}, \bibinfo {author} {\bibfnamefont {A.}~\bibnamefont {Buonanno}}, \ and\ \bibinfo {author} {\bibfnamefont {A.}~\bibnamefont {{Le Tiec}}},\ }\href@noop {} {\bibfield  {journal} {\bibinfo  {journal} {Phys. Rev. D}\ }\textbf {\bibinfo {volume} {87}},\ \bibinfo {pages} {024030} (\bibinfo {year} {2013})},\ \Eprint {http://arxiv.org/abs/1211.1060} {arXiv:1211.1060 [gr-qc]} \BibitemShut {NoStop}%
\bibitem [{\citenamefont {Le~Tiec}(2014)}]{ALT.14}%
  \BibitemOpen
  \bibfield  {author} {\bibinfo {author} {\bibfnamefont {A.}~\bibnamefont {Le~Tiec}},\ }\href@noop {} {\bibfield  {journal} {\bibinfo  {journal} {Classical and Quantum Gravity}\ }\textbf {\bibinfo {volume} {31}},\ \bibinfo {pages} {097001} (\bibinfo {year} {2014})},\ \Eprint {http://arxiv.org/abs/1311.3836} {arXiv:1311.3836 [gr-qc]} \BibitemShut {NoStop}%
\bibitem [{\citenamefont {{Le Tiec}}(2015)}]{Le.15}%
  \BibitemOpen
  \bibfield  {author} {\bibinfo {author} {\bibfnamefont {A.}~\bibnamefont {{Le Tiec}}},\ }\href@noop {} {\bibfield  {journal} {\bibinfo  {journal} {Phys. Rev. D}\ }\textbf {\bibinfo {volume} {92}},\ \bibinfo {pages} {084021} (\bibinfo {year} {2015})},\ \Eprint {http://arxiv.org/abs/1506.05648} {arXiv:1506.05648 [gr-qc]} \BibitemShut {NoStop}%
\bibitem [{\citenamefont {Ramond}\ and\ \citenamefont {Le~Tiec}(2022)}]{RaLe.22}%
  \BibitemOpen
  \bibfield  {author} {\bibinfo {author} {\bibfnamefont {P.}~\bibnamefont {Ramond}}\ and\ \bibinfo {author} {\bibfnamefont {A.}~\bibnamefont {Le~Tiec}},\ }\href {\doibase 10.1103/PhysRevD.106.044057} {\bibfield  {journal} {\bibinfo  {journal} {Phys. Rev. D}\ }\textbf {\bibinfo {volume} {106}},\ \bibinfo {pages} {044057} (\bibinfo {year} {2022})},\ \Eprint {http://arxiv.org/abs/2202.09345} {arXiv:2202.09345 [gr-qc]} \BibitemShut {NoStop}%
\bibitem [{\citenamefont {Gonzo}\ \emph {et~al.}(2025)\citenamefont {Gonzo}, \citenamefont {Lewis},\ and\ \citenamefont {Pound}}]{GoLePo.25}%
  \BibitemOpen
  \bibfield  {author} {\bibinfo {author} {\bibfnamefont {R.}~\bibnamefont {Gonzo}}, \bibinfo {author} {\bibfnamefont {J.}~\bibnamefont {Lewis}}, \ and\ \bibinfo {author} {\bibfnamefont {A.}~\bibnamefont {Pound}},\ }\href {\doibase 10.1103/s85p-gh7b} {\bibfield  {journal} {\bibinfo  {journal} {Phys. Rev. Lett.}\ }\textbf {\bibinfo {volume} {135}},\ \bibinfo {pages} {131401} (\bibinfo {year} {2025})},\ \Eprint {http://arxiv.org/abs/2409.03437} {arXiv:2409.03437 [gr-qc]} \BibitemShut {NoStop}%
\bibitem [{\citenamefont {{Maggiore}}\ \emph {et~al.}(2020)\citenamefont {{Maggiore}} \emph {et~al.}}]{Mag.al.20}%
  \BibitemOpen
  \bibfield  {author} {\bibinfo {author} {\bibfnamefont {M.}~\bibnamefont {{Maggiore}}} \emph {et~al.},\ }\href {\doibase 10.1088/1475-7516/2020/03/050} {\bibfield  {journal} {\bibinfo  {journal} {Journal of Cosmology and Astroparticle Physics}\ }\textbf {\bibinfo {volume} {2020}},\ \bibinfo {eid} {050} (\bibinfo {year} {2020})},\ \Eprint {http://arxiv.org/abs/1912.02622} {arXiv:1912.02622} \BibitemShut {NoStop}%
\bibitem [{\citenamefont {Penrose}(1960)}]{Pen.60}%
  \BibitemOpen
  \bibfield  {author} {\bibinfo {author} {\bibfnamefont {R.}~\bibnamefont {Penrose}},\ }\href {\doibase 10.1016/0003-4916(60)90021-X} {\bibfield  {journal} {\bibinfo  {journal} {Annals Phys.}\ }\textbf {\bibinfo {volume} {10}},\ \bibinfo {pages} {171} (\bibinfo {year} {1960})}\BibitemShut {NoStop}%
\bibitem [{\citenamefont {M{\"u}ller}\ \emph {et~al.}(1999)\citenamefont {M{\"u}ller}, \citenamefont {Schubert},\ and\ \citenamefont {van~de Ven}}]{MuScVdV.99}%
  \BibitemOpen
  \bibfield  {author} {\bibinfo {author} {\bibfnamefont {U.}~\bibnamefont {M{\"u}ller}}, \bibinfo {author} {\bibfnamefont {C.}~\bibnamefont {Schubert}}, \ and\ \bibinfo {author} {\bibfnamefont {A.~M.~E.}\ \bibnamefont {van~de Ven}},\ }\href@noop {} {\bibfield  {journal} {\bibinfo  {journal} {Gen. Rel. Grav.}\ }\textbf {\bibinfo {volume} {31}},\ \bibinfo {pages} {1759} (\bibinfo {year} {1999})},\ \Eprint {http://arxiv.org/abs/gr-qc/9712092} {arXiv:gr-qc/9712092} \BibitemShut {NoStop}%
\bibitem [{\citenamefont {Hatzinikitas}(2000)}]{Hatz.00}%
  \BibitemOpen
  \bibfield  {author} {\bibinfo {author} {\bibfnamefont {A.}~\bibnamefont {Hatzinikitas}},\ }\href@noop {} {\  (\bibinfo {year} {2000})},\ \Eprint {http://arxiv.org/abs/hep-th/0001078} {arXiv:hep-th/0001078} \BibitemShut {NoStop}%
\bibitem [{\citenamefont {Vines}\ \emph {et~al.}(2016)\citenamefont {Vines}, \citenamefont {Kunst}, \citenamefont {Steinhoff},\ and\ \citenamefont {Hinderer}}]{ViKuStHi.16}%
  \BibitemOpen
  \bibfield  {author} {\bibinfo {author} {\bibfnamefont {J.}~\bibnamefont {Vines}}, \bibinfo {author} {\bibfnamefont {D.}~\bibnamefont {Kunst}}, \bibinfo {author} {\bibfnamefont {J.}~\bibnamefont {Steinhoff}}, \ and\ \bibinfo {author} {\bibfnamefont {T.}~\bibnamefont {Hinderer}},\ }\href@noop {} {\bibfield  {journal} {\bibinfo  {journal} {Physical Review D}\ }\textbf {\bibinfo {volume} {93}},\ \bibinfo {pages} {103008} (\bibinfo {year} {2016})},\ \Eprint {http://arxiv.org/abs/1601.07529} {arXiv:1601.07529} \BibitemShut {NoStop}%
\bibitem [{\citenamefont {Dixon}(1973)}]{Di.73}%
  \BibitemOpen
  \bibfield  {author} {\bibinfo {author} {\bibfnamefont {W.~G.}\ \bibnamefont {Dixon}},\ }\href@noop {} {\bibfield  {journal} {\bibinfo  {journal} {Gen. Rel. Grav.}\ }\textbf {\bibinfo {volume} {4}},\ \bibinfo {pages} {199} (\bibinfo {year} {1973})}\BibitemShut {NoStop}%
\bibitem [{\citenamefont {Steinhoff}\ and\ \citenamefont {Puetzfeld}(2010)}]{StPu.10}%
  \BibitemOpen
  \bibfield  {author} {\bibinfo {author} {\bibfnamefont {J.}~\bibnamefont {Steinhoff}}\ and\ \bibinfo {author} {\bibfnamefont {D.}~\bibnamefont {Puetzfeld}},\ }\href@noop {} {\bibfield  {journal} {\bibinfo  {journal} {Phys. Rev. D}\ }\textbf {\bibinfo {volume} {81}},\ \bibinfo {pages} {044019} (\bibinfo {year} {2010})},\ \Eprint {http://arxiv.org/abs/0909.3756} {arXiv:0909.3756 [gr-qc]} \BibitemShut {NoStop}%
\bibitem [{\citenamefont {{Harte}}(2020)}]{Ha.20}%
  \BibitemOpen
  \bibfield  {author} {\bibinfo {author} {\bibfnamefont {A.~I.}\ \bibnamefont {{Harte}}},\ }\href {\doibase 10.1103/PhysRevD.102.124075} {\bibfield  {journal} {\bibinfo  {journal} {Physical Review D}\ }\textbf {\bibinfo {volume} {102}},\ \bibinfo {eid} {124075} (\bibinfo {year} {2020})},\ \Eprint {http://arxiv.org/abs/2011.00110} {arXiv:2011.00110 [gr-qc]} \BibitemShut {NoStop}%
\bibitem [{\citenamefont {Witzany}(2022)}]{WitzAA.22}%
  \BibitemOpen
  \bibfield  {author} {\bibinfo {author} {\bibfnamefont {V.}~\bibnamefont {Witzany}},\ }\href@noop {} {\enquote {\bibinfo {title} {{Action-angle coordinates for black-hole geodesics I: Spherically symmetric and Schwarzschild}},}\ } (\bibinfo {year} {2022}),\ \Eprint {http://arxiv.org/abs/2203.11952} {arXiv:2203.11952 [gr-qc]} \BibitemShut {NoStop}%
\end{thebibliography}%

\clearpage
\section*{End Matter}

\emph{Couplings at $N=5$.---} We write $\nabla_s:=s^a\nabla_a$, $\nabla_t:=t^a\nabla_a$, $\mcB_{ss}:=\mcB_{ab}s^as^b$, $\mcE^c{}_s:=\mcE^c{}_as^a$, $(\mcE\nabla_s\mcB)_{ss}:=\mcE_{ac}\,(\nabla_s\mcB^c{}_b)\,s^as^b$ and $\mathrm{tr}(\mcE\nabla_s\mcB):=\mcE_{ab}\nabla_s\mcB^{ab}$, derivatives acting on the curvature only, and likewise with any of $\mcE_{ab}$, $\mcB_{ab}$ in each slot. A multipole$\times$curvature coupling of order $N$ is a scalar built from $\nabla^kR$, $t^a$, $s^a$, $g_{ab}$ and $\varepsilon_{abcd}$ with $N$ derivatives of the metric, written with the power of $s^2$ that makes it of degree $N$ in the spin. In vacuum, $\mcE_{ab}$ and $\mcB_{ab}$ are trace-free and, with $\varepsilon_{abc}:=\varepsilon_{dabc}t^d$, the Bianchi identities give $\nabla^a\mcE_{ab}=0=\nabla^a\mcB_{ab}$, $\nabla_t\mcE_{ab}=\varepsilon_{cad}\nabla^c\mcB^d{}_b$ and $\nabla_t\mcB_{ab}=-\varepsilon_{cad}\nabla^c\mcE^d{}_b$, while Penrose's wave equation \cite{Pen.60} makes $\nabla^e\nabla_eR_{abcd}$ quadratic in the curvature (Sec.~\ref{sec:couplings}). We keep the couplings even under time reversal (non-dissipative). At $N=5$, every contraction then reduces to the couplings $X_1,\ldots,X_{24}$ of Eq.~\eqref{O5cond}, with
\begin{equation}\label{EM:X}
\begin{aligned}
X_1&=\nabla_s\nabla_s\nabla_s\mcB_{ss}, & X_2&=s^2\nabla_t\nabla_t\nabla_s\mcB_{ss},\\
X_3&=\mcE_{ss}\nabla_s\mcB_{ss}, & X_8&=\mcB_{ss}\nabla_s\mcB_{ss},\\
X_4&=s^2(\mcE\nabla_s\mcB)_{ss}, & X_9&=s^2(\mcB\nabla_s\mcB)_{ss},\\
X_5&=s^4\mathrm{tr}(\mcE\nabla_s\mcB), & X_{10}&=s^4\mathrm{tr}(\mcB\nabla_s\mcB),\\
X_6&=s^2\mcE^c{}_s\nabla_c\mcB_{ss}, & X_{11}&=s^2\mcB^c{}_s\nabla_c\mcB_{ss},\\
X_7&=s^4\mcE^{cd}\nabla_c\mcB_{ds}, & X_{12}&=s^4\mcB^{cd}\nabla_c\mcB_{ds},
\end{aligned}
\end{equation}
and $X_{12+i}$ obtained from $X_i$ by exchanging $\mcE_{ab}$ and $\mcB_{ab}$. $X_1$, $X_2$, $X_{13}$, $X_{14}$ are linear in the curvature (spin-induced $2^5$-pole), the others quadratic. The ten couplings $X_3,\ldots,X_7$ and $X_{15},\ldots,X_{19}$ are related by an invertible linear map to the ten quadratic structures of the worldline effective theory of \cite{ScheopnerVines.24}, written with the octupolar tidal tensors and $\nabla_t\mcE_{ab}$, $\nabla_t\mcB_{ab}$ (Sec.~\ref{sec:couplings}).

\emph{Reading Fig.~\ref{fig:data}.---} The numerical part of our work has a single input: the average integrals \eqref{harm}, from which we extract the resonant harmonics $\langle\sigma_n\rangle_m(\varphi_0)$ entering \eqref{test}. Turning them into the points of Fig.~\ref{fig:data} takes three steps.

\emph{(i) Averages.} For a given torus, order $n$ and constant ($Q$ or $K$), we compute $\langle\sigma_n\rangle_m(\varphi_0)$ for every resonant harmonic $m$ such that $mw\in\mathbb Z$, $0\le m\le n$ (notice $m<0$ is redundant, since $\sigma_n\in\mathbb{R}$ and $\langle\sigma_n\rangle_{-m}=\overline{\langle\sigma_n\rangle_m}$) and for the initial polar phases $\varphi_0=2\pi i/(9n_r)$, $i=0,\ldots,4$ (five phases; six on $T_2$, where $i=8$ is added). In practice, we evaluate the average \eqref{harm} at eleven equally spaced initial spin phases $\psi_{0}=\psi_{0,k}:=2\pi k/11$, $k=0,\ldots,10$, and obtain the harmonics by a discrete Fourier transform, $\langle\sigma_n\rangle_m(\varphi_0)=\frac1{11}\sum_{k=0}^{10}e^{-2\ui m k \pi/11}\langle\sigma_n\rangle(\varphi_0,\psi_{0,k})$, which is exact since $\sigma_n$ has degree $n\le5$ in the spin. If the constant extends, all of them vanish. Instead of plotting all of them, they are condensed into their root-mean-square,
\begin{equation}\label{EM:Nn}
\mathcal N_n:=\Big(\frac1{2N_\varphi}\sum_{\varphi_0}\ \sum_{m\ge0,\,mw\in\mathbb Z}\big|\langle\sigma_n\rangle_m(\varphi_0)\big|^2\Big)^{1/2},
\end{equation}
which vanishes if and only if every condition \eqref{test} holds.

\emph{(ii) Scale.} The values of $\mathcal N_n$ are all small: about $10^{-26}$ at $n\le4$ on $T_2$, where $Q$ and $K$ do extend, but also $10^{-7.5}$ for $Q$ at $n=5$ on the same torus. This could seem promising, but whether a number is small has no meaning without a scale. A natural scale is  size of the integrand itself, $\langle|\sigma_n\Sigma|\rangle$: the mean of $|\sigma_n\Sigma|$ over the integration nodes, the eleven initial spin phases, and the sampled $\varphi_0$, for $\kappa_n=1$ and all other couplings zero. This scale drops quickly with $n$: on $T_2$, $\langle|\sigma_n\Sigma|\rangle=0.32$, $0.17$, $2.3\times10^{-3}$, $1.7\times10^{-4}$ for $Q$ at $n=1,\ldots,4$, and it lies between $2\times10^{-6}$ and $1.6\times10^{-4}$ for $Q$ at $n=5$, depending on the torus. A small $\mathcal N_n$ can thus simply reflect a small integrand. The points of Fig.~\ref{fig:data} are therefore
\begin{equation}\label{EM:Rn}
\mathcal{R}_n := \mathcal{N}_n / \langle|\sigma_n\Sigma|\rangle \,,
\end{equation}
the fraction of the integrand that survives the average. Since an average never exceeds the mean absolute value, $\mathcal{R}_n\lesssim1$. On a given torus and for a given constant, the same denominator is used for all points (circles, diamonds and squares), so that they can be compared.

\emph{(iii) Numerical zeros.} With finite precision, an average that vanishes analytically does not come out as $0$, and $\mathcal R_n\ll1$ does not mean that a constant extends. We therefore measure, on each torus, the value of $\mathcal R_n$ for averages that we know to vanish exactly. For any function $P$ on phase space, the derivative $\ud P/\ud s_h$ along $\scL_0$ has the form $\{P,H_0\}^{\rm D}$, like the term $\{Q_5,H_0\}^{\rm D}_5$ of \eqref{O5cond}: by the same argument, its one-period average is $\mu[P(T)-P(0)]/T$, whose resonant harmonics vanish identically on a closed orbit. We take three generic monomials of the Kerr scalars,
\begin{equation}\label{EM:ctrl}
P\in\{B^2Z_{\xi s}\mcZ_{\xi s}\mcZ_{st}/\mcC^2, \mcZ_{tt}\,|Z_{\xi t}|^2/\mcC, s^2E_s\,\xi^2 / \bar{\mcC}^3 \},
\end{equation}
and run their derivatives through exactly the same pipeline as $\sigma_n$ (orbit, nodes, spin phases, quadrature, 40-digit arithmetic, resonant harmonics, Eqs.~\eqref{EM:Nn}--\eqref{EM:Rn} with the denominator of $Q$). The largest of the three results is the numerical zero of the torus, i.e., what a true zero looks like for us; it is shown by the red lines on the right of Fig.~\ref{fig:data}. It lies between $10^{-24}$ and $10^{-20}$ on nine tori, and between $10^{-16.5}$ and $10^{-15}$ on $T_9$, $T_{11}$ and $T_{12}$. We checked that it is the truncation error of the trapezoidal rule used for Eq.~\eqref{harm} (Sec.~\ref{sec:num}), not a limit of the arithmetic or of the orbit closure. At $n\le4$, the Kerr values of $\mathcal R_n$, which must vanish, serve directly as the numerical zero (left red line). $K$ is not shown at $n\le2$ on $T_2$, where only the harmonic $m=0$ is resonant, nor on the five tori whose only nonzero resonant harmonic leaves it at the numerical zero; $K$ is insensitive to $m=0$ (Sec.~\ref{sec:num}). 

\emph{Constants of motion through $O(S^4)$.---} At the integrable point \eqref{O4result}, with $\lambda_4=0$, $B=Z_{st}$, the other Kerr scalars of Sec.~\ref{sec:kerr}, and $\xi^2=\xi_a\xi^a$, the Hamiltonian $H=\sum_{i=0}^4H_i$, the Rüdiger constant $K=\sum_{i=1}^4K_i$ and the Carter constant $Q=\sum_{i=0}^4Q_i$ read as follows:
\begin{widetext}\allowdisplaybreaks

\begin{align*}
H_0 &= -\frac{\mu^2}{2}, \quad 
H_1=0, \quad 
H_2 = M\,\mathrm{Im}\Big[\frac{1}{\mcC^3}\Big( \frac{3 B^2}{2} -\frac{s^2}{2}\Big)\Big], \quad
H_3 = -\frac{M}{\mu }\,\mathrm{Im}\Big[\frac{1}{\mcC^4}\Big( \frac{5 B^2 Z_{\xi s}}{2} -\frac{Z_{\xi s} s^2}{2} +B E_0 s^2\Big)\Big], \\
H_4 &= \frac{M^2}{\mu^2}\,\mathrm{Re}\Big[\frac{1}{\mcC^4 \bar{\mcC}^2}\Big( -\frac{5 B^2 \mcZ_{ss}}{16} +\frac{1}{8} B \bar B s^2 +\frac{\mcZ_{ss} s^2}{16}\Big) + \frac{1}{\mcC^6}\Big( -\frac{3}{16} B^2 s^2 +\frac{s^4}{16}\Big)\Big]\\
&\quad -\frac{M}{\mu^2}\,\mathrm{Im}\Big[\frac{1}{\mcC^5}\Big( -\frac{35}{8} B^2 Z_{\xi s}^2 +\frac{5}{8} B^2 E_s^2 +\frac{5}{8} Z_{\xi s}^2 s^2 -\frac{5}{2} B Z_{\xi s} E_0 s^2 -\frac{E_s^2 s^2}{8}   -\frac{ E_0^2 s^4}{4} -\frac{5}{8} B^2 s^2 \xi^2 +\frac{s^4 \xi^2}{8} \Big)\Big]; \\ 
& \\
K_1 &= \mathrm{Re}[2 \bar B \bar\mcC], \quad 
K_2 = 0, \quad  
K_3 =\frac{2M}{\mu^2}\,\mathrm{Im}\!\Big[\frac{B(B^2-s^2)}{\mcC^2}+\frac{B\bar\mcC(B\bar B-\mcZ_{ss})}{2\mcC^3}\Big], \\ 
K_4 &= -\frac{M}{\mu^3}\,\mathrm{Im}\Big[\frac{\bar\mcC}{\mcC^4}\Big( \frac{5}{8} \bar B \bar Z_{\xi s} \mcZ_{ss} +\frac{5 \mcZ_{\xi s} \mcZ_{st} \mcZ_{ss}}{8} -\frac{5 \mcZ_{\xi t} \mcZ_{ss}^2}{8} +\frac{3}{8} \bar B Z_{\xi s} s^2 -\frac{5}{8} \bar B^2 \mcZ_{\xi t} s^2 \\  
& \qquad -\frac{5}{8} Z_{\xi t} B \mcZ_{st} s^2 -\frac{3}{8} E_s \mcZ_{st} s^2 +\frac{5}{8} \mcZ_{\xi t} \mcZ_{st}^2 s^2 +\frac{11}{16} E_0 \mcZ_{ss} s^2 -\frac{5}{8} \mcZ_{\xi t} \mcZ_{tt} \mcZ_{ss} s^2 +\frac{1}{4} \mcZ_{\xi t} s^4 +\frac{1}{16} E_0 \mcZ_{tt} s^4\Big)\Big]\\
&\qquad -\frac{M}{\mu^3}\,\mathrm{Im}\Big[\frac{1}{\mcC^3}\Big( \frac{15}{2} Z_{\xi t} \bar B^3 \mcZ_{st} -\frac{15}{4} \bar B \bar Z_{\xi s} \mcZ_{st}^2 -\frac{15}{2} Z_{\xi t} \bar B \mcZ_{st}^3 -\frac{15 \mcZ_{\xi s} \mcZ_{st}^3}{4} +\frac{15}{8} \bar B Z_{\xi s} \mcZ_{ss}\\
&\qquad -\frac{15}{4} \bar B^2 \mcZ_{\xi t} \mcZ_{ss} -\frac{15}{4} E_s \mcZ_{st} \mcZ_{ss} +\frac{15}{2} \mcZ_{\xi t} \mcZ_{st}^2 \mcZ_{ss} +\frac{15}{2} Z_{\xi t} \bar B \mcZ_{st} \mcZ_{tt} \mcZ_{ss} +\frac{15 E_0 \mcZ_{ss}^2}{8}  -\frac{15}{4} \mcZ_{\xi t} \mcZ_{tt} \mcZ_{ss}^2\\
&\qquad -\frac{15}{8} \bar B \bar Z_{\xi s} s^2 -\frac{21}{8} \bar B^2 E_0 s^2 -\frac{45}{8} Z_{\xi t} \bar B \mcZ_{st} s^2\\
&\qquad -\frac{15}{4} \mcZ_{\xi s} \mcZ_{st} s^2 +\frac{3}{4} E_0 \mcZ_{st}^2 s^2 -\frac{15}{4} Z_{\xi t} B \mcZ_{st} \mcZ_{tt} s^2 +\frac{45}{16} \mcZ_{\xi t} \mcZ_{ss} s^2 -\frac{21}{8} E_0 \mcZ_{tt} \mcZ_{ss} s^2 +\frac{15}{16} \mcZ_{\xi t} \mcZ_{tt} s^4\Big)\Big]; \\
& \\
Q_0 &= \frac{\mu^2}{2} \Bigl(|\mcC|^2 \mcZ_{tt} - \mathrm{Re}\left[\bar{\mcC}^2\right]\Bigr), \quad
Q_1= \mu \, \mathrm{Re}\Bigl[\bar\mcC \left(2 \bar Z_{\xi s}-2 \bar B E_0\right)\Bigr], \\
Q_2&= M\,\mathrm{Im}\Big[\frac{\bar{\mcC}^2}{\mcC^3} \frac{3 B^2-s^2}{4} - \frac{\bar{\mcC}}{\mcC^2} |B|^2 + \frac{1}{\mcC} \frac{B^2+s^2}{4}\Big] +\mathrm{Re}\left[Z_{\xi s}^2+2 B Z_{\xi s} E_0-2 Z_{\xi t} B E_s-Z_{\xi t}^2 s^2-B^2 \xi^2\right]  +\xi^2 s^2, \\
Q_3 &= -\frac{M}{\mu }\,\mathrm{Im}\Big[\frac{\bar\mcC}{\mcC^3}\Big( -B \bar B Z_{\xi s} +\frac{3}{2} B^2 \bar Z_{\xi s} +B E_s \mcZ_{st} -B E_0 \mcZ_{ss} -\frac{1}{2} \bar Z_{\xi s} s^2\Big)\Big]\\
&\quad -\frac{M}{\mu } \, \mathrm{Re}[\mcC^2] \,\mathrm{Im}\Big[ \frac{1}{\mcC^4}\Big( \frac{5 B^2 Z_{\xi s}}{2} -\frac{Z_{\xi s} s^2}{2} + B E_0 s^2\Big)\Big],  \\ 
Q_4&= -\frac{M}{\mu^2}\,\mathrm{Im}\Big[\frac{1}{\mcC^3}\Big(-\frac{25}{16} B^2 Z_{\xi s}^2 -\frac{5}{16} B^2 E_s^2 +\frac{5}{16} Z_{\xi s}^2 s^2 -\frac{3}{4} B Z_{\xi s} E_0 s^2 -\frac{7}{16} Z_{\xi t} B E_s s^2\\
&\qquad +\frac{1}{16} \bar Z_{\xi t} \bar B E_s s^2 -\frac{1}{16} E_s \mcZ_{\xi t} \mcZ_{st} s^2 +\frac{1}{16} \mcZ_{\xi s} E_s \mcZ_{tt} s^2 -\frac{1}{8} Z_{\xi t}^2 s^4 -\frac{1}{16} B^2 s^2 \xi^2 +\frac{1}{16} s^4 \xi^2\Big)\Big]\\
&\quad -\frac{M}{\mu^2}\,\mathrm{Im}\Big[\frac{\bar\mcC}{\mcC^4}\Big( -\frac{15}{4} B^2 Z_{\xi s} \bar Z_{\xi s} +\frac{1}{4} B^2 \mcZ_{\xi s} E_s -\frac{1}{4} B \bar B E_s^2 +\frac{1}{4} B Z_{\xi s} E_s \mcZ_{st} -\frac{1}{4} Z_{\xi t} B E_s \mcZ_{ss}\\
&\qquad +\frac{1}{2} Z_{\xi s} \bar Z_{\xi s} s^2 +\frac{3}{2} \bar B Z_{\xi s} E_0 s^2 +\frac{5}{4} B Z_{\xi s} \mcZ_{\xi t} s^2 -\frac{3}{2} E_0 E_s \mcZ_{st} s^2 +\frac{3}{2} E_0^2 \mcZ_{ss} s^2 -\frac{5}{4} E_0 \mcZ_{\xi t} s^4 +\frac{1}{4} B \bar B s^2 \xi^2\Big)\Big]\\
&\quad -\frac{M}{\mu^2}\,\mathrm{Im}\Big[\frac{\bar{\mcC}^2}{\mcC^5}\Big( \frac{5}{16} B^2 E_s^2 -\frac{35}{16} B^2 Z_{\xi s}^2 +\frac{5}{16} Z_{\xi s}^2 s^2 -\frac{5}{4} B Z_{\xi s} E_0 s^2 -\frac{1}{16} E_s^2 s^2 -\frac{1}{8} E_0^2 s^4 -\frac{5}{16} B^2 s^2 \xi^2 +\frac{1}{16} s^4 \xi^2\Big)\Big]\\
&\quad +\frac{M^2}{\mu^2}\,\mathrm{Re}\Big[\frac{1}{\mcC^4}\Big( -\frac{5 B^2 \mcZ_{ss}}{32} -\frac{B^2 s^2}{16} +\frac{1}{8} B \bar B s^2 +\frac{1}{32} \bar B^2 s^2 -\frac{\mcZ_{st}^2 s^2}{32} +\frac{1}{32} \mcZ_{tt} \mcZ_{ss} s^2 +\frac{1}{32} \mcZ_{tt} s^4\Big)\Big]\\
&\quad +\frac{M^2}{\mu^2}\,\mathrm{Re}\Big[\frac{1}{\mcC^3 \bar\mcC}\Big( -\frac{1}{8} B^2 s^2 +\frac{1}{16} \bar B^2 s^2 -\frac{\mcZ_{st}^2 s^2}{16} +\frac{1}{16} \mcZ_{tt} \mcZ_{ss} s^2\Big)\Big] +\frac{M^2}{\mu^2}\,\mathrm{Re}\Big[\frac{\bar{\mcC}^2}{\mcC^6}\Big( -\frac{3B^2 s^2}{32}  +\frac{s^4}{32}\Big)\Big] \\
&\quad +\frac{M^2}{\mu^2}\,\mathrm{Re}\Big[\frac{\bar\mcC}{\mcC^5}\Big( -\frac{5 B^2 \mcZ_{ss}}{16} +\frac{1}{4} B \bar B s^2 +\frac{1}{16} \mcZ_{tt} s^4\Big)\Big] -\frac{M^2}{\mu^2}\mathrm{Re}\left[\frac{5 B^2 \mcZ_{ss}}{32 |\mcC|^4}\right] + \frac{M^2 s^2}{8\mu^2|\mcC|^4}\Big(|B|^2+\frac14\mcZ_{tt}\,s^2\Big).
\end{align*}
\end{widetext}

\clearpage
\onecolumngrid
\setcounter{equation}{0}\setcounter{section}{0}\setcounter{table}{0}\setcounter{figure}{0}
\renewcommand{\theequation}{S\arabic{equation}}
\renewcommand{\thesection}{S\Roman{section}}
\renewcommand{\thetable}{S\arabic{table}}
\renewcommand{\thefigure}{S\arabic{figure}}
\allowdisplaybreaks
\setcounter{secnumdepth}{3}

\clearpage

\begin{center}
{\large\bfseries Supplemental Material}
\end{center}

\vspace{5mm}
\noindent The following sections follow the order \blue{of the main text}. Sec.~\ref{sec:conv} fixes conventions. Sec.~\ref{sec:dyn} collects the ingredients of the multipolar dynamics (tensor extensions, moments, potential), and Sec.~\ref{sec:geo} the test-gyroscope limit in Kerr used throughout. Sec.~\ref{sec:mass} proves the expressions of the force and torque and the conservation of mass and spin, and Sec.~\ref{sec:ham} shows the equivalence between the usual MPD equations and their Hamiltonian formulation. Sec.~\ref{sec:kerr} defines the algebra of Kerr scalars, Sec.~\ref{sec:couplings} the multipolar couplings and their counting, and Sec.~\ref{sec:KQ} the construction of the constants of motion. Sec.~\ref{sec:num} explains the implementation of the resonant average method, and Sec.~\ref{sec:schw} details the Schwarzschild limit of our results. The 24 couplings at $N=5$, the explicit Hamiltonian and constants of motion, and the definitions of $\mathcal R_n$ and of the numerical zero are given in the End Matter.

\section{Conventions}\label{sec:conv}
We use geometric units $G=c=1$, signature $(-,+,+,+)$, and the conventions of Wald \cite{Wald}: $\nabla_a\nabla_b\omega_c-\nabla_b\nabla_a\omega_c=R_{abc}{}^d\omega_d$, $R_{ab}:=R_{acb}{}^c$. Abstract indices are $a,b,\ldots$, coordinate components carry Greek indices $\alpha,\beta,\ldots$. Symmetrization and antisymmetrization have unit weight, $T_{(ab)}=\frac12(T_{ab}+T_{ba})$, and indices between vertical bars are excluded, e.g., $T_{(a|b|c)}=\frac12(T_{abc}+T_{cba})$. The Hodge dual of an antisymmetric pair is $f_{ab}^\star:=\frac12\varepsilon_{abcd}f^{cd}$, and $R^\star_{abcd}:=\frac12R_{ab}{}^{ef}\varepsilon_{efcd}$. In Kerr, the background is vacuum, $R_{ab}=0$, so that $R_{abcd}$ is the Weyl tensor and its left and right duals coincide.

The body has momentum $p_a$, spin tensor $S^{ab}$, kinematic mass $\mu:=\sqrt{-p_ap^a}$, unit vector $t^a:=p^a/\mu$, spin vector $s_a:=\frac1{2\mu}\varepsilon_{bcda}S^{bc}p^d$ and spin norm $s^2:=\frac12S_{ab}S^{ab}=s_as^a$. The Tulczyjew-Dixon condition is $C^a:=S^{ab}p_b=0$, under which $S^{ab}=\varepsilon^{abcd}t_cs_d$ \cite{Ra.Iso.Dru.IntegO2.26} and one has 
\begin{equation}\label{SS}
S^{ac}S_c{}^b=s^as^b-s^2\,h^{ab}=-s^2P^{ab},\qquad \textrm{where} \qquad 
\left\{\begin{aligned}
h^{ab}&:=g^{ab}+t^at^b,\\
P^{ab}&:=h^{ab}-s^as^b/s^2.
\end{aligned}\right. 
\end{equation}
Here, $h^{ab}$ projects on the 3-plane orthogonal to $p^a$ and $P^{ab}$ on the 2-plane orthogonal to both $p^a$ and $s^a$. 
In the body's frame $t^a$, the electric and magnetic parts of the curvature are $\mcE_{ab}:=R_{acbd}t^ct^d$ and $\mcB_{ab}:=R^\star_{acbd}t^ct^d$, both symmetric, trace-free and orthogonal to $t^a$. Contractions with $t^a$ and $s^a$ are written with indices $t$ and $a$, e.g., $\mcE_{ss}:=\mcE_{ab}s^as^b$, $\nabla_s:=s^a\nabla_a$, and $\nabla_t:=t^a\nabla_a$, where the derivative acts on the curvature only, $t^a$ and $s^a$ being held fixed.

\section{Multipolar dynamics}\label{sec:dyn}
\vspace{5mm}\paragraph{MPD equations and generalized force.} In the MPD equations and Harte's generalized force (\blue{Eqs.~\eqref{MPD}--\eqref{genF}}), the force and torque are defined through their action on generalized Killing fields $k^a$ \cite{Ha.12}, which satisfy $\nabla_{(a}k_{b)}=0$ at the body's position (and are otherwise determined by their value and first derivative there). Since $k^a$ and $\nabla_{[a}k_{b]}$ are arbitrary at the point, \blue{Eq.~\eqref{genF}} determines $F_a$ and $N^{ab}$ uniquely. They are found by expanding the Lie derivative on the right hand side in terms of covariant derivative and identifying terms $\propto k_a$ (force) and $\propto \nabla_ak_b$ (torque).

\vspace{5mm}\paragraph{Tensor extensions.} The extensions $g_{ab,c_1\cdots c_n}$ are $n!$ times the coefficients of the Taylor expansion of the metric in Riemann normal coordinates centered on the body \cite{Di.74,Ha.12}. They are symmetric in $ab$, symmetric in $c_1\cdots c_n$, and satisfy $g_{a(b,c_1\cdots c_n)}=0$. Through $n=5$, in vacuum, they read \cite{MuScVdV.99,Hatz.00}
\begin{align}
g_{ab,c_1c_2}&=-\tfrac23\,R_{a(c_1|b|c_2)}\,,\\
g_{ab,c_1c_2c_3}&=-\nabla_{(c_1}R_{|a|c_2|b|c_3)}\,,\\
g_{ab,c_1c_2c_3c_4}&=-\tfrac65\,\nabla_{(c_1}\nabla_{c_2}R_{|a|c_3|b|c_4)}+\tfrac{16}{15}\,R_{a(c_1|e|c_2}R^{e}{}_{c_3|b|c_4)}\,,\\
g_{ab,c_1c_2c_3c_4c_5}&=-\tfrac43\,\nabla_{(c_1}\nabla_{c_2}\nabla_{c_3}R_{|a|c_4|b|c_5)}+\tfrac83\,R_{a(c_1|e|c_2}\nabla_{c_3}R^{e}{}_{c_4|b|c_5)}+\tfrac83\,R_{b(c_1|e|c_2}\nabla_{c_3}R^{e}{}_{c_4|a|c_5)}\,,
\end{align}
where all the indices $c_1,\ldots,c_n$ are symmetrized, the ones between vertical bars excluded. The first two are Dixon's \cite{Di.74}; the last two follow from the expansion of the metric in normal coordinates to fifth order, Eq.~(17) of Ref.~\cite{Hatz.00}, which agrees with Ref.~\cite{MuScVdV.99}.
For $N\ge4$, the extensions contain both derivatives of curvature (linear in it) as well as products of curvatures. See also \cite{Scheopner:2026sok} for a recurrence relation giving $g_{ab,c_1 \ldots c_n}$ to all orders.

\vspace{5mm}\paragraph{Symmetries of the moments.} Writing the $2^n$-pole as $J^{e_1\cdots e_{n-2}abcd}$, with $(e_1,\ldots,e_{n-2})=(c_1,\ldots,c_{n-2})$ and $(a,b,c,d)=(c_{n-1},a,c_n,b)$ in the index placement of \blue{Eq.~\eqref{genF}}, the moments satisfy
\begin{equation}
J^{e_1\cdots e_{n-2}abcd}=J^{e_1\cdots e_{n-2}[ab]cd}=J^{e_1\cdots e_{n-2}ab[cd]}=J^{(e_1\cdots e_{n-2})abcd}\,,
\end{equation}
as well as
\begin{equation}
J^{e_1\cdots e_{n-2}a[bcd]}=0\,,\qquad J^{e_1\cdots e_{n-3}[e_{n-2}ab]cd}=0\,,\qquad t_{e_1}J^{e_1\cdots e_{n-2}abcd}=0\,,
\end{equation}
the last two for $n\ge3$. The quadrupole $J^{abcd}$ thus has the symmetries of the Riemann tensor. All these algebraic symmetries and the orthogonality condition imply that the $2^n$-pole has $(n+2)(3n-1)$ independent components \cite{Di.74}. 

\vspace{5mm}\paragraph{Potential and graded moments.} Conservative moments derive from a potential, $J^{(n)}=-(\partial U/\partial g_{(n)})/a_n$ with $a_n=2(n-1)/(n+1)!$, the derivative being taken on the space of tensors with the symmetries of $g_{(n)}$. The moment of degree $j$ in the extensions, $J^{(n)}_{[j]}$, comes from the part of $U$ of degree $j+1$, so that Euler's theorem gives \blue{Eq.~\eqref{potential}}. At quadrupolar and octupolar orders, using the cyclic identity, $J^{cadb}g_{ab,cd}=-\frac12J^{abcd}R_{abcd}$ for a moment with Riemann symmetries, so that
\begin{equation}
U_2=\tfrac16\big(J^{abcd}_{[0]}+\tfrac12J^{abcd}_{[1]}\big)R_{abcd}\,,\qquad U_3=\tfrac1{12}\big(J^{abcde}_{[0]}+\tfrac12J^{abcde}_{[1]}\big)\nabla_aR_{bcde}\,.
\end{equation}
The four examples quoted \blue{in the main text} are: (i)
spin-induced quadrupole ($n=2$, $j=0$), as in Refs.~\cite{ViKuStHi.16,Ra.Iso.Dru.IntegO2.26}:
\begin{equation}
J^{abcd}_{[0]}=\frac{3\kappa_2}{\mu}\,t^{[a}\Theta^{b][c}t^{d]},\qquad \Theta^{ab}:=S^{ac}S_c{}^b,\qquad U_2=-\frac{\kappa_2}{2\mu}\mcE_{ab}s^as^b\,;
\end{equation}
(ii) spin-induced octupole ($n=3$, $j=0$), as in Ref.~\cite{Ma.15}:
\begin{equation}
J^{abcde}_{[0]}=\frac{\kappa_3}{4\mu^2}\Big(\Theta^{a[b}S^{c][d}t^{e]}-\Theta^{a[b}t^{c]}S^{de}+S^{a[b}\Theta^{c][d}t^{e]}+\Theta^{a[d}S^{e][b}t^{c]}-\Theta^{a[d}t^{e]}S^{bc}+S^{a[d}\Theta^{e][b}t^{c]}\Big),
\end{equation}
with $U_3=\frac{\kappa_3}{6\mu^2}\nabla_a\mcB_{bc}s^as^bs^c$; (iii) tidal quadrupole ($n=2$, $j=1$), linear response $Q^{ab}=-\lambda_2^{\mcE}\mcE^{ab}$ \cite{Hen.al.20,Ra.NoCarter.26}:
\begin{equation}
U=-\frac{\lambda_2^{\mcE}}{4}\mcE_{ab}\mcE^{ab}\,,\qquad J^{abcd}_{[1]}=6\,\frac{\partial U}{\partial R_{abcd}}=-3\lambda_2^{\mcE}\,t^{[a}\mcE^{b][c}t^{d]}\,,
\end{equation}
and likewise $-\frac{\lambda_2^{\mcB}}{4}\mcB_{ab}\mcB^{ab}$ for the magnetic response; (iv) tidal octupole ($n=3$, $j=1$), linear response to the gradient of the tidal field \cite{Hen.al2.20}: $U=-\frac{\lambda_3}{12}\mcE_{abc}\mcE^{abc}$ with $\mcE_{abc}:=\nabla^\perp_{(a}\mcE_{bc)}$ and $\nabla^\perp_a:=(\delta_a{}^d+t_at^d)\nabla_d$, i.e.\ the term $\frac{\mu^{(3)}}{12}G_{abc}G^{abc}$ of Ref.~\cite{Hen.al2.20} with $\lambda_3=\mu^{(3)}$ (the same normalization as $\frac{\mu^{(2)}}{4}G_{ab}G^{ab}$ for the quadrupole above), and $J^{abcde}_{[1]}=12\,\partial U/\partial(\nabla_aR_{bcde})$. In the counting of Sec.~\ref{sec:couplings}, the first two are $N=2,3$ couplings and the last two $N=4,6$ couplings.

\section{Geodesic and test-gyroscope limits in Kerr}\label{sec:geo}
\vspace{5mm}\paragraph{Geodesics.} Setting $F_a=N^{ab}=0$ and dropping the spin-curvature term, the momenta satisfy $\nabla_v p_a=0$ and $\nabla_v S^{ab}=0$. This implies $p^a\propto v^a$, and the corresponding worldline $\scL_0$ is a geodesic with $S^{ab}$ parallel-transported along it. In Kerr, using Boyer-Lindquist coordinates $(t,r,\theta,\phi)$, $z:=\cos\theta$, $\Sigma:=r^2+a^2z^2$, and Mino time $\lambda$, $\ud\lambda=\ud\tau/\Sigma$, the radial, polar and spin evolutions decouple \cite{Carter.68,Marck.83}. A bound geodesic is labeled by the constants $(E_0,L_0,Q_0)$ (per unit mass), or equivalently by $(p,e,x)$ \cite{BHPToolkit}: the semi-latus rectum $p$ and eccentricity $e$ give the turning points $r_{p,a}=p/(1\pm e)$, and $x:=L_0/\sqrt{L_0^2+Q_0}$ is the cosine of the inclination ($x>0$ prograde, $x<0$ retrograde), the polar motion oscillating in $|z|\le z_{\max}:=\sqrt{1-x^2}$. In Mino time, $r$ and $z$ are periodic with periods $\Lambda_r$ and $\Lambda_\theta$ and frequencies $\Upsilon_{r,\theta}=2\pi/\Lambda_{r,\theta}$, and $t$ and $\phi$ advance at mean rates $\Upsilon_t$, $\Upsilon_\phi$. We write the motion with the radial and polar phases in Mino time, $q_r:=\Upsilon_r\lambda$ and $\varphi:=\Upsilon_\theta\lambda$ (up to initial values), $r=r(q_r)$ and $z=z(\varphi)$, $2\pi$-periodic with $r(0)=r_p$ and $z(0)=z_{\max}$, as implemented in the BHP Toolkit \cite{BHPToolkit}.

\vspace{5mm}\paragraph{Marck tetrad and spin precession.} With $u^a$ the four-velocity and $f_{ab}$ the KY tensor (Sec.~\ref{sec:kerr}), the vector $\ell^a:=f^a{}_bu^b/\sqrt{\mathcal K}$, $\mathcal K:=Q_0+(L_0-aE_0)^2$, is unit and parallel transported along the geodesic \cite{Marck.83,VdM.20}. Completing $(u^a,\ell^a)$ with two unit vectors $e_1^a,e_2^a$, algebraic in $(r,\theta)$ and the constants, gives Marck's tetrad. A parallel-transported spin keeps constant components along $\ell^a$ and in the plane $(e_1,e_2)$, and rotates in that plane by the Marck angle $\psi$:
\begin{equation}
s^a=s_\parallel\,\ell^a+s_\perp\big[\cos\psi\,e_1^a+\sin\psi\,e_2^a\big],\qquad
\frac{\ud\psi}{\ud\lambda}=\sqrt{\mathcal K}\left[\frac{E_0(r^2+a^2)-aL_0}{r^2+\mathcal K}+\frac{a\big(L_0-aE_0(1-z^2)\big)}{\mathcal K-a^2z^2}\right].
\end{equation}
The angle $\psi$ advances at the mean rate $\Upsilon_\psi=2\pi/\Lambda_\psi$. The motion is quasi-periodic with frequencies $(\Upsilon_r,\Upsilon_\theta,\Upsilon_\phi,\Upsilon_\psi)$ (\blue{Fig.~\ref{fig:torus}}). The legs $e_1^a$ and $e_2^a$ are given in Ref.~\cite{VdM.20}.

\vspace{5mm}\paragraph{Resonant tori and closed orbits.} An $n_r{:}n_\theta$ resonance, $n_r\Lambda_r=n_\theta\Lambda_\theta=:T$, fixes one relation between $(p,e,x)$. After $T$, position and momentum have returned to their initial values, and the spin is rotated by $2\pi N_\psi$ with $N_\psi:=T/\Lambda_\psi=n_r\Upsilon_\psi/\Upsilon_r$; we write $w$ for the fractional part of $N_\psi$. At fixed $(a,x,n_r{:}n_\theta)$, we tune $e$ (and thus $p$) so that $N_\psi$ takes a prescribed rational value, $w=p_w/q$ in lowest terms; the full state then closes after $qT$. The twelve tori used \blue{in the main text} are listed in Table~\ref{tab:tori} with their associated orbital and spin elements.

\vspace{5mm}\paragraph{Initial conditions.} All closed orbits start at periapsis ($r=r_p$, $\chi_r=0$) and with $t=\phi=\lambda=0$ without loss of generality. On a resonant torus, these closed orbits form a family labeled by the initial polar phase $\varphi_0$, i.e., the polar motion is started a Mino time $\varphi_0/\Upsilon_\theta$ after its turning point $z=z_{\max}$. Since one radial period advances $\varphi$ by $2\pi n_\theta/n_r$, distinct closed orbits have distinct $\varphi_0$ modulo $2\pi/n_r$. The spin has fixed norm $s$, and its tilt with respect to $\ell^a$, $s_\parallel/s_\perp$, is held fixed in all computations, with $s_\parallel=3/10$ and $s_\perp=2/5$ (in units where $s=1/2$; only the tilt $s_\parallel/s_\perp=3/4$ matters, the spin degree being extracted exactly); the only remaining spin variable is the initial Marck phase $\psi_0:=\psi(\lambda=0)$. A closed orbit (position + momentum + spin) is thus labeled by a pair $(\varphi_0,\psi_0)$.

\section{Force, torque, mass and spin}\label{sec:mass}
In this section $U(p_a,S^{ab};g_{(n)})$ is any scalar built by contracting the extensions $g_{(n)}$ with the momentum, the spin, the metric and the Levi-Civita tensor, $J^{(n)}$ the moments it generates, and $V:=\mu U$. We write $G^a:=\partial V/\partial p_a$ and $G_{ab}:=2\,\partial V/\partial S^{ab}$ (antisymmetric), as \blue{in the main text}, and $w^a:=\partial U/\partial p_a$, $W_{ab}:=2\,\partial U/\partial S^{ab}$, so that $G^a=\mu w^a-Ut^a$ and $G_{ab}=\mu W_{ab}$. We also write $\nabla_aU:=\sum_n(\partial U/\partial g_{(n)})\cdot\nabla_ag_{(n)}$ for the gradient of $U$ through the extensions only: a sum of ordinary covariant derivatives of background tensors, contracted with $p$'s and $S$'s. Since $\mu$ does not depend on the extensions, $\nabla_aV=\mu\nabla_aU$. The worldline $\scL$ is described parametrically by four functions $x^\alpha(\tilde s)$; its tangent is the vector $v^a$, with components $\ud x^\alpha/\ud\tilde s$, and $\nabla_v:=v^a\nabla_a$ is the derivative along $\scL$.

\vspace{5mm}\paragraph{Force and torque.} \blue{Eq.~\eqref{genF}} defines $F_a$ and $N^{ab}$: following \cite{Ha.12,Ha.15}, a generalized Killing field $k^a$ is characterized, at any worldline point, by the ten free parameters $k^a$ and $\nabla_{[a}k_{b]}$, its symmetrized derivative vanishing there, $\nabla_{(a}k_{b)}=0$, exactly like a Killing vector. Writing the right-hand side of Eq.~\eqref{genF} in the form $(\ldots)_ak^a+(\ldots)_{ab}\nabla^{[a}k^{b]}$ and matching coefficients determines $F_a$ and $N^{ab}$ uniquely. By definition of the moments (Sec.~\ref{sec:dyn}), that right-hand side equals $-D_k U$, where
\begin{equation}
D_k U:=\sum_{n\ge2}\frac{\partial U}{\partial g_{(n)}}\cdot\Lie_k\, g_{(n)}
\end{equation}
is the variation of $U$ produced by Lie-dragging the background alone along $k^a$, at fixed $(p_a,S^{ab})$. To compute it, expand each Lie derivative in terms of covariant ones with the standard formula \cite{Wald},
\begin{equation}
\Lie_k\, g_{ab,c_1\cdots c_n}=k^e\nabla_e\, g_{ab,c_1\cdots c_n}+g_{eb,c_1\cdots c_n}\nabla_ak^e+g_{ae,c_1\cdots c_n}\nabla_bk^e+\sum_i g_{ab,c_1\cdots e\cdots c_n}\nabla_{c_i}k^e\,,
\end{equation}
a transport term (the first term) plus one rotation term per index (the second and third terms and the sum over $i$). At the worldline point, $\nabla_{(a}k_{b)}=0$, so that $\nabla_ak_b$ is antisymmetric there: it is an infinitesimal Lorentz transformation of the local frame, and each rotation term is its action on one index of the tensor (see \cite{Ha.12,Ha.15} for details). Summed over $n$, the transport terms give exactly $k^a\nabla_aU$. The rotation terms are handled by the following identity: in each monomial of $U$, every index is contracted, and if \emph{every} factor of the monomial (extensions, $p$'s, $S$'s, inverse metrics and volume forms) is replaced in turn by its rotation and the terms are summed, the result vanishes at the point. Indeed, an upper index contracted directly with a lower one produces two rotation terms that cancel identically; a pair of indices contracted through $g^{ab}$ produces $\nabla^{(e}k^{f)}=0$; and the rotation of $\varepsilon_{abcd}$ is $\varepsilon_{abcd}\nabla_ek^e=0$. Hence the rotation terms of the extensions equal \emph{minus} those of $p_a$ and $S^{ab}$, which are read off from $\Lie_k p_a=k^e\nabla_ep_a+p_e\nabla_ak^e$ and $\Lie_k S^{ab}=k^e\nabla_eS^{ab}-S^{eb}\nabla_ek^a-S^{ae}\nabla_ek^b$:
\begin{equation}\label{DkU}
D_k U=k^a\nabla_aU-w^a\,p_e\nabla_ak^e+W_{ab}\,S^{eb}\nabla_ek^a
=k^a\nabla_aU-\nabla_ak_b\,\big(w^{[a}p^{b]}+W^{[a}{}_cS^{b]c}\big)\,,
\end{equation}
the last equality using the antisymmetry of $\nabla_ak_b$ at the point. Matching coefficients in \blue{Eq.~\eqref{genF}} then gives
\begin{equation}\label{FNexp}
F_a=-\nabla_aU\,,\qquad N^{ab}=2w^{[a}p^{b]}+2W^{[a}{}_cS^{b]c}\,.
\end{equation}
Since $G^{[a}p^{b]}=\mu w^{[a}p^{b]}$ (the $-Ut^a$ part of $G^a$ cancels $p^{b]}$) and $G_{ab}=\mu W_{ab}$, these are \blue{Eqs.~\eqref{FN}}.

\vspace{5mm}\paragraph{Velocity.} The MPD equations determine $\nabla_vp_a$ and $\nabla_vS^{ab}$ once $v^a$ is known, but they do not prescribe $v^a$. It is however determined once the Tulczyjew-Dixon spin supplementary condition $p_a S^{ab}=0$ is chosen, up to the normalization of the parameter. This is called the momentum-velocity relation \cite{CoNa.15}, and obtained as follows. Differentiating $p_a S^{ab}=0$ along $\scL$: $0=\nabla_v(S^{ab}p_b)=(\nabla_vS^{ab})p_b+S^{ab}\nabla_vp_b$, inserting \blue{Eq.~\eqref{MPD}} and expanding $2p^{[a}v^{b]}p_b=(p_bv^b)p^a+\mu^2v^a$ gives
\begin{equation}\label{vel}
\big(\mu^2\delta^a{}_e+S^{ab}R_{bcde}S^{cd}\big)v^e=-p_bv^bp^a-N^{ab}p_b-S^{ab}F_b,
\end{equation}
a linear equation for $v^a$, from which everything follows by contractions. (i) Contracting with $p_a$, using $p_aS^{ab}=0$ and the antisymmetry of $N^{ab}$, both sides reduce to $\mu^2(p_av^a)$: the component of $v^a$ along $p^a$ is not determined, which is the freedom of reparameterizing $\tilde s$. A particularly useful choice is $v^ap_a=-\nu$ with $\nu:=\mu-t_aG^a$. (ii) Define $\delta^a:=\mu v^a-p^a-G^a$; then $p_a\delta^a=-\mu\nu+\mu^2-\mu\,t_aG^a=0$. (iii) Contracting \eqref{vel} with $s_a$, using $s_aS^{ab}=0$ and $s_ap^a=0$, gives $\mu^2(s_av^a)=-s_aN^{ab}p_b$; inserting the torque of \blue{Eq.~\eqref{FN}}, $-s_aN^{ab}p_b=\mu\,s_aG^a$ (the other three terms cancel from $s_ap^a=0$, $S^{bc}p_b=0$ and $s_aS^{ac}=0$), so that $s_a(\mu v^a)=s_aG^a$, hence $s_a\delta^a=0$. 
(iv) $\delta^a$ is thus orthogonal to both $p^a$ and $s^a$, so that $P^a{}_b\delta^b=\delta^a$. Eq.~\eqref{SS} then gives $S^{ab}S_{bc}\delta^c=-s^2\delta^a$, i.e.\ $\delta^a=S^{ab}\zeta_b$ with $\zeta_b:=-S_{bc}\delta^c/s^2$. Altogether, we have established that
\begin{equation}\label{vsol}
\mu\,v^a=p^a+G^a+\delta^a\,,\qquad \delta^a=S^{ac}\zeta_c\,,\qquad p_a\delta^a=s_a\delta^a=0.
\end{equation}
The derivations below only use \eqref{vsol}, never the explicit form of $\delta^a$ or $\zeta_a$, which is fixed by the projection $P^a{}_b$ of \eqref{vel} provided $\mcD:=\mu^2+\frac14R_{abcd}S^{ab}S^{cd}\neq0$.

\vspace{5mm}\paragraph{Equations of motion in the parameter $s_h$.} We now use \blue{the parameter $s_h$ of the main text}, defined by $\ud\tilde s=\mu\,\ud s_h$, and write $\dot T:=\Dl{T}=\mu\nabla_vT$ for any tensor $T$ defined along $\scL$. Since $S^{bc}$ is antisymmetric and $R_{[abc]d}=0$, the spin-curvature coupling in \blue{Eq.~\eqref{MPD}} reads $R_{abcd}S^{bc}v^d=-\frac12R_{abcd}v^bS^{cd}$. With $\mu F_a=-\nabla_aV$, the MPD equations multiplied by $\mu$ then read
\begin{equation}\label{MPDlam}
\dot p_a=-\tfrac12R_{abcd}\,(\mu v^b)\,S^{cd}-\nabla_aV\,,\qquad
\dot S^{ab}=2p^{[a}(\mu v^{b]})+\mu N^{ab}=2G^{[a}{}_cS^{b]c}+2p^{[a}\delta^{b]}\,,
\end{equation}
the last equality from inserting \eqref{vsol} and the torque of \blue{Eq.~\eqref{FN}}, the terms $2p^{[a}G^{b]}$ and $2G^{[a}p^{b]}$ cancelling each other.

\vspace{5mm}\paragraph{Conservation of the spin.} From \eqref{MPDlam},
\begin{equation}
\frac{\ud s^2}{\ud s_h}=S_{ab}\dot S^{ab}=2S_{ab}G^a{}_cS^{bc}+2S_{ab}p^a\delta^b=0\,.
\end{equation}
Now, the first term on the right hand side is the contraction of the symmetric tensor $S_{ab}S^{bc}$ with the antisymmetric $G^a{}_c$, and the second term is proportional to the spin supplementary condition. 
Both thus vanish and $s^2$ is exactly conserved, for any $U$.

\vspace{5mm}\paragraph{Conservation of the mass.} Here $m^2:=\mu^2-2\mu U=\mu^2-2V$. Along $\scL$, $V$ is a function of $s_h$ through its arguments only, so differentiating each of its monomials factor by factor, and using $\nabla_a g_{bc}=\nabla_a \varepsilon_{bcde}=0$, gives
\begin{equation}\label{Leib}
\frac{\ud V}{\ud s_h}=(\mu v^a)\,\nabla_aV+G^a\,\dot p_a+\tfrac12\,G_{ab}\,\dot S^{ab}\,,
\end{equation}
the three terms collecting the extensions, momentum and spin factors, respectively. For the spin-induced quadrupole, for instance, $V\propto J^{abcd}(p,S)R_{abcd}$ and \eqref{Leib} is the Leibniz rule $(J\cdot R)^{\textstyle\cdot}=\dot J\cdot R+J\cdot\dot R$, with $\dot J$ itself expanded into the $\dot p$ and $\dot S$ terms, as presented, e.g., in \cite{Ra.Iso.Dru.IntegO2.26}. With $\ud\mu^2/\ud s_h=-2p^a\dot p_a$ and \eqref{Leib}, the derivative of $m^2$ is
\begin{equation}
\frac{\ud m^2}{\ud s_h}=-2\Big((p^a+G^a)\,\dot p_a+(\mu v^a)\,\nabla_aV+\tfrac12G_{ab}\dot S^{ab}\Big)\,.
\end{equation}
By \eqref{vsol}, $p^a+G^a=\mu v^a-\delta^a$, and by contracting \eqref{MPDlam} with $v^a$, $\mu v^a\dot p_a=-\mu v^a\nabla_aV$ (using $R_{abcd}v^av^b=0$): the ``work'' of the force exactly cancels the variation of $V$ through the background, leaving
\begin{equation}
\frac{\ud m^2}{\ud s_h}=2\delta^a\dot p_a-G_{ab}\dot S^{ab}=0\,,
\end{equation}
the last equality following from the
two terms being equal. Indeed, for the second, inserting \eqref{MPDlam} and using the antisymmetry of $G_{ab}$, $G_{ab}\dot S^{ab}=2G_{ab}G^a{}_cS^{bc}+2G_{ab}p^a\delta^b$; the first term vanishes by symmetry, so that $G_{ab}\dot S^{ab}=2p^aG_{ab}\delta^b$. For the first, differentiating $p_aS^{ab}=0$ along $\scL$ once more gives $S^{ca}\dot p_a=-\dot S^{ca}p_a$, so that, with $\delta^a=S^{ac}\zeta_c$,
\begin{equation}
\delta^a\dot p_a=-\zeta_cS^{ca}\dot p_a=\zeta_c\,\dot S^{ca}p_a=\zeta_c\big(-p_aG^a{}_eS^{ce}+\mu^2\delta^c\big)=p^aG_{ae}\delta^e\,,
\end{equation}
where $\dot S^{ca}p_a$ was evaluated with \eqref{MPDlam} and used $S^{ae}p_a=0$ and $p_a\delta^a=0$.  

Note that, with the moments $J^{(n)}$ inserted as they are, no approximation enters these calculations: $m$ and $s$ are exactly conserved, for any $U$. We stress that the conservative assumption (i.e., the specific form of $U$) is, ultimately, what is responsible for these exact conservation laws. 

\vspace{5mm}\paragraph{Reparametrization invariance.} The reader familiar with Dixon's formalism, and in particular with the integral formulae for the moments, may worry that giving the moments as prescribed functions breaks the invariance of the MPD equations under a change of parameter along $\scL$. For moments defined by Dixon's integrals, $p_a$ and $S^{ab}$ are hypersurface integrals and do not depend on the parameter, whereas the moments are densities along $\scL$: under $\tilde s\to\tilde s'$ they transform as $J'=(\ud\tilde s/\ud\tilde s')\,J$, exactly like $\nabla_vp_a$ and $\nabla_vS^{ab}$, so that $F_a$ and $N^{ab}$ rescale consistently and the MPD equations keep their form. Moments given as functions $J^{(n)}(p,S,g,\varepsilon,g_{(m)})$ do not transform in this way, and inserting them in \blue{Eq.~\eqref{MPD}} implicitly fixes the parameter. A manifestly invariant formulation is obtained by taking $-(v_ap^a/\nu)\,J^{(n)}$ as the Dixon moments: since $v^a\to(\ud\tilde s/\ud\tilde s')\,v^a$, these transform as densities, and they reduce to $J^{(n)}$ for $v^ap_a=-\nu$. With this prescription, the MPD equations hold for any parameter, and $m$ and $s$ are exactly conserved for any choice of $v^a$. See \cite{Di.73,Ha.12,Ha.15,Di.15,StPu.10} for further discussions on this.

\vspace{5mm}\paragraph{Killing charges.} For any Killing vector $k^a$, $p_ak^a+\frac12S^{ab}\nabla_ak_b$ is conserved, since $\Lie_kg_{(n)}=0$ makes the right-hand side of \blue{Eq.~\eqref{genF}} vanish \cite{Di.74,Ha.15}. In Kerr, this provides two constants of motion, the energy $E$ and axial angular momentum $L_z$ via the two Killing fields $(\partial_t)^a$ and $(\partial_\phi)^a$ in Boyer-Lindquist coordinates.

\section{Hamiltonian formulation}\label{sec:ham}
\vspace{5mm}\paragraph{Derivative on phase space.} A function $f(x^\alpha,p_\alpha,S^{\alpha\beta})$ on phase space $\mcM$ is not a tensor field on spacetime, so $\nabla_\alpha f$ is a priori undefined, and $\partial f/\partial x^\alpha$ alone is not covariant, as it depends on the coordinates through the components $p_\alpha$ and $S^{\alpha\beta}$. The covariant notion is defined by parallel transport. Given a vector $u^\alpha$ at $x$, let $x(\varsigma)$ be any curve through $x$ with tangent $u^\alpha$ there, and let $\big(p_\alpha(\varsigma),S^{\alpha\beta}(\varsigma)\big)$ be the parallel transport of $(p_\alpha,S^{\alpha\beta})$ along it. We define an operator $\rD_\alpha$ by
\begin{equation}\label{Ddef}
u^\alpha\,\rD_\alpha f:=\frac{\ud}{\ud\varsigma}\,f\big(x(\varsigma),p_\alpha(\varsigma),S^{\alpha\beta}(\varsigma)\big)\Big|_{\varsigma=0}\,.
\end{equation}
The right-hand side is linear in $u^\alpha$, so that \eqref{Ddef} defines $\rD_\alpha f$ unambiguously, from the connection alone. The chain rule, with $\ud p_\beta/\ud\varsigma=\Gamma^\sigma_{\alpha\beta}p_\sigma u^\alpha$ and $\ud S^{\beta\gamma}/\ud\varsigma=-(\Gamma^\beta_{\alpha\sigma}S^{\sigma\gamma}+\Gamma^\gamma_{\alpha\sigma}S^{\beta\sigma})u^\alpha$ along the transport, gives its explicit form,
\begin{equation}\label{Dcoord}
\rD_\alpha f=\frac{\partial f}{\partial x^\alpha}+\Gamma^\sigma_{\alpha\beta}\,p_\sigma\,\frac{\partial f}{\partial p_\beta}-\big(\Gamma^\beta_{\alpha\sigma}S^{\sigma\gamma}+\Gamma^\gamma_{\alpha\sigma}S^{\beta\sigma}\big)\frac{\partial f}{\partial S^{\beta\gamma}}\,,
\end{equation}
the sixteen components of $S^{\beta\gamma}$ being treated as independent in $\partial f/\partial S^{\beta\gamma}$. Two properties follow directly from the definition \eqref{Ddef}. (i) If $f$ does not depend on $(p_\alpha,S^{\alpha\beta})$, i.e.\ if $f$ is a scalar field, then $\rD_\alpha f=\nabla_\alpha f$: the operator extends the spacetime covariant derivative to $\mcM$. (ii) If $f$ is built by contracting background tensors with $p_\alpha$ and $S^{\alpha\beta}$, then $\rD_\alpha f$ differentiates the background tensors only, e.g., $\rD_\alpha(T^{\beta\gamma}p_\beta p_\gamma)=(\nabla_\alpha T^{\beta\gamma})\,p_\beta p_\gamma$, since the transported $(p_\alpha,S^{\alpha\beta})$ contribute nothing by construction. In particular, $\rD_\alpha\mu=0$, $\rD_\alpha s^2=0$, $\rD_\alpha C^a=0$, and $\rD_\alpha U=\sum_n(\partial U/\partial g_{(n)})\cdot\nabla_\alpha g_{(n)}$ is the gradient of $U$ through the extensions only, written $\nabla_\alpha U$ \blue{in the main text}.

\vspace{5mm}\paragraph{Brackets and Hamilton's equations.} On $\mcM\ni(x^\alpha,p_\alpha,S^{\alpha\beta})$, the Poisson brackets $\{\cdot,\cdot\}$ are those presented in Refs.~\cite{WiStLu.19,Ra.Iso.IntegO1.26,Ra.Iso.Dru.IntegO2.26}. With them, for any $H(x,p,S)$, the flow $\dot f=\{f,H\}$ reads \cite{Ra.Iso.Dru.IntegO2.26}
\begin{equation}\label{Hamgen}
\dot x^\alpha=\frac{\partial H}{\partial p_\alpha}\,,\qquad \Dl{p_\alpha}=-\tfrac12R_{\alpha\beta\gamma\delta}\,\dot x^\beta S^{\gamma\delta}-\rD_\alpha H\,,\qquad \Dl{S^{\alpha\beta}}=2H^{S[\alpha}{}_\gamma S^{\beta]\gamma}\,,
\end{equation}
where $H^S_{\alpha\beta}:=2\,\partial H/\partial S^{\alpha\beta}$, and $\dot x^\alpha$ are the derivatives of the coordinates of the flow's trajectory, i.e., the components of its tangent vector. Notice that the connection terms of the brackets (e.g., $\{p_\alpha,S^{\beta\gamma}\} = 2\Gamma^{[\gamma}_{\alpha\delta} S^{\beta]\delta}$) are exactly those that enter \eqref{Dcoord} to turn $\partial H/\partial x^\alpha$ into $\rD_\alpha H$, and combine with $\ud p_\alpha/\ud s_h$, $\ud S^{\alpha\beta}/\ud s_h$ into the covariant derivatives $\Dl{p_\alpha}$ and $\Dl{S^{\alpha\beta}}$ of the left-hand sides.

\vspace{5mm}\paragraph{Poisson bracket on $\mcP$.} On the constraint surface $C^\alpha=0$, one has $\{C^\alpha,C^\beta\}=-\mcD\,S^{\alpha\beta}$ \cite{WiStLu.19,Ra.Iso.IntegO1.26}. Only two of the four $C^\alpha$ are independent ($p_\alpha C^ \alpha \equiv 0$ and $s_\alpha C^\alpha \equiv 0$), and since $S^\alpha{}_\beta S^\beta{}_\gamma S^\gamma{}_\delta=-s^2S^\alpha{}_\delta$ by \eqref{SS}, the matrix $S_{\alpha\beta}/(\mcD s^2)$ is an inverse of $\{C^\alpha,C^\beta\}$ under the spin supplementary condition as was noticed already in \cite{WiStLu.19}. This gives \blue{Eq.~\eqref{Dirac}}.

\vspace{5mm}\paragraph{Hamilton's equations on $\mcP$.} The Dirac correction in \blue{Eq.~\eqref{Dirac}} accounts for the Tulczyjew-Dixon condition, which is not an evolution equation but an algebraic condition on the phase-space variables: the Poisson flow of $H$ does not preserve $C^\alpha=0$ in general, while the Poisson bracket $\{\cdot,\cdot\}^{\rm D}$ is built so that it does. We follow Ref.~\cite{Ra.Iso.Dru.IntegO2.26}, which gives in detail the minimal theory needed, and why this bracket is required. With \blue{Eq.~\eqref{Dirac}}, Hamilton's equations read
\begin{equation}\label{DiracFlow}
\frac{\ud f}{\ud s_h}=\{f,H\}^{\rm D}=\{f,H\}-\{f,C^\gamma\}\,\zeta_\gamma\,,\qquad \zeta_\gamma:=\frac{S_{\gamma\delta}\{C^\delta,H\}}{\mcD s^2}\,,
\end{equation}
i.e.\ the flow \eqref{Hamgen} plus a correction, which we compute for $f=x^\alpha,p_\alpha,S^{\alpha\beta}$ from the brackets \blue{of the main text} and $C^\gamma=S^{\gamma\delta}p_\delta$. Define $\delta^\alpha:=S^{\alpha\gamma}\zeta_\gamma$, so that $p_\alpha\delta^\alpha=s_\alpha\delta^\alpha=0$; since $S_{\gamma\delta}p^\gamma=S_{\gamma\delta}s^\gamma=0$, $\zeta_\gamma$ is itself orthogonal to $p^\gamma$ and $s^\gamma$, and it coincides with (the components of) the vector $\zeta_c$ of Sec.~\ref{sec:mass}.
\begin{itemize}
\item Position: $\{x^\alpha,C^\gamma\}=S^{\gamma\delta}\{x^\alpha,p_\delta\}=S^{\gamma\alpha}$, so the correction to $\ud x^\alpha/\ud s_h$ is $-S^{\gamma\alpha}\zeta_\gamma=\delta^\alpha$.
\item Momentum: $\{p_\alpha,C^\gamma\}=\{p_\alpha,S^{\gamma\delta}\}p_\delta+S^{\gamma\delta}\{p_\alpha,p_\delta\}$. Contracted with $-\zeta_\gamma$, the first term gives $\Gamma^\delta_{\alpha\varepsilon}\,p_\delta\,\delta^\varepsilon$ (the other term vanishes from $S^{\delta\varepsilon}p_\delta=0$), which is exactly the Christoffel term needed to turn $\ud p_\alpha/\ud s_h$ into $\Dl{p_\alpha}$ once $\delta^\alpha$ is added to the velocity. The second gives $R_{\alpha\varepsilon\eta\delta}S^{\varepsilon\eta}\delta^\delta=-\frac12R_{\alpha\delta\varepsilon\eta}\,\delta^\delta S^{\varepsilon\eta}$, i.e., the spin-curvature coupling evaluated on $\delta^\alpha$. 
\item Spin: $\{S^{\alpha\beta},C^\gamma\}=\{S^{\alpha\beta},S^{\gamma\delta}\}p_\delta+S^{\gamma\delta}\{S^{\alpha\beta},p_\delta\}$. Contracted with $-\zeta_\gamma$, the second term gives the Christoffel terms that turn $\ud S^{\alpha\beta}/\ud s_h$ into $\Dl{S^{\alpha\beta}}$ once $\delta^\alpha$ is added to the velocity. Expanding $\{S^{\alpha\beta},S^{\gamma\delta}\}=g^{\alpha\delta}S^{\gamma\beta}-g^{\alpha\gamma}S^{\delta\beta}-g^{\beta\delta}S^{\gamma\alpha}+g^{\beta\gamma}S^{\delta\alpha}$, and using $p_\alpha S^{\alpha\beta}=0$ the non-vanishing terms are $2p^{[\alpha}\delta^{\beta]}$.
\end{itemize}
Altogether, the flow of $\{\cdot,\cdot\}^{\rm D}$ is simply \eqref{Hamgen} with $\dot x^\alpha\to\dot x^\alpha+\delta^\alpha$ in the velocity and in the spin-curvature coupling, and with $2p^{[\alpha}\delta^{\beta]}$ added to $\Dl{S^{\alpha\beta}}$.

\vspace{5mm}\paragraph{Equivalence with MPD.} For $H=-\frac12\mu^2+\mu U$, with $\partial\mu/\partial p_\alpha=-t^\alpha$, $\partial H/\partial p_\alpha=p^\alpha+G^\alpha$, $\rD_\alpha H=\mu\,\rD_\alpha U$ and $H^S_{\alpha\beta}=G_{\alpha\beta}$, the flow of $\{\cdot,\cdot\}^{\rm D}$ reads
\begin{equation}
\dot x^\alpha=p^\alpha+G^\alpha+\delta^\alpha\,,\quad \Dl{p_\alpha}=-\tfrac12R_{\alpha\beta\gamma\delta}\,\dot x^\beta S^{\gamma\delta}-\mu\,\rD_\alpha U\,,\quad \Dl{S^{\alpha\beta}}=2p^{[\alpha}\dot x^{\beta]}+2G^{[\alpha}p^{\beta]}+2G^{[\alpha}{}_\gamma S^{\beta]\gamma}\,.
\end{equation}
Dividing by $\mu$, i.e.\ using the parameter $\tilde s$ with $\ud\tilde s=\mu\,\ud s_h$, these are \blue{the MPD equations \eqref{MPD} with the force and torque \eqref{FN}}. The flow preserves $C^a=0$ by construction. The conservation laws for $(m,s)$ of Sec.~\ref{sec:mass} now follow simply from $\{H,H\}^{\rm D}=0$ and $s^2$ being a Casimir of the brackets.

\section{Algebra of Kerr scalars}\label{sec:kerr}
\vspace{5mm}\paragraph{Killing-Yano tensor and descendants.} We follow the conventions of Ref.~\cite{Ra.Iso.Dru.IntegO2.26}. The Kerr spacetime admits a non-degenerate Killing-Yano tensor $f_{ab}$, $\nabla_{(a}f_{b)c}=0$, with Hodge dual $f^\star_{ab}$, from which descend the Killing vector $\xi_a:=\frac13\nabla^bf^\star_{ab}$, which is $\xi^a=(\partial_t)^a$ in Kerr \cite{Ra.Iso.Dru.IntegO2.26}, so that $E_0=-\xi_at^a$ is the specific energy of geodesics, and the self-dual complex conformal Killing-Yano tensor
\begin{equation}
\mcC Z_{ab}:=f_{ab}-\ui f^\star_{ab}\,,\qquad Z^\star_{ab}=\ui Z_{ab}\,,\qquad Z_{ab}Z^{ab}=-4\,,
\end{equation}
which defines the complex scalar $\mcC$ and the self-dual bivector $Z_{ab}$. In Boyer-Lindquist coordinates, $\mcC=a\cos\theta+\ui r$. We also use the real symmetric tensor $\mcZ_{ab}:=Z_{ac}\bar Z^c{}_b$. The Weyl scalar is $\Psi_2=\ui M\mcC^{-3}$, so that $\mcC\propto\Psi_2^{-1/3}$.

\vspace{5mm}\paragraph{Algebraic relations.} In vacuum, one has \cite{Ha.20,Ra.Iso.Dru.IntegO2.26}
\begin{equation}
g_{ab}=Z_{ac}Z^c{}_b,\qquad
\varepsilon_{abcd}=-\ui\big(Z_{ab}Z_{cd}+2Z_{c[a}Z_{b]d}\big), \quad 
R_{abcd}=\mathrm{Re}\Big[\frac{\ui M}{\mcC^3}\big(3Z_{ab}Z_{cd}+2g_{a[c}g_{d]b}-\ui\varepsilon_{abcd}\big)\Big]\,,
\end{equation}
so that the metric, Levi-Civita and Riemann tensors are algebraic in $(\mcC,Z_{ab})$ \cite{Ra.Iso.Dru.IntegO2.26}. The symmetric Killing-St\"ackel tensor $K^{ab}$, which generates the Carter constant $Q_0:=K^{ab}p_a p_b$, is given by $K_{ab}=f_{ac}f^c{}_b=\frac12\mathrm{Re}[\mcC^2]g_{ab}+\frac12|\mcC|^2\mcZ_{ab}$.

\vspace{5mm}\paragraph{Closure relations.} The derivatives close on $(\mcC,\xi^a,Z_{ab})$ (see \cite{ComDru.22,Ra.Iso.IntegO1.26} for detailed proofs):
\begin{equation}
\nabla_a\mcC=-\ui Z_{ab}\xi^b\,,\qquad
\nabla_a\xi_b=-M\,\mathrm{Re}\Big[\frac{Z_{ab}}{\mcC^2}\Big]\,,\qquad
\nabla_a(\mcC Z_{bc})=\varepsilon_{abcd}\xi^d+2\ui\,g_{a[b}\xi_{c]}\,.
\end{equation}
The first follows from the last by contraction with $Z^{bc}$; the last is $\nabla_af_{bc}=\varepsilon_{abcd}\xi^d$ and $\nabla_af^\star_{bc}=-2g_{a[b}\xi_{c]}$  combined \cite{Ra.Iso.Dru.IntegO2.26}. Consequently, the covariant derivative of any scalar built from $(\mcC,\xi^a,Z_{ab},g_{ab},\varepsilon_{abcd},g_{ab,c_1\ldots c_n})$ and the body's momentum vectors $(t^a,s^a)$ is again such a scalar.

\vspace{5mm}\paragraph{The nineteen scalars.} With the vectors $u,v\in\{\xi^a,t^a,s^a\}$, every full contraction reduces to $u\cdot T\cdot v$ with $T\in\{g,Z,\bar Z,\mcZ\}$, plus $\mcC$, $\bar\mcC$ and $\mu$. Since $Z_{ab}$ is antisymmetric, $t_at^a=-1$ and $t_as^a=0$, the non-trivial contractions are
\begin{itemize}
\item contractions with $g_{ab}$: $E_0:=-\xi_at^a$, $E_s:=-\xi_as^a$, $\xi^2:=\xi_a\xi^a$, $s^2=s_a s^a$;
\item contractions with $Z_{ab}$ (complex): $Z_{\xi t}$, $Z_{st}=:B$, $Z_{\xi s}$, and their complex conjugates;
\item contractions with $\mcZ_{ab}$ (real): $\mcZ_{\xi\xi}$, $\mcZ_{\xi t}$, $\mcZ_{\xi s}$, $\mcZ_{tt}$, $\mcZ_{st}$, $\mcZ_{ss}$;
\item $\mcC$, $\bar\mcC$ and $\mu$;
\end{itemize}
with $Z_{uv}:=Z_{ab}u^av^b$ and $\mcZ_{uv}:=\mcZ_{ab}u^av^b$. Their spin degrees are $1$ for $E_s,B,Z_{\xi s},\mcZ_{\xi s},\mcZ_{st}$ (and conjugates), $2$ for $s^2,\mcZ_{ss}$, and $0$ otherwise. Counting real and imaginary parts, they are nineteen real functions on the reduced phase space, which is nine-dimensional ($s^2$ included): only nine of them are thus functionally independent, but we never need such a set. Since all spacetime tensors are algebraic in $(\mcC,\xi^a,Z_{ab})$, every scalar function on $\mcP$ of interest to us (masses, norms, Hamiltonians, potentials, constants of motion, etc.) is a function of these scalars. In all cases considered in this work, they are a polynomials in the scalars and in $\mcC^{-1}$, $\bar\mcC^{-1}$ and $\mu^{-1}$. 

\vspace{5mm}\paragraph{Bracket closure.} The Poisson bracket $\{\cdot,\cdot\}^{\rm D}$ of any two of the nineteen scalars, computed from \blue{Eq.~\eqref{Dirac}} and the relations above, is a polynomial in the scalars, in $\mcC^{-1}$, $\bar\mcC^{-1}$, $\mu^{-1}$, and in $\mcD^{-1}$, with
\begin{equation}
\mcD = \mu^2 - M \,\mathrm{Im}\left[ \frac{s^2-3 B^2}{\mcC^3} \right]
,
\end{equation}
which follows from $\mcD=\mu^2+\tfrac14 R_{abcd}S^{ab}S^{cd}$ and the decomposition of the Riemann tensor above. There are 153 nonzero brackets among them, all computed symbolically and checked against a direct evaluation of the Poisson bracket $\{\cdot,\cdot\}^{\rm D}$ in Boyer-Lindquist components at random phase-space points. An example is \blue{Eq.~\eqref{CCbar}}: the bracket $\{\mcC,\bar\mcC\}$ vanishes, since $\mcC$ depends on the position only, while the bracket $\{\mcC,\bar{\mcC}\}^{\rm D}$ does not. The bracket of any two rational functions of the scalars then follows from the Leibniz rule, $\{F,G\}^{\rm D}=\sum_{i,j}(\partial F/\partial X_i)(\partial G/\partial X_j)\{X_i,X_j\}^{\rm D}$.

\section{Multipolar couplings}\label{sec:couplings}
\blue{In this section, a \emph{coupling} means a multipole$\times$curvature coupling, i.e.\ a scalar; its coefficient in $U$ is its coupling constant.} We discuss how the finite space of possible multipole couplings was constructed at each order, and how it compares with the literature.

\vspace{5mm}\paragraph{Notation.} The curvature enters through $\mcE_{ab}$ and $\mcB_{ab}$ (Sec.~\ref{sec:conv}) and their covariant derivatives, the vectors $t^a$ and $s^a$ being held fixed. We write
\begin{equation}
\begin{gathered}
\mcE_{ss}:=\mcE_{ab}s^as^b\,,\qquad (\mcE^2)_{ss}:=\mcE_{ac}\mcE^c{}_bs^as^b\,,\qquad (\mcE\mcB)_{ss}:=\mcE_{ac}\mcB^c{}_bs^as^b\,,\qquad \mcE^c{}_s:=\mcE^c{}_as^a\,,\\
\mathrm{tr}\,\mcE^2:=\mcE_{ab}\mcE^{ab}\,,\qquad \mathrm{tr}(\mcE\mcB):=\mcE_{ab}\mcB^{ab}\,,\qquad \nabla_s:=s^a\nabla_a\,,\qquad \nabla_t:=t^a\nabla_a\,,
\end{gathered}
\end{equation}
and likewise with $\mcB_{ab}$ in place of $\mcE_{ab}$. In expressions such as $(\mcE\nabla_s\mcB)_{ss}:=\mcE_{ac}\,s^d\nabla_d\mcB^c{}_b\,s^as^b$, the derivative acts on the curvature only.

\vspace{5mm}\paragraph{Order of a coupling.} A coupling is a scalar built from $\nabla^kR$, $t^a$, $s^a$, $g_{ab}$ and $\varepsilon_{abcd}$, entering $U$ \cite{Ha.15}. Its order $N$ is its number of derivatives of the metric: each Riemann tensor counts two, each additional $\nabla$ one. It is written with the power of $s^2$ that makes it of degree $N$ in the spin, and with the factor $\mu^{1-N}$ in $U$ (as in $U_2$, $U_3$ and \blue{Eq.~\eqref{H4}}), so that a coupling of order $N$ contributes at $O(S^N)$. A moment $J^{(n)}_{[j]}$ with $j$ extensions of total derivative order $\ell$ gives couplings of order $N=n+\ell$: spin-induced moments have $N=n$, the tidal quadrupole ($\mathrm{tr}\mcE^2$) has $N=4$, and the tidal octupole ($\nabla \mcE\,\nabla \mcE$) $N=6$. Grading by $N$ makes the list of couplings at each order finite, in contrast with a bare spin-grading. 

\vspace{5mm}\paragraph{Reduction.} In vacuum, $R_{abcd}$ is the Weyl tensor, so $\mcE_{ab}$ and $\mcB_{ab}$ are trace-free. The vanishing divergences and the propagation equations for $\mcE_{ab}$ and $\mcB_{ab}$ given in the End Matter follow from the contracted Bianchi identities $\nabla^aR_{abcd}=0=\nabla^aR^\star_{abcd}$. Contractions of two derivatives reduce to products of curvatures through Penrose's wave equation for the vacuum Weyl tensor \cite{Pen.60}, $\nabla^e\nabla_e R_{abcd}=-R_{ab}{}^{ef}R_{efcd}-2R_{aecf}R_b{}^e{}_d{}^f+2R_{aedf}R_b{}^e{}_c{}^f$. This gives, for example, $s^as^b\,\nabla^e\nabla_e\mcE_{ab}=6(\mcE^2-\mcB^2)_{ss}-2s^2\,\mathrm{tr}(\mcE^2-\mcB^2)$. We keep only time-reversal even couplings (in accordance with our conservative assumption). Parity is not assumed at $N=5$: there, the parity-odd couplings, i.e., with an odd total number of $\mathcal B_{ab}$'s, $\varepsilon_{abcd}$'s and $s^a$'s, are included (12 of the 24). At $N=4$, the parity-odd quadratic couplings are included as well, and come out zero; at $N=2,3$ and for the linear couplings at $N=4$, only the parity-even ones are kept.

\vspace{5mm}\paragraph{Couplings at $N=2,3$.} One coupling each, $\mcE_{ss}$ and $\nabla_s\mcB_{ss}$, with potentials $U_2,U_3$ \blue{of Eq.~\eqref{Ukerr}}. 

\vspace{5mm}\paragraph{Couplings at $N=4$.} The parity-even couplings are $\ell_1=\nabla_s\nabla_s\mcE_{ss}$, $\ell_2=s^2\nabla_t\nabla_t\mcE_{ss}$ and $I_{1,\ldots,6}=\mcE_{ss}^2$, $s^2(\mcE^2)_{ss}$, $s^4\mathrm{tr}\mcE^2$, $\mcB_{ss}^2$, $s^2(\mcB^2)_{ss}$, $s^4\mathrm{tr}\mcB^2$ (\blue{Eq.~\eqref{H4}}). Every other contraction of $\nabla\nabla R$ with four spins reduces to these. The parity-odd quadratic couplings $\mcE_{ss}\mcB_{ss}$, $s^2(\mcE\mcB+\mcB\mcE)_{ss}/2$ and $s^4\mathrm{tr}(\mcE\mcB)$ were included in the solve and come out zero.

\vspace{5mm}\paragraph{The coupling $\ell_2$ at $N=4$.} The integrable point \eqref{O4result} leaves $\lambda_4$ free, because the coupling $\ell_2$ only amounts to a change of phase-space coordinates, as we explain now. Along the unperturbed flow, $t^a$ and $s^a$ are parallel transported and $\mu$ is constant, so that $\{F,H_0\}^{\rm D}=\ud F/\ud s_h=\mu\,t^a\nabla_aF$ for any function $F$ of the scalars; hence $\ell_2=\{\chi,H_0\}^{\rm D}+O(S^5)$ with $\chi:=\frac{s^2}{\mu}\,t^a\nabla_a\mcE_{ss}$. Now define new coordinates $z'=z-\frac{\lambda_4}{\mu^2}\{z,\chi\}^{\rm D}$, with $z=(x^\alpha,p_\alpha,S^{\alpha\beta})$: this map preserves the Poisson brackets to this order, and by the chain rule $H(z')=H(z)-\frac{\lambda_4}{\mu^2}\{H,\chi\}^{\rm D}=H(z)+\lambda_4\ell_2/\mu^2+O(S^5)$. The Hamiltonian with the term $\lambda_4\ell_2/\mu^3$ in $U$, in the variables $z$, is thus the Hamiltonian without it, in the variables $z'$: the two describe the same dynamics, and a constant of motion of one, expressed in the variables $z'$, is a constant of motion of the other. Integrability through $O(S^4)$ is therefore expected to hold for every $\lambda_4$, which is what we found (its value for a black hole is known: $\lambda_4=-\frac1{40}$ \cite{ScheopnerVines.24}). The $O(S^5)$ remainder is a scalar of order $N=5$, hence a combination of the 24 couplings below, so that the $O(S^5)$ result is also independent of $\lambda_4$.

\vspace{5mm}\paragraph{Couplings at $N=5$.} The 24 possible couplings $X_1,\ldots,X_{24}$ are listed in \blue{the End Matter}. In vacuum, the spatial gradient $\nabla_c\mcB_{de}$ splits into its symmetric trace-free part and a curl part, equal to $\nabla_t\mcE$ by the Bianchi identity (and likewise for $\nabla\mcE$). The ten quadratic couplings of Ref.~\cite{ScheopnerVines.24} at this order are written instead with the symmetric trace-free octupolar tidal tensors $\mcE_{abc}$, $\mcB_{abc}$ and the time derivatives $\dot\mcE_{ab}$, $\dot\mcB_{ab}$. Denoting by $\mathcal O_{5a},\ldots,\mathcal O_{5j}$ the structures multiplying their $D_{5a},\ldots,D_{5j}$ [their Eq.~(8.67)], without numerical pre-factors, we find
\begin{equation}\label{O5map}
\begin{aligned}
&\mathcal O_{5a}=X_{15}, \quad \mathcal O_{5b}=\tfrac13(2X_{16}+X_{18}), \quad \mathcal O_{5c}=\tfrac13(X_{17}+2X_{19}), \quad \mathcal O_{5d}=X_3, \quad \mathcal O_{5e}=\tfrac13(2X_4+X_6),\\
&\mathcal O_{5f}=\tfrac13(X_5+2X_7), \quad \mathcal O_{5g}=X_6-X_4, \quad \mathcal O_{5h}=X_7-X_5, \quad \mathcal O_{5i}=X_{16}-X_{18}, \quad \mathcal O_{5j}=X_{17}-X_{19},
\end{aligned}
\end{equation}
up to the orientation of $\varepsilon_{abcd}$ in the last four. The map \eqref{O5map} is linear and invertible: $X_3,\ldots,X_7$ and $X_{15},\ldots,X_{19}$ mix the octupolar and time-derivative pieces, but span the same space. 
In practice, the couplings were expanded in the nineteen scalars of Sec.~\ref{sec:kerr} through an equivalent complex basis. Setting $\Omega_{ab}:=\mcE_{ab}+\ui\mcB_{ab}$ (which is algebraic in $(\mcC,Z_{ab})$), that basis consists of the real and imaginary parts of $\nabla_s\nabla_s\nabla_s\Omega_{ss}$, $s^2\nabla_t\nabla_t\nabla_s\Omega_{ss}$ (linear in curvature), and of the five structures $X_{ss}\nabla_sY_{ss}$, $s^2(X\nabla_sY)_{ss}$, $s^4\mathrm{tr}(X\nabla_sY)$, $s^2X^c{}_s\nabla_cY_{ss}$ and $s^4X^{cd}\nabla_cY_{ds}$ with $(X,Y)=(\Omega,\Omega)$ and $(\Omega,\bar\Omega)$ (quadratic in curvature). Since $\Omega\nabla\Omega$ and $\Omega\nabla\bar\Omega$ contain $\mcE\nabla\mcE\mp\mcB\nabla\mcB$ and $\mcE\nabla\mcB\pm\mcB\nabla\mcE$, this basis spans the same 24-dimensional space as $X_1,\ldots,X_{24}$ in Eq.~\eqref{EM:X} (this was also checked numerically).

\vspace{5mm}\paragraph{Matching black hole couplings with the literature at $N=4$.} Our Hamiltonian naturally connects with worldline effective theories through their dynamical mass function $\mathcal M^2$ \cite{ViKuStHi.16,ScheopnerVines.24,Scheopner:2026sok}. Indeed, $\mu^2=m^2+2\mu U$ coincides with it, $m^2$ being its constant part, so that its variable part is $\delta\mathcal M^2=2\mu U$. At fourth order and in vacuum, it contains, with pure-number coefficients (Wilson coefficients) named $C_n$, $D_{n\bullet}$ and $E_{n\bullet}$ in Ref.~\cite{ScheopnerVines.24}, the linear couplings $C_4$ and $E_{4a}$, a coupling $E_{4b}\,s^2s^as^b\,\nabla^e\nabla_e\mcE_{ab}$ that is quadratic in the curvature once Penrose's equation is used, and six quadratic couplings $D_{4a,\ldots,f}$ multiplying $\mcE_{ss}^2$, $s^2(\mcE^2)_{ss}$, $s^4\mathrm{tr}\mcE^2$, $\mcB_{ss}^2$, $s^2(\mcB^2)_{ss}$, $s^4\mathrm{tr}\mcB^2$ [their Eqs.~(8.57), (8.66)]. Term by term:
\begin{itemize}
\item linear: $\kappa_4=C_4$, $\lambda_4=-E_{4a}/40$; the Kerr values $C_4=E_{4a}=1$ [their Eq.~(8.59)] give $\kappa_4=1$ and $\lambda_4=-\frac1{40}$; likewise $\kappa_5=C_5=1$;
\item quadratic: $b_i=\frac12D^{\rm eff}_{4\bullet}$, in the order of $I_{1,\ldots,6}$, with $D^{\rm eff}_{4a,4d}=D_{4a,4d}$, $D^{\rm eff}_{4b}=D_{4b}-\frac12E_{4b}$, $D^{\rm eff}_{4c}=D_{4c}+\frac16E_{4b}$, $D^{\rm eff}_{4e}=D_{4e}+\frac12E_{4b}$, $D^{\rm eff}_{4f}=D_{4f}-\frac16E_{4b}$, from $s^2s^as^b\,\nabla^e\nabla_e\mcE_{ab}=6s^2(\mcE^2)_{ss}-6s^2(\mcB^2)_{ss}-2s^4\mathrm{tr}\mcE^2+2s^4\mathrm{tr}\mcB^2$;
\item spin exponentiation of the Compton amplitude [their Eq.~(8.75)] fixes the $D_{4\bullet}$ such that all $D^{\rm eff}_{4\bullet}$ vanish, for any $E_{4b}$; shift symmetry and Teukolsky matching are consistent with it \cite{GuOcVi.19,BGKV.23,BeSh.23}. Hence $b_1=\cdots=b_6=0$ for a Kerr black hole, which is the \blue{integrable point \eqref{O4result} in the coupling space}.
\end{itemize}

\section{Constants of motion}\label{sec:KQ}
\vspace{5mm}\paragraph{Setting.} We write $H=H_0+H_2+H_3+\cdots$, $K=K_1+K_3+K_4+\cdots$ and $Q=Q_0+Q_1+Q_2+\cdots$, where the subscript is the spin degree (cf. Sec.~\ref{sec:kerr}). At order $n$, the conditions $\{K,H\}^{\rm D}=0$ and $\{Q,H\}^{\rm D}=0$ read
\begin{equation}\label{ordern}
\{K_n,H_0\}^{\rm D}_{n}+\{K_{\le n-1},H_{\le n}\}^{\rm D}_n=0\,,\qquad \{Q_n,H_0\}^{\rm D}_n+\{Q_{\le n-1},H_{\le n}\}^{\rm D}_n=0\,,
\end{equation}
where $\{F,G\}^{\rm D}_n$ is the part of spin degree $n$ of $\{F,G\}^{\rm D}$. Expanding  the denominator $\mcD$ of the Dirac correction in \eqref{Dirac} as \blue{$\mcD^{-1}=\mu^{-2}\sum_k(\mu^2-\mcD)^k\mu^{-2k}$, where $\mcD-\mu^2=\tfrac14R_{abcd}S^{ab}S^{cd}$ has spin degree two}, each bracket splits into finitely many pieces of definite spin degree. The term $\{K_n,H_0\}^{\rm D}_n$ only involves the part of the bracket that preserves the spin degree (the terms in $\{x^\alpha,p_\beta\}$, $\{p_\alpha,S^{\beta\gamma}\}$ and $\{S^{\alpha\beta},S^{\gamma\delta}\}$); the curvature term of $\{p_\alpha,p_\beta\}$ and the Dirac correction raise it by two, so that they contribute to $\{K_n,H_0\}$ only at order $n+2$. Since $H_0=-\frac12\mu^2$ does not depend on the spin, $\{F,H_0\}^{\rm D}_n$ is the derivative of $F$ along the unperturbed flow $\scL_0$ (geodesic, parallel-transported spin), for any $F$ of degree $n$.

\vspace{5mm}\paragraph{Ansatz.} At order $n$, the unknown $K_n$ or $Q_n$ is a real linear combination of monomials of the algebra, with unknown coefficients. Each monomial $\mathfrak{M}$ is written as the real or imaginary part of
\begin{equation}\label{monomial}
\blue{\frac{M^{n_M}}{\mu^{n_\mu}}\,\mcC^{-n_\mcC}\,\bar\mcC^{\,n_{\bar\mcC}}\,\prod_iY_i^{n_{Y_i}}}\,,
\end{equation}
with $Y_i$ the scalars of Sec.~\ref{sec:kerr} other than $(\mcC,\bar\mcC,\mu)$, $n_\mu,n_\mcC,n_{\bar\mcC}\in\mathbb Z$, $n_M,n_{Y_i}\in\mathbb N$, and $\sum_i d_i\,n_{Y_i}=n$ with $d_i$ the spin degree of $Y_i$. 
Of course one should not brute force these exponents to be as large as computationally possible. A careful examination of the quantities involved reveals two important features of the algebra of scalars. Two exact gradings can be used to restrict the monomials further:
\begin{itemize}
\item a binary grade, the sign flip $f_{ab}\to-f_{ab}$ of the Killing-Yano tensor, a convention under which $\xi^a$ and $Z_{ab}$ are odd while $\mcC$, $M$, $\mu$ and all geometric tensors are even; it makes $B$, $\bar B$, $\mcZ_{\xi s}$, $E_0$, $E_s$, $\mcZ_{\xi t}$ odd and the other scalars even. Since $H$ and the Poisson bracket are even, the odd and even parts of a constant of motion are separately conserved; as $K_1$ is odd (linear in $f_{ab}$) and $Q_0$ even (quadratic in $f_{ab}$), $K$ and its corrections may be taken odd, and $Q$ and its corrections even;
\item a weight counting the derivatives of the Killing-Yano tensor ($+1$ per contraction with $\xi^a$, $+2$ per power of $M$, $-1$ per power of the spin), for which every bracket satisfies $w(\{X,Y\})=w(X)+w(Y)+1$. At spin order $n$ and order \blue{$M^{n_M}$}, it fixes the number of $\xi$'s, and bounds $n_M\le n/2$ for $Q$ and $n_M\le(n-1)/2$ for $K$; dimensional analysis and homogeneity in $(p_a,S^{ab})$ then fix the total power of $(\mathcal C,\bar{\mathcal C})$ and the power of $\mu$.
\end{itemize}

Another another way to drastically reduce the Ansatz' length is to only use candidate monomials that are generated from the source, i.e., $\{K_{\le n-1},H_{\le n}\}^{\rm D}_n$ (and from each coupling column, see below). A monomial is kept if its derivative along $\scL_0$ can produce one of them. Allowing terms forbidden by the gradings also provided a useful consistency check: that their coefficients come out zero. 

\vspace{5mm}\paragraph{Exact solve.} The couplings of $H$ at order $n$ (e.g., $\kappa_4,b_1,\ldots$) enter the source linearly and are added to the unknowns (the coefficients of the Ansatz \eqref{monomial}). Evaluating \eqref{ordern} at a phase-space point gives one linear equation for the unknowns, and many points give a linear system of the form $\mathbb A\,\vec x=\vec y$, with $\vec x$ the vector of unknown coefficients and couplings, $\mathbb A$ the matrix of the system and $\vec y$ the inhomogeneous column. The brackets are evaluated at phase space points with rational coordinates $(r,\cos\theta,p_\alpha,S^{\alpha\beta})$, with $S^{\alpha\beta}p_\beta=0$ exactly. To avoid the proliferation of large rational numbers, all equations are computed modulo a large prime number $\mathfrak{p}=2^{31}-1$ and $\mu$ is represented by a square root of $\mu^2$ modulo $\mathfrak p$. 

Phase space points are added to the system until the rank of $\mathbb A$ stops growing. The system is then solvable if and only if appending $\vec y$ to $\mathbb A$ does not raise the rank. The solutions then form an affine space, any one solution plus $\ker\mathbb A$; the couplings are determined when every vector of $\ker\mathbb A$ has vanishing coupling components, i.e., when they take the same value in all solutions. They are first obtained modulo $\mathfrak{p}$, and the rational numbers they represent are recovered by rational reconstruction (extended Euclidean algorithm). The solution is then pasted back into the Ansatz for $K_n$ and $Q_n$, which can then be checked independently (cf. next paragraph). The Ansätze have 288 and 2016 real coefficients for $K_3$ and $Q_3$, 3468 and 5710 for $K_4$ and $Q_4$. At $n=3$ and $n=4$, the systems are solvable only at $\kappa_3=1$ and at \blue{Eq.~\eqref{O4result}}, respectively, which also makes the parity-odd couplings discussed in Sec.~\ref{sec:couplings} vanish. The solution found is unique up to functions of the other constants (e.g., $Q\to Q+E\,K$) and up to the relations among the scalars. The constants are are given explicitly in the End Matter. 

At $n=5$, an Ansatz with up to 6090 unknown coefficients was used for $K_5$, the couplings of $H$ at that order being also added to the unknowns.  The resulting linear system for $K_5$ has rank 2916, and 2917 once the source is appended, whatever the couplings: the system has no solution. This motivated the test of Sec.~\ref{sec:num}, which does not rely on any Ansatz. 

\vspace{5mm}\paragraph{Checks.} The constants of motion $(H,K,Q)$ given in the End Matter were checked independently, with the Kerr geometry and the nineteen scalars built in components from the Killing-Yano tensor and \blue{Eq.~\eqref{Dirac}} evaluated in components (the brackets among the scalars agree to 40 digits): at random phase-space points, with every scalar of spin degree $d$ scaled by $\epsilon^d$, $\{K,H\}^{\rm D}$, $\{Q,H\}^{\rm D}$ and $\{K,Q\}^{\rm D}$ vanish identically through $\epsilon^4$, while controls fail where they should ($\{K_1,H\}^{\rm D}\neq0$ at $\epsilon^3$; $\{K_{\le4},H_{\le3}\}^{\rm D}\neq0$ at $\epsilon^4$). Adding $\lambda_4\ell_2/\mu^3$ to $U$ and transforming the constants by the canonical transformation generated by $-\lambda_4\chi/\mu^2$ (Sec.~\ref{sec:couplings}) restores all brackets through $\epsilon^4$, as stated \blue{in the main text}. Functional independence of $(H,E,L,Q,K)$ follows as in Ref.~\cite{Ra.Iso.Dru.IntegO2.26}: the geodesic constants are independent where $\mathrm{Im}[(Z_{ab}p^a\xi^b)^2]\neq0$, this survives spin corrections perturbatively, and $K$ depends on the direction of the spin, absent from the geodesic phase space; the argument is unchanged at $O(S^3)$ and $O(S^4)$.

\section{Resonant averages}\label{sec:num}
\vspace{5mm}\paragraph{What is computed.} On a resonant torus (Sec.~\ref{sec:geo}), for a closed orbit labeled by $(\varphi_0,\psi_0)$, we compute the one-period average of \blue{Eq.~\eqref{harm}},
\begin{equation}\label{avg}
\langle\sigma_n\rangle(\varphi_0,\psi_0)=\frac1T\int_0^T\sigma_n\,\Sigma\,\ud\lambda\,,
\end{equation}
where $\sigma_n$ is the source of \blue{Eq.~\eqref{O5cond}} at spin order $n$ (the couplings of order $N$ enter $\sigma_n$ for $n\ge N$), for $Q$ or $K$, evaluated along $\scL_0$, and $\Sigma\,\ud\lambda=\ud\tau=\mu\,\ud s_h$ along $\scL_0$ (since $\ud x^a/\ud s_h=p^a$ there), so that \eqref{avg} is $\mu$ times the average over the Hamiltonian time. Since $\sigma_n$ is affine in the order-$n$ couplings, one orbit gives its average for every Hamiltonian of the family: the coupling-independent part and each coupling column are averaged once, then combined linearly.

\vspace{5mm}\paragraph{Spin harmonics.} The spin vector $s^a$ enters $\sigma_n$ polynomially, with degree $n$, and along $\scL_0$ its components in the Marck tetrad depend on $\psi_0$ only through $\cos(\psi_0+\psi(\lambda))$ and $\sin(\psi_0+\psi(\lambda))$. Hence $\langle\sigma_n\rangle$ is a trigonometric polynomial of degree $n$ in $\psi_0$, which is written
\begin{equation}
\langle\sigma_n\rangle(\varphi_0,\psi_0)=\sum_{|m|\le n}\langle\sigma_n\rangle_m(\varphi_0)\,e^{\ui m\psi_0}\,,
\end{equation}
with $2n+1$ unknown amplitudes $\langle\sigma_n\rangle_m(\varphi_0)$ with $m=-n,\ldots,+n$. At every order $n\le5$, we compute \eqref{avg} at the eleven equally spaced phases $\psi_{0,k}=2\pi k/11$, $k=0,\ldots,10$, and take the discrete Fourier transform,
\begin{equation}
\langle\sigma_n\rangle_m(\varphi_0)=\frac1{11}\sum_{k=0}^{10}e^{-\ui m\psi_{0,k}}\,\langle\sigma_n\rangle(\varphi_0,\psi_{0,k})\,,
\end{equation}
which gives the amplitudes exactly, since no harmonic beyond $|m|=n$ exists and $11\ge2n+1$. Eleven one-period integrations per orbit thus give the test for all $\psi_0$ at once. Since $\sigma_n$ is real, $\langle\sigma_n\rangle_{-m}=\overline{\langle\sigma_n\rangle_m}$, and only $m\ge0$ is kept. A resonant harmonic ($mw\in\mathbb Z$) gives one real condition for $m=0$ and two for $m>0$.

\vspace{5mm}\paragraph{Numerical integration.} The geodesic is the exact Kerr solution in Mino time obtained from the Black Hole Perturbation Toolkit \cite{BHPToolkit}, and the spin is parallel-transported along it by numerical integration, both in 40-digit arithmetic with relative and absolute error goals $10^{-26}$ (both the orthonormality $s_a p^a=0$ and the parallel-transport closure residuals are below $2\times10^{-23}$); the explicit Marck precession formula \cite{VdM.20} is used to tune the tori via orbital eccentricity. The integral \eqref{avg} is evaluated with a classical trapezoidal rule on $N_{\rm n}=128$ equally spaced nodes in $[0,T]$. 

For a resonant harmonic ($mw\in\mathbb Z$), the integrand returns to its initial value after $T$, the spin phase having advanced by $2\pi mN_\psi$, an integer multiple of $2\pi$. The trapezoidal rule then converges exponentially with the number of nodes. On the torus $T_2$, the error relative to the size of the integrand falls from $10^{-1}$ with 16 nodes to $4\times10^{-19}$ with 128 nodes, as estimated by comparison with 256 nodes. We use 128 nodes throughout, as a compromise between \blue{computing cost and our ability to reach a numerical zero far below the tested values}.

\vspace{5mm}\paragraph{Relative norm and numerical zero.} The relative norm $\mathcal R_n$ plotted in \blue{Fig.~\ref{fig:data}} and \blue{the numerical zero} of each torus are defined in \blue{the End Matter}. Their values are given in Table~\ref{tab:vals}.

\vspace{5mm}\paragraph{Benchmark and fits.} At $n\le4$, on the torus $T_2$, the resonant amplitudes of $Q$ ($n=2,3,4$) and $K$ ($n=3,4$) at the six sampled $\varphi_0$ and $m=0,3$ give 90 real equations, affine in $\delta\kappa_k:=\kappa_k-1$ ($k=2,3,4$). For each $\kappa_k$ \blue{set to the non-black-hole value} $2$ (Kerr values are $\kappa_k=1)$, their least-squares solution returns $\kappa_k=1$ to $10^{-20}$ and $\mathcal R_n$ to \blue{the numerical zero} (diamonds on the left of \blue{Fig.~\ref{fig:data}}). 

At $n=5$, the resonant amplitudes of all tori, sampled phases and both constants give 486 real equations, which form the linear system $\blue{\vec z+\mathbb L\,\vec c=\vec 0}$ for the unknown vector $\vec c:=(c_1,\ldots,c_{24})\in\RR^{24}$ of the $N=5$ couplings ($\kappa_5$ among them). Here $\blue{\vec z}\in\RR^{486}$ collects the averages of the source for the integrable Hamiltonian through $O(S^4)$ ($\kappa_2=\kappa_3=\kappa_4=1$, $b_1=\cdots=b_6=0$, $\lambda_4=0$), with all $N=5$ couplings set to zero, and the columns of the $486\times24$ matrix $\blue{\mathbb L}$ are the averages of the 24 coupling terms. The black-hole values of the $N=5$ couplings are not settled (see main text); the circles on the right of \blue{Fig.~\ref{fig:data}} correspond to the natural choice $\kappa_5=1$, all other $N=5$ couplings zero. 

Like $\lambda_4$ (Sec.~\ref{sec:couplings}), the coefficient $\lambda_5$ of the linear coupling $s^2\nabla_t\nabla_t\nabla_s\mcB_{ss}$ plays no role: along $\scL_0$, this coupling is $\{\chi_5,H_0\}^{\rm D}+O(S^6)$ with $\chi_5:=\frac{s^2}{\mu}\nabla_t\nabla_s\mcB_{ss}$, so that it only redefines the constants of motion by a canonical transformation, and its averages vanish on every closed orbit (the same holds for its parity-odd variant with $\mcE_{ab}$). The matrix $\mathbb L$ has rank 20. Two of its four null directions are the columns of these two couplings; the other two are combinations of couplings whose averages also vanish on every orbit, and none of them affects the conclusion.

Appending $\vec z$ to $\mathbb L$ adds a singular value $4\times10^8$ times above the numerical zero, so that $\vec z$ is not in the span of the columns of $\mathbb L$. The held-out points of \blue{Fig.~\ref{fig:data}} use the same fit on eleven tori, evaluated on the twelfth. Fitting the rows of $Q$ alone, or of $K$ alone, is equally inconsistent, while the rows of any single torus alone can be fitted to the numerical zero.

\vspace{5mm}\paragraph{Functions of the constants.} Through $O(S^4)$, $Q$ and $K$ are defined only up to functions of the other constants (e.g., $Q\to Q+E\,K$), so the relevant question at $O(S^5)$ is whether \emph{some} function $f(Q,K,E,L,m,s)$ extends. We can answer this already with our available data on tests for $K$ and $Q$. Indeed, since $E$, $L$, $m$ and $s$ are exactly conserved, one has $\{X,H\}^{\rm D}=0$ for any $X\in(E,L,m,s)$, and the chain rule for any function $f:\mcP\rightarrow\RR$ reads 
\begin{equation}\label{LeibQK}
\{f,H\}^{\rm D}=\frac{\partial f}{\partial Q}\,\{Q,H\}^{\rm D}+\frac{\partial f}{\partial K}\,\{K,H\}^{\rm D},
\end{equation}
whose leading part is $\frac{\partial f}{\partial Q}\,\sigma_5(Q)+\frac{\partial f}{\partial K}\,\sigma_5(K)$. On each torus $T$, the two partial derivatives in \eqref{LeibQK} take constant values, so the resonant condition for $f$ is that of $Q+\alpha_T K$, with $\alpha_T$ the ratio of these values on $T$ (the case $\partial f/\partial Q=0$ is the fit to $K$ alone). We therefore fitted the 24 couplings, common to all tori, together with one free $\alpha_T$ per torus where $K$ is tested; letting the $\alpha_T$ vary independently only relaxes the condition. Minimizing over the $\alpha_T$ leaves on every torus a residual at least $10^{4}$ times \blue{the numerical zero}, and at least $10^{7}$ times on the tori where $K$ is tested: no function of $Q$, $K$ and the exact constants extends to $O(S^5)$.

\vspace{5mm}\paragraph{Tori.} \blue{We use the twelve tori\footnote{Only three tori ($T_2$, $T_4$, $T_8$) would have been enough to make the fit inconsistent and thus reach our conclusions.} of Table~\ref{tab:tori}, with several orbits and spin windings each, so that the conclusion does not depend on the choice of data. On every torus, the relative norm $\mathcal R_5$ of $Q$, and of $K$ where it is tested, lies at least eight decades above the numerical zero of that torus, both for Kerr's linear couplings alone and for the couplings fitted to all tori, and no torus is needed on its own. Three features are worth mentioning about these different tori. \emph{(i) Harmonics.} On a resonant torus, only the spin harmonics with $mw\in\mathbb Z$ survive the average over the closed orbit $[0,qT]$; for them, this average equals the one-period average \eqref{avg}, which is the one we compute. The other harmonics are discarded, so that each torus probes only the harmonics allowed by its $w$. \emph{(ii) Constant $K$.} The harmonic $m=0$ carries no information on $K$: at leading order, $K$ is the action conjugate to the Marck angle $\psi$ \cite{WitzHJ.19,WitzAA.22}, so that its bracket with any function $X$ is $\partial X/\partial\psi$, whose $m=0$ harmonic vanishes; numerically, the whole source of $K$ at $m=0$ (constant term and all 24 coupling columns) is at the numerical zero. On $T_3$, $T_6$, $T_7$, $T_9$ and $T_{12}$, the only other resonant harmonic is $m=4$ or $5$, at which the source of $K$ is also at the numerical zero. These five tori therefore test only $Q$, and $K$ is not shown for them in Fig.~2. \emph{(iii) Precision.} All twelve tori are shown, without selection. On $T_9$, $T_{11}$ and $T_{12}$, the numerical zero is higher ($10^{-16.5}$ to $10^{-15}$ of the integrand, against $10^{-24}$ to $10^{-20}$ on the other tori), because their 128-node quadrature is not yet converged: the numerical zero still falls by about ten decades between 64 and 128 nodes, while the closure residual of orbit and spin is below $4\times10^{-23}$ on every torus. These three tori are thus weaker tests, but none of them is needed for the conclusion.}

\begin{table}[h!]
\centering
\begin{tabular}{cccccccccc}
\hline\hline
Torus & $a$ & $x$ & $n_r{:}n_\theta$ & $N_\psi$ & $w$ & $e$ & $p$ & $m>0$ & Tested \\
\hline
$T_{1}$  & $0.6$ & $-0.5$ & $2{:}5$ & $11/3$ & $2/3$ & $0.5565886$ & $8.552486$ & $3$ & $Q,K$ \\
$T_{2}$  & $0.6$ & $0.5$  & $2{:}5$ & $11/3$ & $2/3$ & $0.4904538$ & $5.999897$ & $3$ & $Q,K$ \\
$T_{3}$  & $0.6$ & $0.5$  & $2{:}5$ & $15/4$ & $3/4$ & $0.1447617$ & $5.760590$ & $4$ & $Q$ \\
$T_{4}$  & $0.6$ & $0.8$  & $1{:}3$ & $11/5$ & $1/5$ & $0.2023094$ & $4.804310$ & $5$ & $Q,K$ \\
$T_{5}$  & $0.6$ & $0.8$  & $2{:}5$ & $11/3$ & $2/3$ & $0.5422138$ & $5.351101$ & $3$ & $Q,K$ \\
$T_{6}$  & $0.6$ & $0.8$  & $2{:}5$ & $15/4$ & $3/4$ & $0.2566068$ & $5.112347$ & $4$ & $Q$ \\
$T_{7}$  & $0.9$ & $-0.5$ & $2{:}5$ & $15/4$ & $3/4$ & $0.1367973$ & $8.554932$ & $4$ & $Q$ \\
$T_{8}$  & $0.9$ & $0.5$  & $2{:}3$ & $5/2$  & $1/2$ & $0.5391344$ & $7.861822$ & $2,4$ & $Q,K$ \\
$T_{9}$  & $0.9$ & $0.5$  & $2{:}5$ & $18/5$ & $3/5$ & $0.3366981$ & $4.759406$ & $5$ & $Q$ \\
$T_{10}$ & $0.9$ & $0.5$  & $3{:}4$ & $7/2$  & $1/2$ & $0.5099527$ & $10.36894$ & $2,4$ & $Q,K$ \\
$T_{11}$ & $0.9$ & $0.5$  & $3{:}5$ & $4$    & $0$   & $0.5659255$ & $6.667993$ & $1,\ldots,5$ & $Q,K$ \\
$T_{12}$ & $0.9$ & $0.8$  & $2{:}5$ & $18/5$ & $3/5$ & $0.4286523$ & $3.643120$ & $5$ & $Q$ \\
\hline\hline
\end{tabular}
\caption{The twelve resonant tori: Kerr spin $a$ (units of $M$), inclination $x$, orbital resonance $n_r{:}n_\theta$, $N_\psi=T/\Lambda_\psi$ and its fractional part $w$, eccentricity $e$ and semi-latus rectum $p$ (units of $M$, given to seven digits), resonant harmonics $m>0$ ($m=0$ is always resonant), and constants tested. $T_2$ is the torus of \blue{Fig.~\ref{fig:torus}}, also used at $n\le4$.}
\label{tab:tori}
\end{table}

\begin{table}[h!]
\centering
\begin{tabular}{cccccccc}
\hline\hline
Torus & $Q$ Kerr & $Q$ best & $Q$ held-out & $K$ Kerr & $K$ best & $K$ held-out & \,\,\blue{Num. zero}\,\,\\
\hline
$T_{1}$  & $-3.1$ & $-2.4$ & $-2.3$ & $-3.7$ & $-3.0$ & $-2.9$ & $-23.1$\\
$T_{2}$  & $-2.7$ & $-3.7$ & $-2.6$ & $-3.1$ & $-4.5$ & $-3.0$ & $-20.3$\\
$T_{3}$  & $-5.6$ & $-6.9$ & $-6.6$ & --     & --     & --     & $-22.2$\\
$T_{4}$  & $-3.1$ & $-4.4$ & $-3.4$ & $-3.8$ & $-5.1$ & $-4.1$ & $-23.6$\\
$T_{5}$  & $-2.9$ & $-3.8$ & $-2.5$ & $-3.1$ & $-4.8$ & $-2.8$ & $-22.2$\\
$T_{6}$  & $-4.6$ & $-6.4$ & $-5.7$ & --     & --     & --     & $-23.5$\\
$T_{7}$  & $-6.2$ & $-7.3$ & $-7.2$ & --     & --     & --     & $-21.4$\\
$T_{8}$  & $-1.5$ & $-2.4$ & $-1.6$ & $-2.3$ & $-3.1$ & $-3.1$ & $-23.5$\\
$T_{9}$  & $-4.3$ & $-7.0$ & $-5.0$ & --     & --     & --     & $-15.4$\\
$T_{10}$ & $-2.0$ & $-2.8$ & $-2.8$ & $-2.6$ & $-3.8$ & $-3.7$ & $-23.0$\\
$T_{11}$ & $-1.4$ & $-2.2$ & $-1.7$ & $-2.0$ & $-2.7$ & $-2.5$ & $-15.1$\\
$T_{12}$ & $-3.8$ & $-7.3$ & $-4.8$ & --     & --     & --     & $-16.5$\\
\hline\hline
\end{tabular}
\caption{$\log_{10}\mathcal R_5$ plotted on the right of \blue{Fig.~\ref{fig:data}}: \blue{Kerr's linear couplings alone ($\kappa_5=1$, all others zero)}, best fit of the 24 couplings to all tori (equal weight), and the same fit with the torus left out (held-out). A dash marks $K$ at \blue{the numerical zero} at the only nonzero resonant harmonic. Last column: numerical zero of the torus, $\log_{10}$ of the control relative to the same integrand scale (red lines of \blue{Fig.~\ref{fig:data}}).}
\label{tab:vals}
\end{table}

\FloatBarrier
\section{Schwarzschild limit}\label{sec:schw}
\vspace{5mm}\paragraph{Carter constant.} At $a=0$, the three rotational Killing vectors $k_{(i)}^a$ of Schwarzschild give exactly conserved charges $L_i=p_ak_{(i)}^a+\frac12S^{ab}\nabla_ak_{(i)b}$ for any conservative moments, and $L^2:=L_x^2+L_y^2+L_z^2$ Poisson-commutes with any Hamiltonian built from the Schwarzschild geometry. We verified that $Q_{\le4}=-L^2$ at $a=0$, order by order in the spin: $Q_n+L^2_n$ vanishes for $n=0,1,2$, as found at $n\le2$ in \cite{Ra.Iso.Dru.IntegO2.26}, and $Q_3$, $Q_4$ vanish identically at $a=0$. Hence, in Schwarzschild, $Q$ extends at every order ($Q_5=0$) for any couplings, and only $K$ remains to be tested. As a consistency check of our pipeline, the computed source $\sigma_5$ of $Q$, together with its 24 coupling columns, vanishes at every node of the orbits below.

\vspace{5mm}\paragraph{Resonances at $a=0$.} The integrand $\sigma_5$ is then invariant under rotations and does not depend on the initial polar phase $\varphi_0$. On a torus with resonance $n_r:n_\theta$ and spin winding $N_\psi=T/\Lambda_\psi$, a spin harmonic $m$ therefore survives the average only if $mN/n_r\in\mathbb Z$ (instead of $mN\in\mathbb Z$), and the polar phase $\varphi_0$ is redundant (we tested two values of $\varphi_0$ on each run, giving the same amplitudes to $10^{-21}$). We sample instead the orientation of the spin with respect to the orbital angular momentum, i.e., the component $s_\parallel$ along $\ell^a$. This changes the torus of the unperturbed flow $\scL_0$, through the value of the exactly conserved $K_1\propto s_\parallel$, but not the orbit being averaged.

\vspace{5mm}\paragraph{Tori and result.} We use the three tori of Table~\ref{tab:schw}, at $x=1/2$ (irrelevant at $a=0$) with 128 nodes, five spin orientations on $T_{13},T_{14}$ and three on $T_{15}$, and enough values of $\psi_0$ to resolve every surviving harmonic. The source $\sigma_5$ of $K$ has even harmonics $m$ only; odd harmonics constrain the couplings but not the source. We condense the resonant amplitudes into $\mathcal R_5$ exactly as in Kerr, the sum over $\varphi_0$ being replaced by a sum over the spin orientations, and fit the 24 couplings to all tori, each weighted by the norm of its own source. The linear couplings of a Kerr test body alone ($\kappa_5=1$, all others zero) give $\mathcal R_5$ between $10^{-4.5}$ and $10^{-3.2}$; the best fit gives $\mathcal R_5$ between $10^{-6.8}$ and $10^{-5.0}$, and couplings fitted on two tori fail on the third by the same amount. Every torus thus lies at least seventeen decades above its own numerical zero. Hence, in Schwarzschild as in Kerr, $K$ does not extend to $O(S^5)$, whatever the couplings at that order, while $Q=-L^2$ does.

\begin{table}[h!]
\centering
\begin{tabular}{cccccccccc}
\hline\hline
Torus & $n_r{:}n_\theta$ & $N_\psi$ & $e$ & $p$ & $m$ & $K$, $\kappa_5{=}1$ & $K$ best & $K$ held-out & \blue{zero}\\
\hline
$T_{13}$ & $2{:}7$  & $5$  & $0.3760085$ & $6.883850$ & $2,4$        & $-4.2$ & $-6.4$ & $-6.1$ & $-24.1$\\
$T_{14}$ & $3{:}8$  & $6$  & $0.2307633$ & $7.073153$ & $1,\ldots,5$ & $-4.5$ & $-6.8$ & $-6.4$ & $-24.3$\\
$T_{15}$ & $5{:}14$ & $10$ & $0.6437946$ & $7.510931$ & $1,\ldots,5$ & $-3.2$ & $-5.0$ & $-4.4$ & $-24.0$\\
\hline\hline
\end{tabular}
\caption{Schwarzschild ($a=0$) resonant tori and $\log_{10}\mathcal R_5$ for $K$: orbital resonance $n_r{:}n_\theta$, $N_\psi=T/\Lambda_\psi$ (integer on all three tori), eccentricity $e$ and semi-latus rectum $p$ (units of $M$), surviving harmonics $m>0$ ($mN_\psi/n_r\in\mathbb Z$), then the \blue{linear couplings of a Kerr test body alone} ($\kappa_5=1$, all others zero), best fit of the 24 couplings to all tori, the same fit with the torus left out (held-out), and \blue{the numerical zero} of the torus, as in Table~\ref{tab:vals}.}
\label{tab:schw}
\end{table}

\FloatBarrier

\end{document}